\documentclass[12pt,a4paper]{article}

\usepackage{inputenc,authblk,comment,color,multicol,natbib}
\usepackage{graphics}
\usepackage{graphicx}
\usepackage{colortbl}
\usepackage{xcolor}
\usepackage{lscape}
\usepackage{rotating}
\usepackage{siunitx}
\usepackage{amsmath,amssymb}

\usepackage[hidelinks]{hyperref}

\usepackage{tikz}
\usepackage{academicons} 

\definecolor{lime}{HTML}{A6CE39}
\DeclareRobustCommand{\orcidicon}{
	\begin{tikzpicture}
	\draw[lime, fill=lime] (0,0) 
	circle [radius=0.16] 
	node[white] {{\fontfamily{qag}\selectfont \tiny ID}};
	\draw[white, fill=white] (-0.0625,0.095) 
	circle [radius=0.007];
	\end{tikzpicture}
	\hspace{-2mm}
}

\bibpunct{(}{)}{;}{a}{}{,}

\begin{document}

\begin{titlepage}
   \begin{center}
       \vspace*{1cm}

       \Large
       \textbf{Sciences with Thai National Radio Telescope}

       \vspace{1cm}
       \normalsize
       \textbf{Editors}
       
       \vspace{0.3cm}
        Jaroenjittichai, Phrudth\href{https://orcid.org/0000-0002-9676-5424}{\hspace{-1.5mm}\orcidicon}$^{1,\ast}$,
        Sugiyama, Koichiro\href{https://orcid.org/0000-0002-6033-5000}{\hspace{-1.5mm}\orcidicon}$^{1,2}$, 
        Kramer, H. Busaba\href{https://orcid.org/0000-0001-8168-5141}{\hspace{-1.5mm}\orcidicon}$^{3,1}$,
        \\and Soonthornthum, Boonrucksar\href{https://orcid.org/0000-0001-5511-7183}{\hspace{-1.5mm}\orcidicon}$^{1}$

       \vspace{0.7cm}
       \textbf{Authors}

       \vspace{0.3cm}
        Akahori, Takuya\href{https://orcid.org/0000-0001-9399-5331}{\hspace{-1.5mm}\orcidicon}$^{2,4}$,
        Asanok, Kitiyanee\href{https://orcid.org/0000-0002-4125-0941}{\hspace{-1.5mm}\orcidicon}$^{1}$,
        Baan, Willem\href{https://orcid.org/0000-0003-3389-6838}{\hspace{-1.5mm}\orcidicon}$^{5}$,
        Bran, Sherin Hassan$^{1,6}$,
        \\
        Breen, L. Shari\href{https://orcid.org/0000-0002-4047-0002}{\hspace{-1.5mm}\orcidicon}$^{7}$,
        Cho, Se-Hyung\href{https://orcid.org/0000-0002-2012-5412}{\hspace{-1.5mm}\orcidicon}$^{8}$,
        Chanapote, Thanapol\href{https://orcid.org/0000-0002-6992-6755}{\hspace{-1.5mm}\orcidicon}$^{1}$,
        Dodson, Richard\href{https://orcid.org/0000-0003-0392-3604}{\hspace{-1.5mm}\orcidicon}$^{9}$,
        \\
        Ellingsen, P. Simon\href{https://orcid.org/0000-0002-1363-5457}{\hspace{-1.5mm}\orcidicon}$^{10}$,
        Etoka, Sandra$^{11}$,
        Gray, D. Malcolm$^{1,11}$,
        Green, A. James$^{12}$, 
        \\
        Hada, Kazuhiro\href{https://orcid.org/0000-0001-6906-772X}{\hspace{-1.5mm}\orcidicon}$^{13}$,
        Halson, Marcus$^{1}$,
        Hirota, Tomoya\href{https://orcid.org/0000-0003-1659-095X}{\hspace{-1.5mm}\orcidicon}$^{2}$,
        Honma, Mareki\href{https://orcid.org/0000-0003-4058-9000}{\hspace{-1.5mm}\orcidicon}$^{13}$,
        \\
        Imai, Hiroshi\href{https://orcid.org/0000-0002-0880-0091}{\hspace{-1.5mm}\orcidicon}$^{14}$,
        Johnston, Simon\href{https://orcid.org/0000-0002-7122-4963}{\hspace{-1.5mm}\orcidicon}$^{12}$,
        Kim, Kee-Tae\href{https://orcid.org/0000-0003-2412-7092}{\hspace{-1.5mm}\orcidicon}$^{8}$,
        Kramer, Michael\href{https://orcid.org/0000-0002-4175-2271}{\hspace{-1.5mm}\orcidicon}$^{3,11}$,
        Li, Di\href{https://orcid.org/0000-0003-3010-7661}{\hspace{-1.5mm}\orcidicon}$^{15}$,
        \\
        Macatangay, Ronald$^{1}$,
        Menten, M. Karl\href{https://orcid.org/0000-0001-6459-0669}{\hspace{-1.5mm}\orcidicon}$^{3}$,
        Minh, Young Chol\href{https://orcid.org/0000-0003-1742-0119}{\hspace{-1.5mm}\orcidicon}$^{8}$,
        Mkrtichian, David\href{https://orcid.org/0000-0001-5094-3910}{\hspace{-1.5mm}\orcidicon}$^{1}$,
        \\
        Pimpanuwat, Bannawit\href{https://orcid.org/0000-0001-8782-0754}{\hspace{-1.5mm}\orcidicon}$^{11}$,
        Richards, M.S. Anita\href{https://orcid.org/0000-0002-3880-2450}{\hspace{-1.5mm}\orcidicon}$^{11}$,
        Rioja, Maria\href{https://orcid.org/0000-0003-4871-9535}{\hspace{-1.5mm}\orcidicon}$^{9,16,17}$,
        \\
        Rujopakarn, Wiphu\href{https://orcid.org/0000-0002-0303-499X}{\hspace{-1.5mm}\orcidicon}$^{1,18}$,
        Sakai, Daisuke\href{https://orcid.org/0000-0003-1515-7230}{\hspace{-1.5mm}\orcidicon}$^{13,1}$,
        Sakai, Nobuyuki\href{https://orcid.org/0000-0002-5814-0554}{\hspace{-1.5mm}\orcidicon}$^{1,8}$,
        Samanso, Nattida$^{1}$,
        \\
        Sanpa-arsa, Siraprapa$^{1}$,
        Semenko, Eugene\href{https://orcid.org/0000-0002-1912-1342}{\hspace{-1.5mm}\orcidicon}$^{1}$,
        Sunada, Kazuyoshi\href{https://orcid.org/0000-0001-6675-0558}{\hspace{-1.5mm}\orcidicon}$^{13}$,
        Surapipith, Vanisa$^{1}$,
        \\
        Thoonsaengngam, Nattaporn$^{1}$, 
        Voronkov, A. Maxim\href{https://orcid.org/0000-0002-4931-4612}{\hspace{-1.5mm}\orcidicon}$^{12}$,
        Wongphecauxson, Jompoj\href{https://orcid.org/0000-0002-7730-4956}{\hspace{-1.5mm}\orcidicon}$^{3}$,
        \\
        Yadav, Ram Kesh$^{1}$,
        Zhang, Bo\href{https://orcid.org/0000-0003-1353-9040}{\hspace{-1.5mm}\orcidicon}$^{19}$,
        Zheng, Xing Wu$^{20}$
        and
        Poshyachinda, Saran$^{1}$

       \vspace{0.6cm}

       \footnotesize
        \begin{flushleft}
        $^{1}$ National Astronomical Research Institute of Thailand (Public Organization), 260 Moo 4, T. Donkaew, A. Maerim, Chiang Mai, 50180, Thailand

        $^{2}$ Mizusawa VLBI Observatory, National Astronomical Observatory of Japan (NAOJ), Mitaka, Tokyo 181-8588, Japan

        $^{3}$ Max Planck Institut f{\"u}r Radioastronomie, Auf dem H{\"u}gel 69, 53121 Bonn, Germany

        $^{4}$ SKA Observatory, Jodrell Bank, Lower Withington, Macclesfield, Cheshire SK11 9FT, UK

        $^{5}$ Netherlands Institute for Radio Astronomy ASTRON, 7991PD Dwingeloo, the Netherlands
        
        $^{6}$ Environmental Science Research Center, Faculty of Science, Chiang Mai University, Chiang Mai, 50200, Thailand
        
        $^{7}$ Sydney Institute for Astronomy (SIfA), School of Physics, University of Sydney, NSW 2006, Australia
        
        $^{8}$ Korea Astronomy and Space Science Institute, 776 Daedeok-daero, Yuseong, Daejeon 34055, Republic of Korea
        
        $^{9}$ International Centre for Radio Astronomy Research, M468, University of Western Australia, 35 Stirling Highwaym Perth 6009, Australia

        $^{10}$ School of Natural Sciences, University of Tasmania, Private Bag 37, Hobart, Tasmania 7001, Australia
        
        $^{11}$ Jodrell Bank Centre for Astrophysics, School of Physics and Astronomy, University of Manchester, M13 9PL, UK

        $^{12}$ Australia Telescope National Facility, CSIRO Space and Astronomy, PO~Box~76, Epping NSW~1710, Australia
        
        $^{13}$ Mizusawa VLBI Observatory, NAOJ, 2-12 Hoshigaoka, Mizusawa, Oshu, Iwate 023-0861, Japan

        $^{14}$ Amanogawa Galaxy Astronomy Research Center, Graduate School of Science and Engineering, Kagoshima University, 1-21-35 Korimoto, Kagoshima 890-0065, Japan

        $^{15}$ National Astronomical Observatories, Chinese Academy of Sciences, Beijing 100012, China
        
        $^{16}$ CSIRO Astronomy and Space Science, 26 Dick Perry Avenue, Kensington WA 6151, Australia
        
        $^{17}$ Observatorio Astron{\'o}mico Nacional, Alfonso XII, 3 y 5, 28014 Madrid, Spain

        $^{18}$ Department of Physics, Faculty of Science, Chulalongkorn University, 254 Phyathai Road, Patumwan, Bangkok Thailand. 10330

        $^{19}$ Shanghai Astronomical Observatory, Chinese Academy of Sciences, Shanghai 200030, China
        
        $^{20}$ School of Astronomy and Space Sciences, Nanjing University, Nanjing 210093, China
        
        \vspace{2mm}
        $^{\ast}$E-mail: phrudth@narit.or.th

        \end{flushleft}

   \end{center}
\end{titlepage}

%
\addtocounter{page}{1}

\vspace{1.5cm}
\large
\begin{center}
\textbf{Preamble}
\end{center}

\normalsize
\noindent
This White Paper summarises potential key science topics to be achieved with Thai National Radio Telescope (TNRT). The commissioning phase has started in mid 2022. The key science topics consist of ``Pulsars and Fast Radio Bursts (FRBs)", ``Star Forming Regions (SFRs)", ``Galaxy and Active Galactic Nuclei (AGNs)", ``Evolved Stars", ``Radio Emission of Chemically Peculiar (CP) Stars", and ``Geodesy", covering a wide range of observing frequencies in L/C/X/Ku/K/Q/W-bands (1--115 GHz). As a single-dish instrument, TNRT is a perfect tool to explore time domain astronomy with its agile observing systems and flexible operation. Due to its ideal geographical location, TNRT will significantly enhance Very Long Baseline Interferometry (VLBI) arrays, such as East Asian VLBI Network (EAVN), Australia Long Baseline Array (LBA), European VLBI Network (EVN), in particular via providing a unique coverage of the sky resulting in a better complete ``uv" coverage, improving synthesized-beam and imaging quality with reducing side-lobes. This document highlights key science topics achievable with TNRT in single-dish mode and in collaboration with VLBI arrays.

\vspace{3mm}
\tableofcontents


\clearpage
\section{Introduction}

Following the successful pathway of the 2.4m optical telescope (Thai National Telescope) in the last decade, National Astronomical Research Institute of Thailand (NARIT) has a 5-year plan to accelerate the development in radio astronomy under the flagship project known as Radio Astronomy Network and Geodesy for Development (RANGD), 2017-2021.
RANGD includes the development of the Thai National Radio Observatory (TNRO) and the establishment of the Radio Astronomy Operation Centre to support the facility and instrumentation development.

TNRT site is located in Huay Hong Krai Royal Development Study Centre, Chiang Mai, Thailand. Its ideal location provides excellent coverage of the sky suitable for conducting large-scale surveys. Two VLBI2010 Global Observing System (VGOS) telescopes have been approved for construction in Chiang Mai (TNRT co-location, under the collaboration with Shanghai Astronomical Observatory) and in Songkla, South of Thailand, for extensive applications in geodesy and tectonic studies in South-East Asia.

TNRT is located at latitude \ang{18;51;52} N and longitude 99$^{\circ}$13$^{\prime}$01$^{\prime \prime}$ E at 450m sea level. Despite being close to the equator, TNRT's site is suitable for K- and Q-band observations during the winter months between October to April, and for W-band, which is feasible during December and January. 

In the Era of large telescopes around the world, TNRT's single dish key science focuses on time domain astronomy, exploring transients and variability, high-cadence monitoring campaigns can be planned for  known sources or as sky surveys, such as pulsars, star-forming regions, AGNs, and Algols and CP stars. As the first large radio telescope in the region, TNRT will be a key international Very Long Baseline Interferometry (VLBI) station to several VLBI arrays.

\begin{figure}[htbp]
    \centering
    \includegraphics[clip,width=18cm]{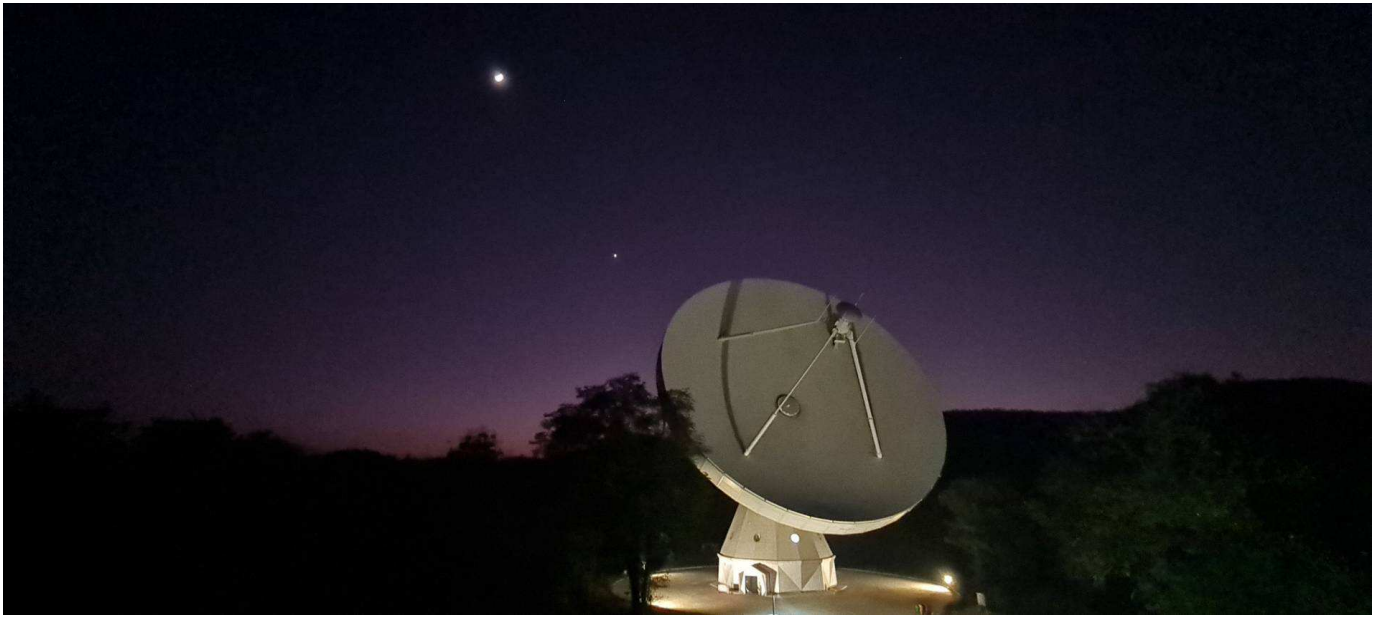}
    \caption{The Moon and Thai National Radio Telescope in December, 2021 (credit: Thanadon Paksin).}
    \label{tnrt}
\end{figure}

\vspace{-5mm}
\subsection{Specifications}\label{subsec1:specifications}
\begin{itemize}
    \item \textbf{Antenna} \\
    TNRT is based on the 40m IGN Yebes Telescope with similar Nasmyth optics (\citealt{2006LNEA....2..257L}). In addition, the Tetrapod Head Unit has been  developed for the installation of prime focus receivers. The specifications are shown in Table  \ref{tableantenna}.
    \begin{table}[htbp]
        \begin{center}
            \caption{Antenna Parameters of TNRT}
               \begin{tabular}{l c}
               \hline
                Parameter & Value \\
                \hline
                Optics & Primary/Nasmyth \\
                Diametre & 40 m \\
                f/D ratio & 0.375 \\
                Pointing accuracy & \ang{;;2} (no wind) \\
                Slew speed  & 180 ($^{\circ}/$min) (AZ) \\
                						   & 60 ($^{\circ}/$min) (EL)\\
                Surface accuracy & 150 $\mu$m (rms) \\
                frequency coverage  & 0.3 - 115 (GHz) \\
                
             \hline
    \end{tabular}
    \label{tableantenna}
    \end{center}
    \end{table}

    \item \textbf{Receivers and Backends} \\
    The L-band (1.0-1.8 GHz), K-band (18-26.5 GHz) and the Universal Software Backend (USB) are being developed under collaboration with Max Planck Insitute for Radioastronomy as the first two receivers for commissioning and early science. Parameters of the receivers and sensitivity are included in Table \ref{tablereceiver}.
    The Telescope Control Software (TCS) is based on ALMA Common Software (\footnote{https://www.eso.org/projects/alma/develop/acs/} under collaboration with Yebes Observatory, IGN. The following observation modes will be available:
    
    \begin{itemize}
        \item Pulsar with coherent dedispersion, baseband and search mode recording
        \item Spectrometer / Continuum
        \item Polarimeter 
        \item VLBI with VDIF format
    \end{itemize}
    
       Future receivers being developed or considered in the next development phase are:
    \begin{itemize}
        \item CXKu (4.5-13.5 GHz) in the design study phase
        \item QW-band (35-50 and 75-115 GHz) to be integrated with existing K-band into a simultaneous quasioptics triband system in design study phase
        \item 0.7-2.1 GHz Phased Array Feed
    \end{itemize}

\vspace{-4mm}
    \begin{table}[h]
        \begin{center}
                \caption{Receiver and Sensitivity}
                \begin{tabular}{l l l }
                \hline
                Parameters & L-band & K-band  \\
                \hline
                RF frequency (GHz) & 1.0-1.8 & 18-26.5 \\
                Centre wavelength (cm) & 21.4 & 1.36 \\
                Beam size (arcmin) & 22 & 1.4 \\
                Polarisation 	& linear & circular \\
                Instantaneous bandwidth (GHz) & 0.8 & 2 \\
                Aperture efficiency & 0.7 & 0.5 \\
                Gain (K/Jy)	& 0.32 & 0.23 \\
                Receiver temperature (K) & 13 & 20 \\
                System temperature (K) & 25 & 70 \\
                System Equivalent Flux Density (Jy) & 78 & 304 \\
                \hline
                \end{tabular}
                \label{tablereceiver}
        \end{center}
    \end{table}
\end{itemize}

\clearpage
\subsection{Impacts on VLBI}\label{subsec1.2:vlbi}

\begin{figure}[htbp]
    \centering
    \includegraphics[clip,width=16cm]{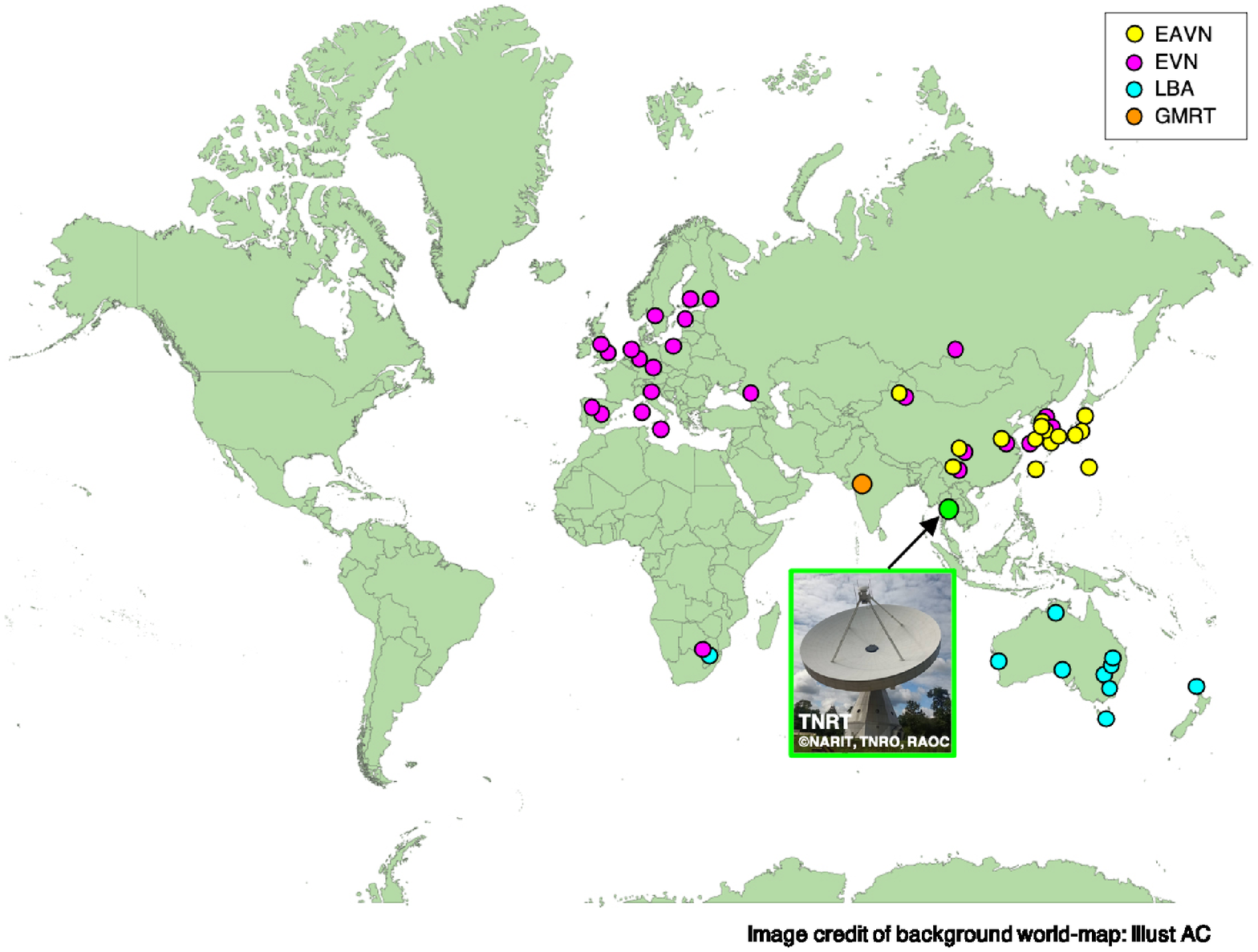}
    \caption{Locations of TNRT and Interferometric/VLBI arrays that will potentially collaborate with. Overlapping spots by different colors shows stations to be official members in multiple VLBI arrays.}
    \label{fig1.2:vlbiarray}
\end{figure}

\begin{figure}[htbp]
 \begin{minipage}[b]{0.50\linewidth}
  \centering
  \includegraphics[clip,width=\textwidth]{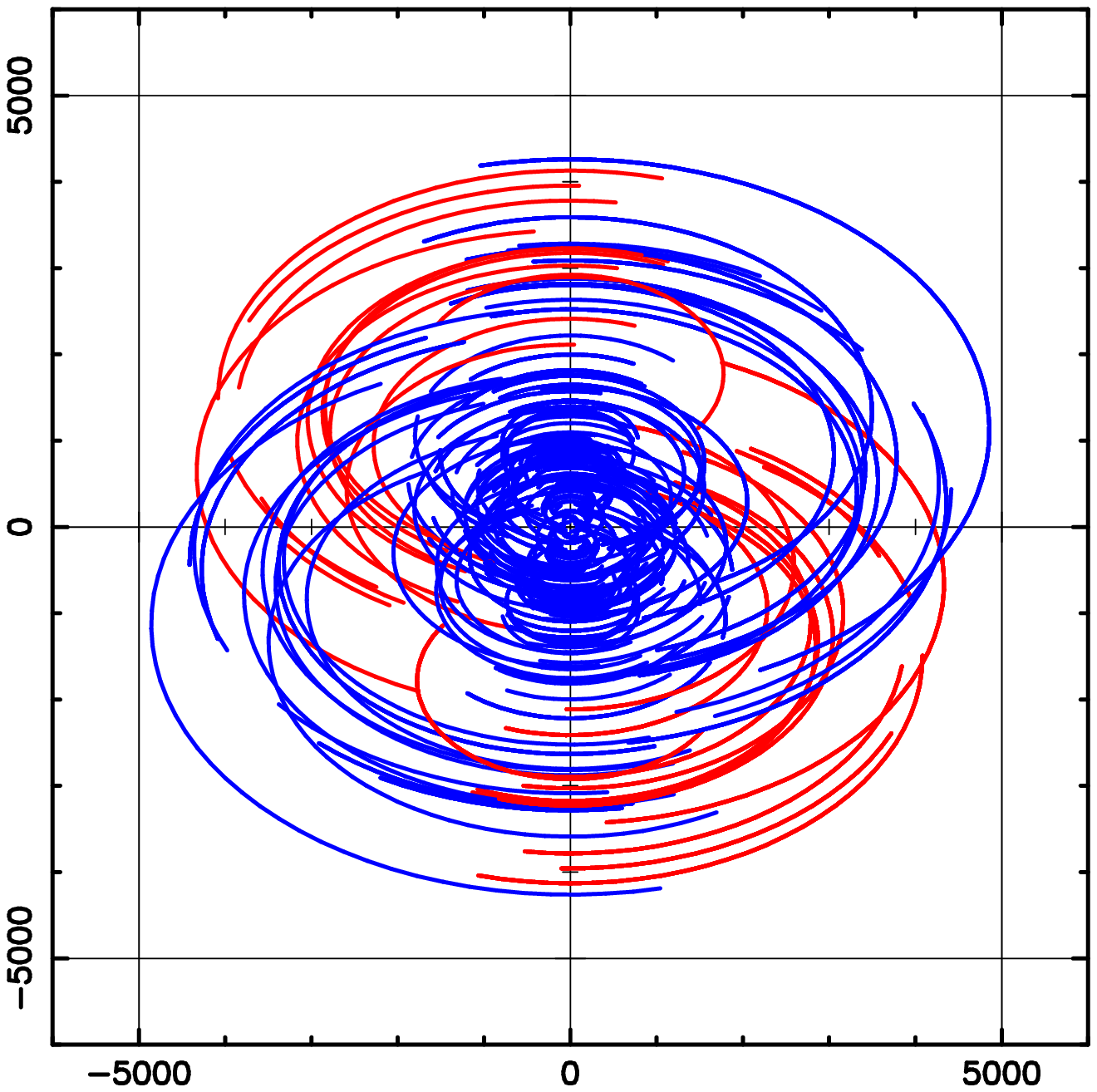}
 \end{minipage}
 \begin{minipage}[b]{0.50\linewidth}
  \centering
  \includegraphics[clip,width=\textwidth]{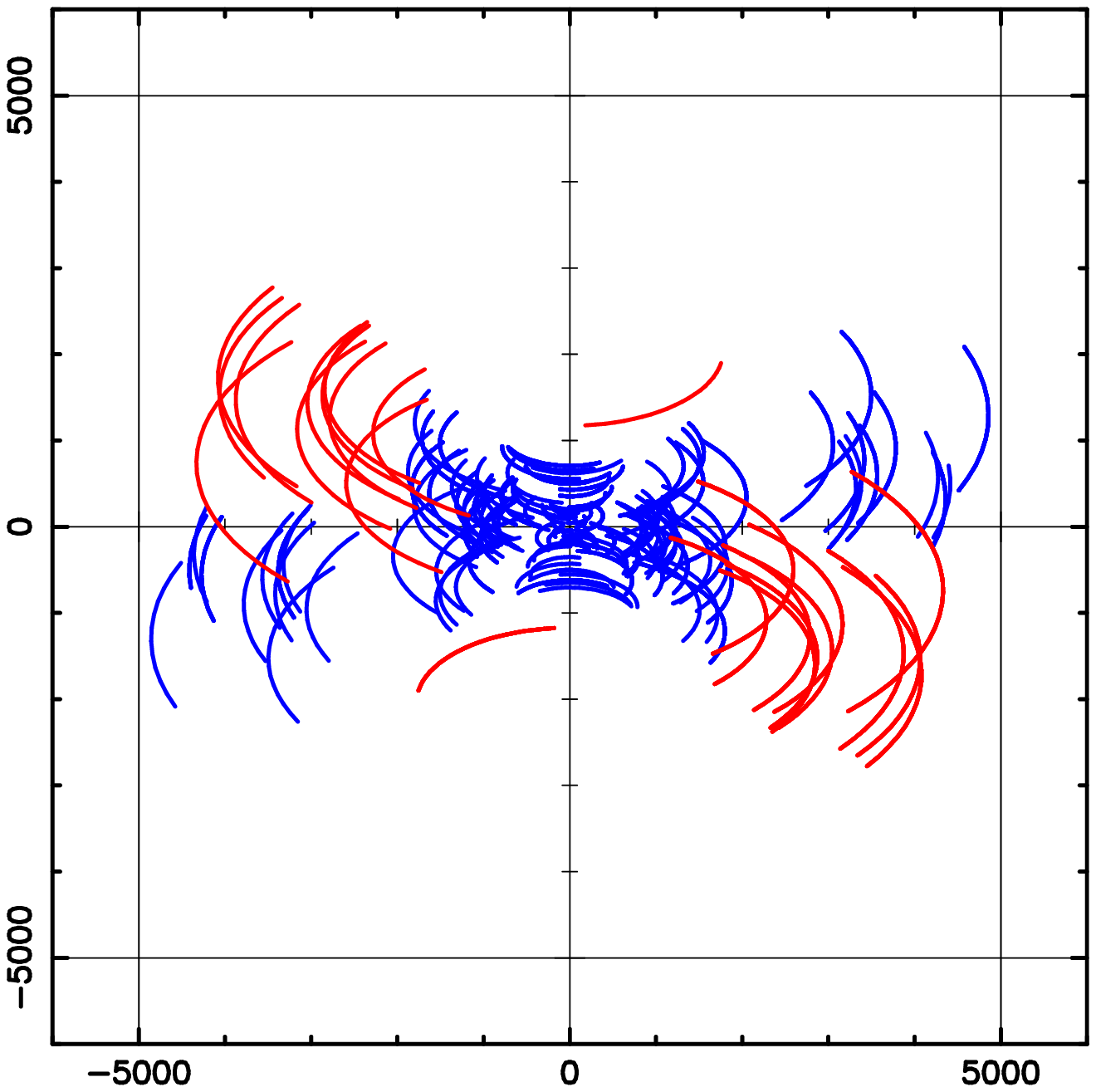}
 \end{minipage}
\caption{Simulated results of uv-coverages for EAVN$+$TNRT in the case of observing sources with declination of $+$40 and $-$29 (toward the Galactic Center) degree in K-band with an elevation limit of 10 degree. The unit of horizontal and vertical axes is kilometer. Contributions of TNRT are highlighted by a color of red.}
\label{fig1.2:uveavn}
\end{figure}

\begin{figure}[htbp]
    \centering
    \includegraphics[clip,width=15.7cm]{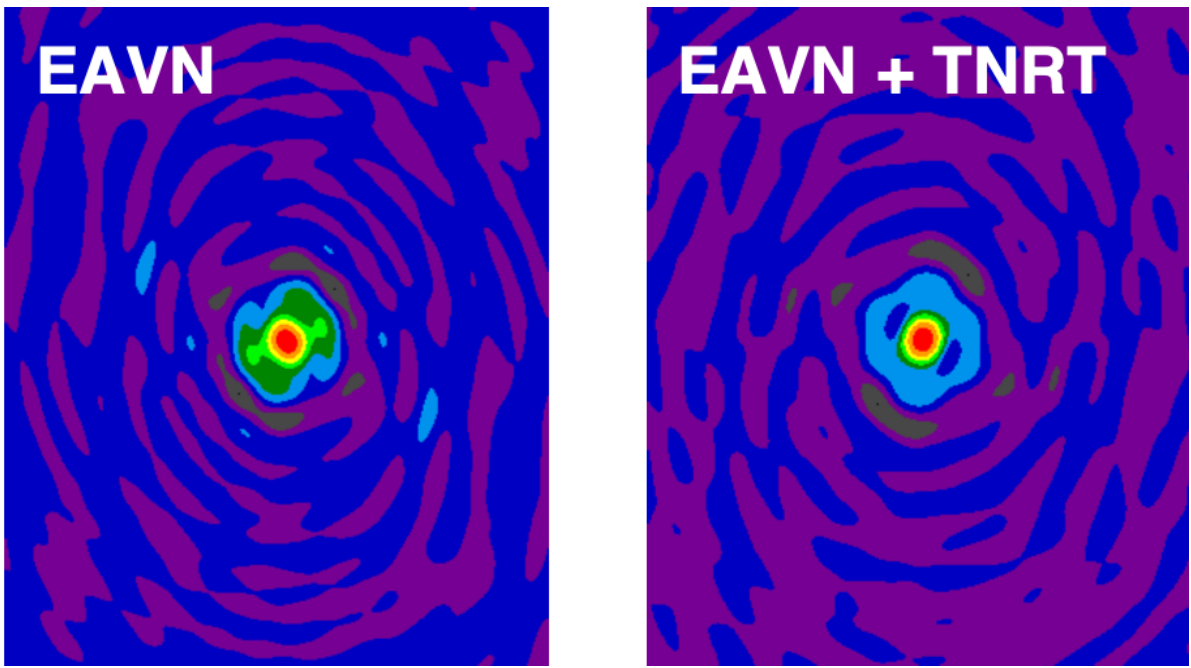}
    \caption{Simulated results of synthesized-beam for EAVN (left panel) and EAVN$+$TNRT (right panel) in the case of observing a source with declination of $+$41.5 degree in K-band. These are plots in the same color scale with normalized peak intensities.}
    \label{fig1.2:beameavn}
\end{figure}

The 40m TNRT is being built at an ideal location to provide unique coverage of the sky, yielding a combination of unique baselines as forming one of the longest baselines (see figure~\ref{fig1.2:vlbiarray}) with a high sensitivity in the wide-range of observable frequencies. For example, here we present results of the simulation for UV-coverages formed via collaborating the 40m TNRT with the East-Asian VLBI Network (EAVN) in K-band that is a legacy band, as shown in figure~\ref{fig1.2:uveavn}. Baselines formed by the TNRT correspond to the 2nd longest baseline up to $\sim$4,500 km but in the northeast-southwest direction uniquely, which is the different direction of ones formed with the Nanshan 26m radio telescope. Even in the case of observing the Galactic Center (GC) locating in the southern hemisphere, the TNRT strongly contributes to form lots of baselines, as shown in the right-hand panel in figure~\ref{fig1.2:uveavn}. These upgrades of the uv-coverage provide improvement of synthesized-beams via reducing side-lobes significantly, as shown in figure~\ref{fig1.2:beameavn}, and yield better quality on VLBI images (see figure~\ref{fig:m87vlbi} in section~\ref{subsec4:jetvlbi} as well for an imaging simulation). 

\vspace{2mm}
Simulations of uv-coverages in other cases for collaborations with the Australia Long Baseline Array (LBA) and the European VLBI Network (EVN) are summarized together with specifications on the collaborations (baseline lengths and baseline sensitivities as 5$\sigma$) in appendix~\ref{specvlbi}. 
In both cases, the TNRT will strongly contribute to form one of the longest baselines with unique directions and fill in a part of ``spaces" on the uv-coverages, so-called uv-hole that is essential factor to achieve better synthesized-beams and imaging quality.


\vspace{1mm}


\vspace{-20mm}

\clearpage
\subsection{Northern Thailand Atmospheric Forecasting System}\label{subsec1-3:forecasting}

\textit{Led by Sherin Hassan Bran, Ronald Macatangay, Vanisa Surapipith, Phrudth Jaroenjittichai, and Koichiro Sugiyama}

\vspace{4mm}
\noindent

One of TNRT's key capabilities is to conduct high-frequency observations. During winter time the weather is ideal for K- and Q-band frequencies. The Driest period is between December and January, which intermittently allows for W-band observation. As the weather condition can evolve rapidly on a timescale of a few days, this allows us to predict the atmospheric conditions and implement dynamic scheduling observation for TNRT to maximise high-frequency observation and science output.

\begin{figure}[h!]
\centering
\includegraphics[width=\linewidth]{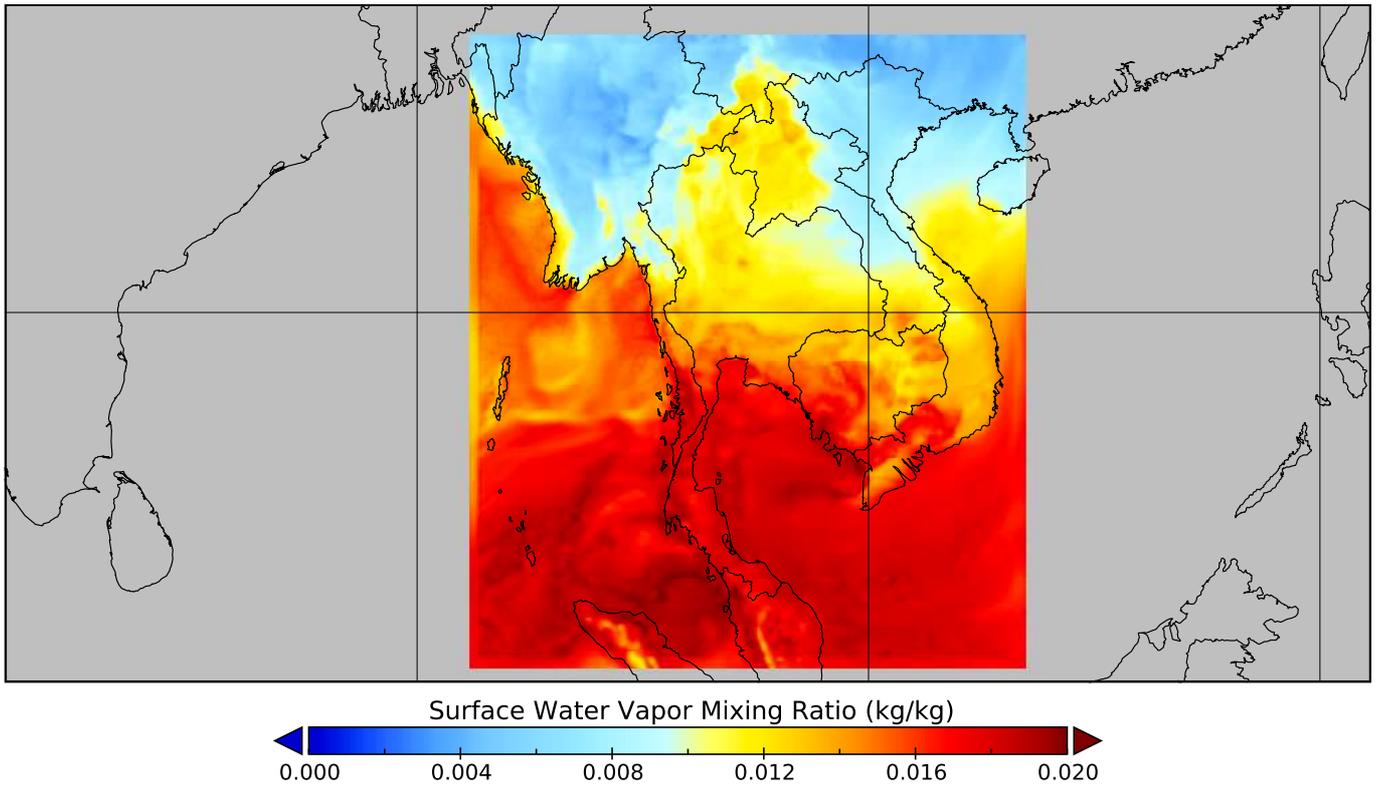}
\caption{Forecast domain showing a sample water vapor mixing ratio forecast.}
\label{fig:forecast1}
\end{figure}

\begin{figure}[h!]
\centering
\includegraphics[width=\linewidth]{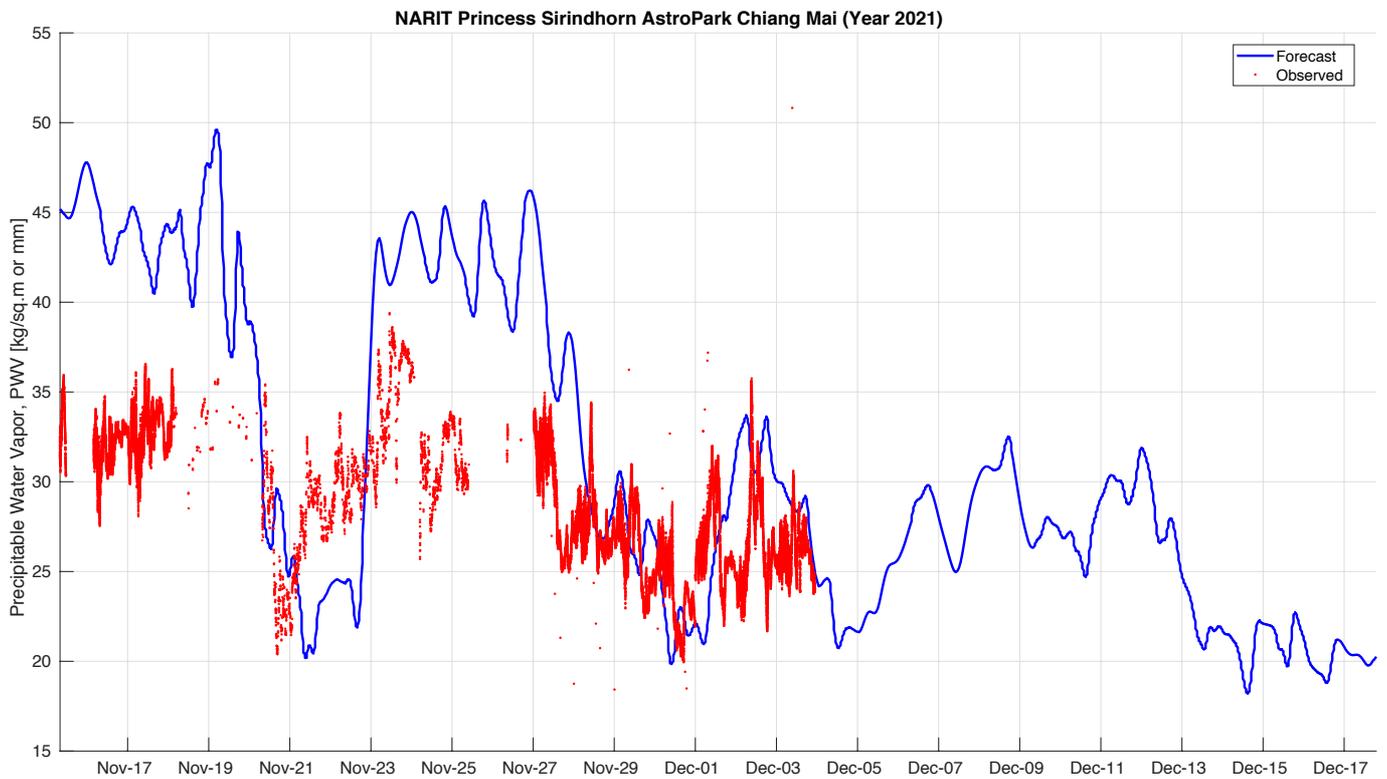}
\caption{Sample precipitable water vapor forecast.}
\label{fig:forecast2}
\end{figure}

\vspace{3mm}
The Northern Thailand Atmospheric Forecasting System (NTAFS) focuses on the spatial domain covering Thailand and adjacent countries (figure~\ref{fig:forecast1}).  The system utilizes an online regional chemical transport model, WRF-Chem v4.3 \citep{2005AtmEn..39.6957G,2006JGRD..11121305F} to analyze the aerosol formation distribution and meteorological parameters over the domain. The forecasts are performed at a horizontal spatial resolution of 9~km with 1~hr temporal resolution, and a vertical profile resolution having 32 layers, with 20 vertical layers being from the surface to 10~km above ground level \citep{2022AtmRe.27706303B}. The land use as well as the terrestrial data set utilized in this study come from the International Geosphere Biosphere Programme (IGBP) -- Modified Moderate Resolution Imaging Spectroradiometer (MODIS) 21 land use categories\footnote{\href{https://www2.mmm.ucar.edu/wrf/users/download/get_sources_wps_geog.html}{https://www2.mmm.ucar.edu/wrf/users/download/get$\_$sources$\_$wps$\_$geog.html}}. National Centers of Environmental Prediction (NCEP) and Global Forecasting System (GFS) forecast data at a horizontal resolution of $0.25^{\circ} \times 0.25^{\circ}$ and temporal resolution of 3 hours\footnote{\href{https://nomads.ncep.noaa.gov/}{https://nomads.ncep.noaa.gov/}} are used as the initial and lateral boundary conditions for the meteorological calculations. The suite of physical, dynamical and radiative schemes employed in the model are summarized in Table 1. The boundary conditions for chemical species are obtained from the National Center for Atmospheric Research (NCAR) Whole Atmosphere Community Climate Model (WACCM)\footnote{\href{https://ncar.ucar.edu/what-we-offer/models/whole-atmosphere-community-climate-model-waccm}{https://ncar.ucar.edu/what-we-offer/models/whole-atmosphere-community-climate-model-waccm}}. A sample forecast for the precipitable water vapor over Chiang Mai is shown in figure~\ref{fig:forecast2}.

\begin{table}[htb]
\small
\centering
\caption{Meteorological parameterizations utilized by the forecast system.}
{\renewcommand\arraystretch{1.5}
\begin{tabular}[t]{|p{48mm}|p{76mm}|p{43mm}|}  \hline \hline
{\bf Process Parameterized} & {\bf Scheme Used} & {\bf Reference} \\ \hline
Microphysics        & Morrison double moment scheme
                    & \citet{2009MWRv..137..991M}  \\ \hline
Convection          & Grell-Freitas scheme          
                    & \citet{2014ACP....14.5233G} \\ \hline
Surface Layer       & Revised MM5 Monin-Obukhov scheme  
                    & \citet{2012MWRv..140..898J} \\ \hline
Land Surface        & NOAH Land Surface model: unified NCEP/NCAR/AFWA scheme              & \citet{2001MWRv..129..569C} \\ \hline
Boundary Layer      & Yonsei University scheme & \citet{2006MWRv..134.2318H}  \\ \hline
Short-wave Radiation & Rapid Radiative Transfer Model for General Circulation Models (RRTMG)\footnote{Aerosol feedback on radiation is enabled.}
                    & \citet{2008JGRD..11313103I} \\ \hline
Long-wave Radiation & RRTMG
                    & \citet{2008JGRD..11313103I} \\ \hline
\multicolumn{3}{l}{$^{\ast}$Aerosol feedback on radiation is enabled.}
\end{tabular}
}
\end{table}

\clearpage
\section{Pulsars and FRBs}\label{sec2:pulsar}

\textit{Led by Phrudth Jaroenjittichai, Takuya Akahori, Thanapol Chanapote, Richard Dodson, Marcus Halson, Simon Johnston, Michael Kramer, Maria Rioja, Siraprapa Sanpa-arsa, Jompoj Wongphecauxson}

\subsection{Timescales \& Emission Physics}\label{subsec2:timescale}

Pulsars exhibit emission variability over 16$^{\textsuperscript{th}}$ orders of magnitude from nano-second pulse structure of the Crab pulsars (\citealt{hkw+03}) to years of intermittent pulsars (e.g.~\citealt{klo+06}, hereafter KLO). However, the radio emission from active regions above the magnetic poles is observed to be stable for most pulsars, except for intermittent and mode-switching pulsars.

The occurrence of radio emission in pulsars, produced from accelerated charged particles in the magnetic open-field-line region above the polar cap, has been shown to be closely connected to pulsar's spin-down ($\dot{\nu}$) via a braking torque caused by the current of the charged particles (KLO). The connection has been made through the study of intermittent pulsars, which switch between the radio-active (ON) and radio-quiet (OFF) states with timescales of months to years. The discovery of this class of pulsars represented by PSR B1931+24 (KLO) was followed later by PSR J1832+0029 (\citealt{llm+12} and PSR J1841+0500 (\citealt{crc+12}). A detailed analysis of another two intermittent pulsars, PSRs J1910+0517 and J1929+1357, has been carried out by \cite{lsf+17} as a part of PALPHA project. Only PSR J1929+1357 has a measured spin-down ratio value. The fact that the $\dot{\nu}$ of the pulsars in the ON state  ($\dot{\nu}_{\mathrm{ON}}$) is larger than that of the OFF state ($\dot{\nu}_{\mathrm{OFF}}$) is interpreted as the pulsars losing energy more slowly in the OFF state because there is no charge current flow in the open-field-line region which would have generated the radio emission. This model of KLO, where the OFF-state spin-down is given by dipolar braking, was later refined by \cite{lst12b,lst12a} who considered the magnetosphere in the open-field-line zone as plasma-rich in the ON state (contributing to the spin-down torque via the plasma current) and vacuum-like in the OFF state, while the closed-filed-line zone is always filled with plasma. The model is able to provide a consistent range of $\dot{\nu}_{\mathrm{ON}}/\dot{\nu}_{\mathrm{OFF}}$ ratio with the observations, and also predicts the $\dot{\nu}$ ratio as a function of $\alpha$. 

We can also consider the possible relationship between intermittent pulsars and those that show the phenomena called ``moding" and ``nulling". While intermittent pulsars switch magnetospheric states on timescales of days, months and years, the latter group shows state changes in the average pulse profile (moding) or temporary shutoffs of radio emission (nulling) on time scales of seconds to hours (e.g.~\citealt{ls06}). It is not clear what causes these changes, but observations by \cite{lhk+10}) demonstrate clearly that long-term variations in pulsar spin-down are related to such profile changes. Even though the switching time-scales for nulling and moding are too fast, so that only weighted averages (or upper limits) for changes in spin-down can be determined, the strong correlation between these long-term averages of profile shapes and spin-down behaviour shows clear similarity to that of the intermittent pulsars. As pointed out by \cite{lhk+10}, these implied partial or complete disruptions or re-distributions in magnetospheric particle supply may not only explain pulsar timing noise to some extent but also unifies a variety of pulsar phenomena to have the same, albeit little understood physical origin. 
With the following strategy with TNRT we can answer the questions of how intermittent and mode-switching pulsars are related and ultimately to a better understanding of pulsar's emission.

\begin{itemize}
    \item Unbiased monitoring of known pulsars
    
Given that a limited number of known population is being monitored and especially with the expected increasing population from MeerKAT (e.g.~\citealt{bja+20}) and SKA (\citealt{ks15}), TNRT will play an important role in monitoring emission variability in a large sample of pulsars. Assuming +10 elevation mask, and a maximum integration time of 1,800s and 30s slew time between sources, TNRT can monitor  1,400 pulsars within 96 hours or 4 days. Similarly, for the integration time of less than 120s, 880 pulsars can be observed within 36 hours. This scheme will also be used to monitor other pulsar properties. 
    
    \item Pulsar search
    
With ~3,300 pulsars known to date\footnote{ \url{https://https://www.atnf.csiro.au/research/pulsar/psrcat/}} compared to ~300,000 detectable pulsars in Galaxy (\citealt{lfl+06}) our knowledge of pulsar properties and the population is very limited. Therefore, searching for new pulsars is very essential and it is also one of the 13 High Priority Science Objectives (HPSOs) for SKA1 (\citealt{bbg+15}).
The pulsar search observation will be performed mainly with L-band receiver which offers the lowest frequency (central frequency of 1.4 GHz) available at the TNRT. At L-band, the observing frequency of 1.4 GHz is high enough to combat a highly frequency-dependent ($\propto f^{-4}$) pulse scattering effect and the Galactic synchrotron background ($\propto f^{-2.6}$) which have a high impact on observing pulsars near the Galactic center and distant pulsars. At this frequency, it is also low enough to still be able to detect pulsars which, in general, have a steep flux spectral index ($\propto f^{-1.41}$, \citealt{blv13}). The 40-m TNRT is capable of searching for new pulsars in spite of the smaller collecting area compared to all-sky pulsar search surveys conducted with larger telescopes such as the 64-m Parkes telescope (HTRU; e.g. \citealt{kjv+10}),
the 100-m Green Bank Telescope (GBT) (GBNCC; e.g. \citealt{slr+14}) or Five-hundred-meter Aperture Spherical Telescope \citep[FAST:][]{2011IJMPD..20..989N,2018IMMag..19..112L}. The advantage of the TNRT is a less competitive observing time; therefore, we have more possibility of finding new pulsars by observing pulsar-like sources over a long period of time (i.e. a targeted search strategy). To determine the minimum detectable flux density, we use the radio equation (\citealt{lk04}) where the signal to noise threshold ($S/N_{min}$) = 8; sky temperature = 12 K; system temperature = 25 K; Gain (G) = 0.32 Jy/K; the number of summed polarization = 2; bandwidth = 800 MHz and expected pulse duty cycle or pulse width over the spin period = 0.1. The resulting minimum detectable flux density is 0.13 mJy for a 1-hour observation.
The standard pulsar search procedures (i.e. RFI removal, dispersion removal acceleration search, periodic search and single-pulse search) will be performed after the data is taken with the pulsar backend. The single-pulse search is the main procedure to search for fast radio bursts (FRBs) (\citealt{lbm+07}) whose origin and population are still unambiguous given that over 800 are detected \footnote{ \url{http://www.wis-tns.org}}. Thus, discovering new FRBs from pulsar search data will help shed light on these mysterious phenomena.

    \item Search for sporadic pulsars
    
Sporadic pulsars are groups of pulsars in which the emission is not detected as periodic pulses due to intrinsic or extrinsic effects. The extrinsic effects such as interstellar scintillation modulate the pulsar's flux on the timescale depending on the observational bandwidth. Some pulsars such as mode-switching pulsars, Rotating Radio Transients (RRAT), intermittent pulsars and magnetars are also sporadic pulsars due to their sporadic intrinsic emission. Only a small number of them are known to show variability in their emissions, despite highly expected numbers (\citealt{mll+06}; \citealt{kk08}; \citealt{lsf+17}).  
Despite several pulsar surveys (.e.g \citealt{mlc+01}; \citealt{cfl+06}; \citealt{kjv+10}; \citealt{dcc+13}; \citealt{clh+14}), some sporadic pulsars might be missing due to the fact that to the observation time on the sky is very limited. 
\par One of the ways to do this is to observe the sky as frequently as possible. Here we present an initiation of a pulsar survey with TNRT (Table \ref{psrsurvey}). We propose to observe the galactic plane (l $\pm$ 60, b $\pm$ 3) using the L-band receiver (central frequency 1.4 GHz), a relatively large bandwidth (800 MHz). With the integration time of 120s per pointing, the expected minimum flux for the S/N of 9 is 0.35 mJy from the radiometer equation. 

\begin{table}[]
\caption{Table of pulsar surveys comparison (HTRU-S, SMIRF (\citealt{vff+20}), TNRT), assuming two polarisations summed and the pulse width of 50 ms and period of 1s}
\label{psrsurvey}
\resizebox{\textwidth}{!}{%
\begin{tabular}{cccccc}
\hline
Parameters                     & HTRUS-HILAT  & HTRUS-MEDLAT & HTRUS-LOWLAT  & SMIRF & TNRT pulsar survey \\ \hline
Instrument                    & Parkes       & Parkes       & Parkes        & UTMOST& TNRT               \\
SEFD                             & 32.65  & 34.28  & 41.63   & 170 & 78                 \\
Central frequency (MHz)          & 1352         & 1352         & 1352     & 835     & 1400               \\
Bandwidth (MHz)                  & 340          & 340          & 340    & 16        & 800                \\
Survey region &
  
  $\delta$\textless{}10 &
  \begin{tabular}[c]{@{}c@{}}-120\textless{}l\textless{}30\\ -15\textless{}b\textless{}+15\end{tabular} &
  \begin{tabular}[c]{@{}c@{}}-80\textless{}l\textless{}30\\ -3.5\textless{}b\textless{}+3.5\end{tabular} &
  \begin{tabular}[c]{@{}c@{}}-115\textless{}l\textless{}40\\ -4\textless{}b\textless{}+4\end{tabular} &
  \begin{tabular}[c]{@{}c@{}}-60\textless{}l\textless{}60\\ -3\textless{}b\textless{}+3\end{tabular} \\
$T_{obs}$                       & 270          & 540          & 4300    & 300      & 120                \\ \hline
Sensitivity limit (mJy)  & 0.16 & 0.12 & 0.05 & 3.6 & 0.35               \\ \hline
\end{tabular}%
}
\end{table}

\par Using the 800~MHz L-band receiver can give us more sensitivity comparable to the HTRU-N high-lat survey and better than SMIRF survey. Moreover, a large bandwidth gives us a better possibility to detect scintillating pulsars, since scintillation affects largely on the bandwidth, and magnetars--a sub-type of pulsars with high magnetic fields--as most of the radio-loud magnetars have a relatively flat spectrum. 

As TNRT with L-band has a beamwidth of 0.11 square degree, we can observe the Galactic plane aforementioned with 6545 pointings with a total observation time of approximately 9 days.

    \item Multi-frequency Astronomy for Sophisticated Statistical Analysis of GRP Events (MASSAGE)\\
The Giant Radio Pulse (GRP) is an enigmatic sporadic intense radio pulse with a typical fluence (kJy ms -- MJy ms) much stronger than pulses of ordinary pulsars. More than ten pulsars have emitted GRPs and the Crab pulsar is the best-known GRP source (see, e.g., \citealt{ag72};\citealt{sbh+99};\citealt{btk08}). \cite{jvk+01} claimed that GRP is defined as the pulses which have at least ten times larger integrated flux or peak flux than the average pulse flux. For Crab pulsar, \citealt{cbh+04} found that GRPs occur only at the phase of main pulse and interpulse. 

It is known that GRPs have a power-law probability distribution (\citealt{mnl+11}), while pulses of ordinary pulsars have a Gaussian (or log-normal) distribution. These different natures imply that the two different pulses originate from different radiation mechanisms, yet the details are not clear. For Crab, the power-law index is about 3 and the GRPs contribute about half of the average pulse flux (e.g., \citealt{lcu+95}). \cite{mat+16} first achieved simultaneous multi-frequency detection of Crab GRPs, unveiling GRP's broadband properties.

\par GRP's occasional, strongest emission implies an explosive event takes place in/on a neutron star or in a magnetosphere. This suggests counterpart emission at high energies such as X-ray and $\gamma$-ray. There is a longstanding discussion about this correlation. Recently, Hitomi collaboration (2018) suggested no apparent correlation between GRPs and the X-ray emission in either the main pulse or interpulse phase for Crab pulsar. The fact that GRPs do not dramatically change the overall properties of neutron stars such as $P$ and $\dot{P}$ suggest that it is a local event rather than a global one. Studying GRPs would be thus useful to explore a higher order topology of the magnetosphere, i.e. beyond the classical dipole magnetic-field magnetosphere.

We propose a multi-wavelength (radio, optical, X-ray) campaign program for observing Crab GRPs. The program aims to measure a statistical difference of GRPs in  different wavelengths. More specifically, we explore the pulse amplitude, width, and shape of GRPs, and synchronization of the pulses among radio, optical, and X-ray. An observing band is L-band (1.4 GHz). 

The 40m TNRT is used as a single dish and collaborating telescopes observe Crab simultaneously. Since GRPs are sufficiently bright, TNRT can safely observe Crab GRPs. Sub-millisecond data recording (P=33 msec) is required. The radio part of this program is useful for science verification and early science operation. The reference data could be available from other telescopes in the world. For instance, the 76m Lovell telescope of Jodrell Bank Observatory is monitoring Crab pulsar.

\end{itemize}

\subsection{Fast Radio Bursts}\label{subsec2:frb}

Radio Transients are astronomical objects or events, which exhibit time-variable radio emissions such as pulsation, cyclic variation, brightening, flash, burst, and outburst. Radio transients are one of the key science objectives of the Square Kilometre Array (see SKA Science Book 2015) \footnote{ \url{https://pos.sissa.it/cgi-bin/reader/conf.cgi?confid=215}}.

There are some aspects of classification for radio transients. For example, they can be classified into fast ($<1$~sec) and slow ($>1$~sec) transients. They can also be classified into galactic and extragalactic transients. Major fast-galactic transients are pulsars and magnetars, while slow-galactic transients are flare stars, magnetars, XRBs, and SNe. Fast extragalactic transients are FRBs and pulsars, while slow extragalactic transients are SN, TDE, AGN, and GW sources (mergers of compact objects). 

Another way of classification for radio transients is based on radiation mechanism, i.e. incoherent and coherent radiation events. Incoherent events are mostly synchrotron radiation and thermal emission. Those radiation mechanisms are normally well-known. The brightness of incoherent events is limited to the brightness temperature $T_{\rm b} < 10^{12}$ K by inverse Compton scattering. Therefore, a more luminous source is a larger source, leading to slower transients (see e.g, Fig. 2 of \citealt{fo15}). These slow transients can be found in multi-epoch images.

Coherent Events are maser and curvature radiation. Those radiation mechanisms are largely unknown. The coherency can break the limit of inverse Compton scattering and coherent events often shows very high brightness temperature, for example, $T_{\rm b} > 10^{20}$~K (pulsars) and $T_{\rm b} > 10^{30}$~K (FRBs) (see e.g, Fig. 4  of \citealt{fo15}). Coherent events are generally fast transients and are found in voltage signals rather than correlated visibility.

Polarized FRB High-precision Understanding by K-band Experiments with TNRT (PHUKET)\\
Fast radio burst (FRB) is stimulating astronomy very much, e.g. at least 10 Nature/Science papers reported FRBs in the last 3 years, yet the origin of FRBs is unknown except extragalactic origins based on their large dispersion measures (DMs) of O(100-1000) pc/cm$^3$. There are 13 linearly-polarized FRB (LPFRB)s as of May 2020. Faraday rotation measure (RM) of linear polarization provides unique information about an environment along the line of sight. For example, a repeating LPFRB121102 showed RM of $O(10^5)$ rad/m$^2$, suggesting an extreme environment like a supermassive black hole. Interestingly, the RM value changed by 10~\% in seven months, suggesting a dynamic environment around the source. Meanwhile, RMs of only 12-14 rad/m$^2$ for one-off LPFRB150807 and LPFRB180924 implied a thin interstellar environment. Moreover, they suggested the first upper limits of O(10) nG magnetic field in the cosmic web along the line of sight.

It has been recognized that the nature of FRB coherent emission is similar to that of Magnetar radio outbursts. For example, the frequency spectrum is relatively flat compared to pulsar coherent emission, and the spectral index is time-variable. Therefore, there is a hypothesis that FRB comes from a young, strongly-magnetized neutron star. A recent discovery of radio outburst from SGR 1935$+$2154, which gave a few kJy ms (CHIME, 400-800 MHz) and 1.5 MJy ms (STARE2, 1.4 GHz), may support this hypothesis.

As of May 2020, there is no FRB detection above 12~GHz. FRB detection of such a high frequency is very useful to constrain radiation mechanism and thus the origin, though a much smaller field of view for higher frequencies. A practical approach is not a blind search of FRBs but a monitor of repeating FRB sources such as FRB121101. Such monitoring is also useful to clarify short-time and long-time variability of RM, which may provide information on the coherent scale of the turbulent magnetic field around the source. High-frequency observation of RM can confirm $O(10^5)$ rad/m$^2$ of LPFRB121101 apart from depolarization effects. Finally, a comparison of pulse profiles among different frequencies is useful to consider a model of radiation mechanism as it has been done for ordinary pulsars.

We propose a long-term monitoring program for observing linear polarization of repeating FRBs. The program aims to measure the variability of linear polarization, particularly the time-dependence of the polarization angle and the rotation measure, with high-cadence (every month) monitoring at K-band (22 GHz). K-band detection for repeating FRBs has never been achieved. An advantage of 22 GHz observation is that it provides almost dispersion-free signals so that it is easy to obtain a clean FRB profile and spectrum as well as a small Faraday rotation effect by the foreground magnetoionic medium. 

TNRT is used as a single dish and it observes the repeating FRBs (Table \ref{frb}) every month. The sensitivity of TNRT is expected to be 3$\sigma$ = 0.7 Jy ms (3 GHz BW, 2bit). Thus, repeating FRBs with relatively radio-loud events can be observed. 

\begin{table}[htbp]
\caption{Information of Repeating FRBs}
\begin{center}
\footnotesize
\begin{tabular}{cccc}
\hline
Name & Right Ascension & Declination & Fluence (Jy/ms) \\
\hline
Arecibo repeater (R1) & 05h31m58.70 s & $+$33$^{\circ}$08$^{\prime}$52.5$^{\prime \prime}$ & $\lesssim 1$ at 1.4 GHz\\
CHIME repeater (R2) & 04h22m22 s & $+$73$^{\circ}$40$^{\prime}$ & $\lesssim 3$ at 600 MHz\\
\hline
\end{tabular}
\end{center}
\label{frb}
\end{table}

\subsection{Gravitational Waves}\label{subsec2:gw}

After the first Gravitational wave, GW150914, was discovered (\citealt{aaa+16}) with LIGO, direct detection of Gravitational Waves (GWs) has become one of the most notable science. Millisecond pulsars play a major role in detecting GWs from the supermassive black hole binary merger at a frequency range of nano-Hz via pulsar timing array (PTA). Pulsar timing is essentially a technique of using observed pulse arrival time from a pulsar to find a model (a timing solution) that accurately predicts future pulse arrival time. Hence, the timing solution contains highly precise measured pulsar parameters and ISM information. The difference between measured and predicted (model) arrival time is called “timing residuals”. When GWs pass pulsars, they leave a unique imprint on the pulsar timing residuals. In order to retrieve this imprint, a collaboration of the international pulsar timing array (IPTA) was established (Manchester \& IPTA 2013). By observing some pulsars in IPTA using the TNRT’s pulsar backend in timing mode and measuring the time of pulse arrival, we can assist in establishing a long baseline of pulsar timing solutions in order to detect GWs signature.

Pulsar Timing Arrays have been known to observe PTA MSPs at most once every 2 weeks (\citealt{vlh+16}). TNRT aims to observe a number of MSPs at \~100ns accuracy every day in order 
to explore high-frequency gravitational waves (stochastic background/individual sources). 
Following \citealt{vlh+16} and assuming $\beta$=1, $S/N$=10, $T_{\mathrm{sys}}$=20 K, $G$=0.32 K/Jy, $N_{\mathrm{pol}}$=2, $BW$=800 MHz and $t_{\mathrm{int}}$=1 hr, the radiometer noise ($\sigma_{\mathrm{TOA}}$) can be determined for five highest precision timing pulsars (Table \ref{timing}).

\begin{table}[!ht]
\centering
\caption{The calculated $S/N$ and $\sigma_{\mathrm{TOA}}$ for TNRT.  } 
\begin{tabular}{l llll lll}
\hline\hline
\# & Bname & Jname & $P$ (s) & $W_{\mathrm{50}}$ (ms) & $S_{\mathrm{1400}}$ (mJy) & $\sigma_{TOA}$ (ns) \\ \hline
1 & J0437-4715 & J0437-4715 & 0.005757 & 0.141 & 149.0 & 4.8 \\
2 & J1713+0747 & J1713+0747 & 0.004570 & 0.110 & 10.2 &  54.4 \\
3 & J1909-3744 & J1909-3744 & 0.002947 & 0.044 & 2.1 &  82.4 \\
4 & B1937+21 & J1939+2134 & 0.001558 & 0.038 & 13.2 &  14.7 \\
5 & J2241-5236 & J2241-5236 & 0.002187 & 0.070 & 4.1 &  99.3 \\
 \hline
\end{tabular}
\label{timing}
\end{table}

\subsection{Pulsar Astrometry}\label{subsec2:astrometry}
\begin{itemize}
\item 
Pulsar astrometry is one of the most promising areas for TRNT to have a significant impact:
Astrometry does not require many antennas to trace the motion on the sky of the source of interest, as these are nearly always point-sources. 
Pulsar astrometry requires multiple observations to derive the astrometric solutions, as scintillation leads to random flux levels and therefore variable SNR.
The TRNT combined with resources from East Asia and Australia provides an ideal configuration, with good resolution in RA and Dec. 

Furthermore, there have been new developments for low-frequency astrometry \citep{2020A&ARv..28....6R}, with the new methods of MultiView (multiple calibrators for the target) \citep{2017AJ....153..105R} and Multi-Frequency Phase Referencing (MFPR; multiple frequencies to solve for the contributions from different atmospheric components \citep{drb+18}. 
These will invigorate the field by increasing the astrometric accuracy by potentially an order of magnitude.

Even with the publication of the VLBA PSR-Pi, which has 60 pulsar parallaxes, there still are relatively few geometric pulsar distances. These are the gold standard for distances in astronomy, as parallaxes are model- and assumption-free. 
Therefore almost any targeted source would have significant contributions to our understanding. 

Taking the sensitivities from Table \ref{tablereceiver}, the list of known pulsars from PSRCAT and including the gains from pulsar gating the observations, we predict that 228 pulsars would be suitable for a TRNT-TianMa65-Parkes64 array (SNR at 1\, min $>$5$\sigma$), of which 167 have no measured parallax (from timing or VLBI).

\item 
MUAY-THAI: Magnetar Unprecedented Astrometry Yielded by Thailand
The characteristic age $\tau_{\rm c}$ given by the rotation period $P$ and period derivative $\dot{P}$ as $\tau_{\rm c}=P/2\dot{P}$ indicates that magnetars are relatively young with $\tau_{\rm c}$ $ \lesssim 10$ kyr. Given a typical transverse velocity (velocity perpendicular to the line of sight) of a neutron star, 200 km/s (see e.g., \citealt{eks19}), and the traditional Sedov solution with typical parameters, the magnetar takes 210 kyr to escape from the shell of the progenitor supernova remnant. In other words, at the typical age of 10 kyr for magnetars, they should stay inside the shells which are still bright and visible in X-rays in general. However, about half of magnetars are not associated with any supernova remnants. Therefore, we are missing some mechanisms and/or misunderstanding the age and the velocity of magnetars. 

There are only four magnetars whose transverse velocities are estimated. This sample is too small to argue whether the average transverse velocity of magnetars is systematically higher than 200 km/s of the average transverse velocity of ordinary pulsars. Therefore, measuring the transverse velocities of magnetars is of great importance. It has been recognized that magnetars are radio-quiet. But they are bright in the radio when outbursts take place. As of May 2020, there are 6 radio-loud outburst events for magnetars. It indicates the possibility of VLBI astrometry which can provide an accurate transverse velocity of a magnetar.

Using the measurement of the precise position and velocity of a magnetar, we can derive the area in which the magnetar was born. If the area is associated with a supernova remnant, the remnant is most likely the candidate of the progenitor. With characteristics of the SNR, we may learn more about the origin of magnetar's strong B-field (e.g. SASI’s dynamo).

We propose a target-of-opportunity (ToO) program for observing magnetar radio outbursts. This program aims to measure the accurate position, velocity, and distance of a target magnetar in an outburst phase which is typically only a couple of months.  An observing band is K-band (22 GHz).  TNRT can be co-operated with East Asia VLBI network (EAVN), where TNRT plays a key role in the VLBI, providing the longest (4529 km) SW-NE baseline of EAVN (with Mizusawa 20 m telescope). The baseline is complementary to the other longest (5100 km) NW-SE baseline of EAVN (between Nanshan 26m telescope and Ogasawara 20m telescope) and provides the world-top-level angular resolution of 0.6 mas for EAVN. This superior angular resolution allows us to measure the typical perpendicular velocity of neutron stars, 200 km/s or 3.4 mas/month (d/kpc)$^{-1}$ for the distance $d$  up to $\sim 6$~kpc, which is double compared to VERA. In other words, TNRT dramatically expands the achievable volume by a factor of $2^3=8$. Since magnetar radio outburst is a rare event (say once per year in the Milky Way), increasing the search volume is a great advantage to carry out this program.

The program continues for a couple of months unless the outburst is faded out. We expect that a magnetar outburst is bright at 22 GHz with the average flux density of $O(10)$~mJy, according to the outburst of XTE J1810-197 (Eie et al. in preparation). The sensitivity of TNRT K-band is expected to be 1.1 mJy (2bit 3 GHz BW, 1hr). Thus, the outburst can be observed with a sufficient signal-to-noise ratio. The sensitivity of EAVN + TNRT is expected to be $10\sigma$ = 0.16 mJy (512 MHz BW, 1hr), promising solid detection and astrometry.

\end{itemize}
\subsection{Exploiting Astrophysical Laboratories}\label{subsec2:lab}
As a result of privilege in observing time, we can dedicate more time to
follow-up on exotic systems and objects. For instance, the triple system, PSR J0337+1715, is the only known galactic three-body system consisting of a millisecond pulsar and two companion white drafts orbiting around at 1.6 days and 327 days (\citealt{rsa+14}). This unique system offers the best test for the Strong Equivalence Principal.

Another puzzling object which we can monitor is Magnetar (See \cite{tzw15} for a review), a neutron star with an extremely high magnetic field (around 100 times stronger than a typical neutron star). The link between magnetars and pulsars is still not confirmed. Given that only 23 magnetars are known, monitoring known magnetars and magnetar candidates will certainly result in interesting scientific results.

The first transitional millisecond pulsar (tMSP), PSR J1023+0038 (hereafter J1023), which is a missing link between low-mass x-ray binary (LMXB) and millisecond pulsar was discovered in 2007 (\citealt{asr+09}) and tMSPs have been fascinating objects to study ever since. Only three such systems were discovered after J1023 (\citealt{pbs+16}, \citealt{sah+14} and \citealt{bph+14}). Before J1023 went back to an accreting stage which is radio-quiet in 2014, it showed a peak in x-ray flux density increase (\citealt{sah+14}). Thus, monitoring J1023 at multi-wavelength frequency would help understand the stage-switching mechanism and possibly predict when the radio emission turns back on. Since the tMSPs can be studied at multiple wavebands, with the 2.4-m optical telescope of Thai National Observation (TNO) we can study both radio and optical counterparts of tMSPs simultaneously in order to study a correlation in variation at both wavebands.

Follow-up observations on known FRBs will as well yield fruitful results. The “repeater”, FRB121102, is the only known FRB that can be observed multiple times with different telescopes (\citealt{ssh+16}); therefore, monitoring known FRBs to investigate whether other FRBs show repeating pulses will be one of the very interesting follow-up projects.


\clearpage
\section{Star Forming Regions}\label{sec3:sfr}

\textit{Led by Koichiro Sugiyama, Busaba H. Kramer, Kitiyanee Asanok, Malcolm D. Gray, Ram Kesh Yadav, Tomoya Hirota, Thanapol Chanapote, Shari L. Breen, James A. Green, Simon P. Ellingsen, Kee-Tae Kim, Kazuyoshi Sunada}

\begin{table}[h]
\centering
 \caption{Summary of basic parameters for each topic in this section.}
  \begin{tabular}{lp{80mm}|lccc} \hline\hline
    Sec. & Topic & \multicolumn{1}{c}{Band} & Single-dish? & VLBI? & Pol.?\\
    \hline
    \ref{subsec3:flux} & High-cadence Flux and Polarization Monitor & LC \hspace{0.3mm} KuKQW & $\bigcirc$ & $\bigcirc$ & $\bigcirc$ \\
    \ref{subsec3:survey} & Unbiased Thermal Molecular and Maser Line Surveys & LC \hspace{0.3mm} KuKQ & $\bigcirc$ & $\times$ & $\times$ \\
    \ref{subsec3:maser} & Address the Fundamental Maser Physics & \hspace{12.9mm} K & $\bigcirc$ & $\bigcirc$ & $\times$ \\
    \ref{subsec3:vlbi} & VLBI Vision with TNRT in High-mass SFRs & LC \hspace{0.3mm} KuKQW & $\times$ & $\bigcirc$ & $\times$ \\
    \hline
  \multicolumn{6}{p{175mm}}{\small Note.-- Columns~1, 2. number and name of sub-sections; Column~3. frequency bands to be used for each topic; Columns~4--6. necessity of ways to observe as single-dish, VLBI, and/or polarization, respectively.}
  \end{tabular}
\end{table}

\subsection{High-cadence Flux and Polarization Monitor}\label{subsec3:flux}

Interstellar masers in various molecules, such as hydroxyl (OH), methanol (CH$_3$OH), water (H$_2$O), silicon monoxide (SiO), and so on, present variability in their flux densities in the evolution of stars both in the formation and evolved phases. There are lots of type in the variability, such as monotonic increase/decrease, anti-correlated, flaring/bursting, random, and so on \citep[e.g.,][]{1973ApJS...25..393S,1988SvAL...14..468M,1999PASJ...51..333O,2000ApJ...534..781L,2004MNRAS.355..553G,2004PASJ...56L..15H,2008PASJ...60.1001S,2011ApJ...739L..59H,2012PASJ...64...17F,2014PASJ...66...78F,2014PASJ...66..109F}.

\subsubsection{Periodic Flux Variability}\label{subsubsec3:period}

One of them is classified into the ``Periodic" flux variability, which was possibly discovered in the Mira variables R Leo and Omicron Ceti via SiO masers at 86~GHz \citep{1979ApJ...234L.199H}. Such a periodic variability was also discovered around the high-mass star G~009.62$+$00.19E via CH$_3$OH masers at 6.7 and 12.2 GHz \citep{2003MNRAS.339L..33G}. Periodic variability in high-mass star-forming regions (SFRs) had been detected from 20 sources as of 2016 \citep{2004MNRAS.355..553G,2009MNRAS.398..995G,2010ApJ...717L.133A,2011A&A...531L...3S,2015MNRAS.448.2284S,2016MNRAS.459L..56S,2012MNRAS.425.1504G,2014PASJ...66...78F}. Their periods show a wide range from a month to over a year. These are classified into two periodic patterns that are continuous like sinusoidal, and intermittent with a quiescent phase. There is a notable characteristic that in some sources all the spectral features are synchronized with the same periodicity but showing various time offsets. This synchronization has been also detected between CH$_3$OH and other masers, such as OH, H$_2$O, and formaldehyde (H$_2$CO) masers \citep{2012MNRAS.425.1504G,2019MNRAS.485.4676G,2021MNRAS.502.5658M,2016MNRAS.459L..56S,2020A&A...634A..41O,2010ApJ...717L.133A}. The periodic variability, thus, has been theoretically suggested to be caused by flux variations related to a common exciting source, such as colliding wind binary \citep[CWB:][]{2009MNRAS.398..961V,2011AJ....141..152V,2019MNRAS.485.2759V}, stellar pulsation \citep{2013ApJ...769L..20I}, and circumbinary disk with a spiral shock \citep{2014MNRAS.444..620P,2016A&A...588A..47V}. The former model is related to increased fluxes of the seed photon, while the latter two models are related to changed physical environments for pumping masers especially the dust temperature. Given the range of periods in the maser flux variability as a month to over a year that for instance can be converted to $\sim$0.1--1 au under the Keplerian rotation condition, the periodic variability enables us to uniquely reach a super-tiny spatial area, such as the surfaces of high-mass protostars (HMPSs) and very close to binary systems. That super-tiny area, corresponding to angular resolutions smaller than 1 milliarcsecond (mas) at typical source distances of high-mass SFRs farther than 1 kpc (equal to 3,260 light year), is impossible to directly resolve by using any interferometers for observing thermal emissions such as VLA (Very Large Array), SMA (Sub-Millimeter Array), ALMA (Atacama Large Millimeter-/submillimeter Array) in radio wavelengths, and VLTI (Very Large Telescope Interferometer) in near-infrared wavelengths due to insufficient spatial resolutions. In terms of understanding the evolution of HMPSs, the accretion rate onto the stellar surface is the most important physical property \citep{2009ApJ...691..823H}. The difference of the accretion rates theoretically determine the extent of their radius just before zero-age main-sequence (ZAMS), e.g., in the case of the rate of 10$^{-3}$ M$_\mathrm{sun}$~yr$^{-1}$ leading to $\sim$100 R$_\mathrm{sun}$ at the maximum. 

\vspace{3mm}
In the three theoretical models, the stellar pulsation model \citep{2013ApJ...769L..20I} must be most attractive to address the evolution of HMPSs, because they have predicted the relation among the period and physical properties, such as stellar luminosity, radius, mass, and accretion rate onto the stellar surface that has been the most essential parameter to understand the evolution. If we verify and establish such a period-luminosity (P-L) relation observationally, the relation will be the only and unique tool to reach the super-tiny spatial area of HMPSs themselves.
Besides recent growth of the number of detection for periodic variations in high-mass SFRs \citep{2018MNRAS.474..219S,2019MNRAS.486.1236O,2019MNRAS.487.2407P}, \citet{2016PASJ...68...74Y} reported that they initiated long-term, highly-frequent, and unbiased flux monitoring toward 442 sources in northern hemisphere (declination $> -$30 deg) with the Hitachi 32-m radio telescope on Dec 30, 2012, to overcome the lack of periodic sample in high-mass SFRs. This monitoring project is entitled ``the Ibaraki 6.7-GHz Methanol Maser Monitor" (iMet). This monitor was designed of 442 sources divided into nine groups, and conducted as daily monitoring that provided us with a spectrum of each with an interval of nine days. From Sep 2015, it was redesigned via extracting 143 sources into four groups, giving us more frequent observations with an interval of 5 days. Those observations have resulted in new detections of periodic variations in more than 30 sources, and the periodic sample has been increased more than twice \citep[][and in prep.]{2015PKAS...30..129S,2017PASJ...69...59S,2018IAUS..336...45S,2019JPhCS1380a2057S}. In these sources, periodic ones at least presenting the continuous pattern can be interpreted to be caused by the stellar pulsation because the pulsation is excited and grown by the kappa mechanism \citep{2013ApJ...769L..20I}, and then used for the observational verification of the P-L relation. But, how about periodic sources with the intermittent pattern? Cannot we invovle those sources to verify the P-L relation?

\vspace{3mm}
To address the issues, with the TNRT we will initiate a flux monitoring project for OH and H$_{2}$O masers in L- and K-bands simultaneously toward the periodic sample compiled in CH$_{3}$OH masers. As mentioned above, a few periodic sources showed synchronization of the periodicity between CH$_{3}$OH and other masers, those were OH, H$_{2}$CO, and H$_{2}$O masers, respectively \citep{2012MNRAS.425.1504G,2019MNRAS.485.4676G,2021MNRAS.502.5658M,2010ApJ...717L.133A,2016MNRAS.459L..56S,2020A&A...634A..41O}. The first of these masers (OH) is radiatively pumped, which is the same to pump CH$_{3}$OH masers \citep{2002MNRAS.331..521C}, while the latter two masers are collisionally pumped \citep[e.g.,][]{2014A&A...562A..68V,1989ApJ...346..983E,1989ApJ...347L..35E}. Multiple masers with different pumping mechanism will provide us a unique opportunity to clearly associate periodic maser sources with the appropriate theoretical model. For example, if all the radiative OH, CH$_{3}$OH, and the collosional H$_{2}$O, H$_{2}$CO masers show synchronization in a periodic variation, the periodic source is expected to be explained by CWB model because this model can produce a weak shock at every periastron in binary system, and the weak shock cause a change in the flux of maser seed photons and simultaneously provide the suitable environment for collisionally pumping masers. This model can be achieved only in relatively later evolutionary phases with formed H{\footnotesize II} regions, which is consistent with detecting OH and H$_{2}$CO masers occurring at this evolutionary phase of high-mass star formation. On the other hand, if a source shows periodic flux variability only in the radiative OH and CH$_{3}$OH masers, that is possibly excited by the stellar pulsation even in the case of its continuous pattern. These will be complementary ways with finding out the synchronization in infrared continuum emissions \citep{2020A&A...634A..41O,2022ApJ...936...31U}.

\vspace{3mm}
This science case can be addressed in the first 5 years with the TNRT because L- and K-bands receivers will be installed at the beginning of the commissioning phase, enabling OH and H$_{2}$O maser transitions, respectively. We have already had the complete list of periodic sources compiled with new detections with the Hitachi 32-m radio telescope toward CH$_{3}$OH masers. From the list, in the first two years, we will pick up sources showing periods shorter than 60 days as targets, to obtain at least three periodic cycle data in a half year (we have to avoid a rainy season in May-October at least for H$_{2}$O maser observations in K-band). Later on in the second three years, we will initiate other sources showing longer periods. This project of simultaneous monitoring for OH and H$_{2}$O masers will provide us a unique opportunity to distinguish completely all the periodic sources in high-mass SFRs in each theoretical model, and to achieve the final goal to understand the evolution of HMPSs via observationally verifying and establishing the period-luminosity relation.

\subsubsection{Bursting Flux Variability}\label{subsubsec3:bursting}

Another characteristic flux variability is ``accretion bursting activity". This activitity is well known in low-mass SFRs as presenting a drastic increase of bolometric luminosities due to an episodic accretion process from a circumstellar disk, and these are classified mainly into two types called EXors and FUors in terms of the magnitude for flux increases, the time-scale for flux rising, and how long lasting the activities \citep[][and reference therein]{2014prpl.conf..387A}. Such a accreting bursting activity was also discovered in high-mass SFRs via flux monitoring observations of the 6.7~GHz CH$_{3}$OH maser in a famous high-mass SFR Sharpless 255 (S255) with the Yamaguchi 32-m radio telescope \citep{2015ATel.8286....1F}, which presented a bursting activity as flux rising with 1-2 orders of magnitude within a few months among all the spectral components. This bursting activity lasted around 4 years, which included the decaying phase \citep{2018A&A...617A..80S}, and their spatial distribution was verified to be expanded from the previous one on the basis of a VLBI follow-up observation \citep{2017A&A...600L...8M}. The follow-up observations in Near IR-bands revealed that such a burst happened in NIR-bands too \citep{2017NatPh..13..276C}. They presented not only a rise of brightness but also showing up the following lines: CO band-heads and Na I lines that originate from the outer layer of the disk, and H$_{2}$ emissions tracing shocks, respectively. These detections are typical signatures of accretion bursting activities in the case of low-mass star formation. 
In addition, other masers, e.g., H$_2$O masers, were followed by rising its flux density as well \citep{2021A&A...647A..23H}.
Triggered by this discovery, this kind of accretion bursting activity has been detected in other high-mass SFRs as well: NGC~6334I \citep{2017ApJ...837L..29H,2018MNRAS.478.1077M,2018ApJ...866...87B,2021ApJ...908..175C}, G~358.93$-$00.03 \citep{2019ATel12446....1S,2020NatAs...4..506B,2020NatAs...4.1170C,2020ApJ...890L..22C,2020AstL...45..764V,2020MNRAS.494L..59V,2021A&A...646A.161S,2022AJ....163...83B,2022A&A...664A..44B} resulting in lots of discovery of new CH$_3$OH maser transitions \citep{2019ApJ...876L..25B,2019ApJ...881L..39B,2019MNRAS.489.3981M}, and G~24.33$+$00.13 \citep{2019ATel13080....1W,2022MNRAS.509.1681M,2022PASJ...74.1234H}.

\vspace{3mm}
This activity could be another probe to reach the super-tiny area close to protostar and/or protostar itself because such a bursting phenomenon is caused by physical processes of accretion onto and ejection from protostars. If multiple-species masers pumped by different pumping mechanism and associated sites are simultaneously monitored for accretion bursting survey, we will understand an entire view of the accretion burst on the basis of changes of luminosity with dust temperature in CH$_{3}$OH and ejections of bursting jets/outflows in OH and H$_{2}$O masers \citep[e.g.,][]{2018MNRAS.478.1077M,2018ApJ...866...87B,2021A&A...647A..23H}. So far, there are only four HMPSs presenting accretion bursting activities due to insufficient survey data to research the bursting activities statistically. Single-dish monitoring with the TNRT to survey accretion bursts in high-mass SFRs will be another good target to kick off scientific activities at NARIT to potentially find out the exact number of accretion bursting sources and the evolutionary phase for their occurrence, and at the beginning of the commissioning phase OH and H$_{2}$O masers will be observable probes in L- and K-bands, respectively. 

\vspace{3mm}
These intense flux monitoring observations are achieved with not only the TNRT but also on the basis of the world-wide collaborations and networks. Such an organization has been established, entitled ``Maser Monitoring Organization" (M2O)\footnote{See M2O website at \href{https://www.masermonitoring.com/}{https://www.masermonitoring.com/}}, since 2017 to coordinate single-dish monitoring of masers and interferometric follow-up measurements. 
Through intense communication with this international organization, the TNRT intense monitoring observations will be greatly enhance the accretion bursting research via prompt alerting of new detections of flaring/bursting activities, achieving immediate follow-up observations in radio and infrared wavelengths, in which interferometric observations promise to enable us to unveil detailed geometrical and physical structures and their variations caused by the accretion bursts, e.g., spiral-arm accretion flow structure on the accretion disk \citep{2020NatAs...4.1170C}, expansion of the CH$_3$OH distribution caused by rising dust temperature surrounding the central protostar due to bursting accretion luminosity \citep{2020NatAs...4..506B}, and so on. These essential works will be accelerated with VLBI arrays that are going to be upgraded with the TNRT soon providing the improvement of uv-coverages/synthesized-beams and imaging quality as introduced in section~\ref{subsec1.2:vlbi}.

\subsubsection{Magnetic Field Variability}\label{subsubsec3:mag}

\vspace{3mm}
Maser variability is well known on timescales of months to decades. Whilst maser variability in flux density is well known on timescales of months to decades, variation in polarisation is less well studied, and an avenue of research ideal for the TNRT. By high-cadence monitoring of maser sources, we therefore mean sampling at higher rates: approximately monthly observations or daily. 
Compared with spatially broader thermal emission, over scales of several thousand AU, that would be blended within single dish observations, at timescales of a few days, variations on the scale of the whole source, some thousands of AU for a massive star forming region, would be blurred out in a single-dish observation, leaving maser sources, of scale a few hundred AU or smaller, as the only sources likely to be bright enough to detect, and to vary quickly enough.

\vspace{3mm}
Polarimetric monitoring of maser flares on a target-of-opportunity basis is an obvious application of the TNRT, with various transitions available at L-, K- and later C-band. Polarization can typically be used to recover magnetic field information in the maser gas. Open-shell molecules, such as OH and CH, are particularly useful in this respect: they have Land\'{e} g-factors sufficient to split helical components of the spectrum by more than the Doppler width, forming Zeeman pairs (and occasionally triplets) from which the magnitude of the magnetic field can be derived directly from the frequency splitting of the left- and right-handed circularly polarized (LHCP and RHCP) spectral components that form the Zeeman pair. The sense of the field, towards or away from the observer, is also immediately recoverable from observation, based on whether the LHCP or RHCP component of the pair has the higher frequency. Such pairs were used by \citet{2006MNRAS.369.1497S} to show that a flaring OH maser object in W75N had the exceptional field strength of 42\,mG, suggesting that compression is likely to be the driver of the flare. With reasonable assumptions about the spatial behaviour of the magnetic field, its value near the surface of the protostar can be estimated (hundreds of Gauss). Long-term monitoring of W33 \citep{2015A&A...575A..49C}, however, showed a more typical magnetic field of only a few mG despite significant variability. \citet{2015A&A...575A..49C} also observed 22.2-GHz H$_2$O masers in W33 with approximately monthly cadence for 35\,yr showing several epochs of flaring activity. Closed shell species, like H$_2$O have Zeeman splittings that are too small to measure the magnetic field directly. Instead, the line of sight component of the magnetic field may be recovered from measurement of Stokes V , for example \citet{2008A&A...484..773V}, or from a combination of the Stokes parameters via a more general model, for example \citep{2019A&A...628A..14L}. Monitoring of a 6.7-GHz CH$_3$OH maser flare in G09.62+0.20 with a cadence of approximately a week \citep{2009A&A...500L...9V} showed a reduction in the line- of-sight magnetic field, with possible sign reversal, in the main flare object, however, in this case, the non-Zeeman effect also needs to be verified by full Stokes polarimetric monitoring. It may therefore be possible to study the relative probability of collisions between clouds of opposite, or the same, magnetic field orientation.

\vspace{3mm}
Most monitoring work in the examples above could be carried out with the TNRT operating as a single dish, but some input interferometric data is desirable to ensure that potential Zeeman pairs in a spectrum are spatially associated on the scale of typical maser spot sizes. Whilst the Zeeman origin of polarization structure is difficult to dispute in the case of molecules like OH, the origin of polarization in cases where the Doppler width dominates may not be related to the magnetic field, or may be related to the field in a manner that differs from the Zeeman interpretation.

\vspace{3mm}
To some extent, polarization can be used to associate spectral features with spatial objects in many maser sources. This allows single-dish observations to extract some spatial information that would otherwise only be available to much more time-intensive interferometic observations. Changes in velocity and polarization structure can be used, for example, to study properties of turbulence in the source region, including extraction of the diffusion coefficient of turbulence and the turbulence viscosity, for example \citet{1999ARep...43..209L}. The presence of a magnetic field generalises the situation to MHD turbulence that is expected to produce significant anisotropy in maser and thermal molecular line spectra \citep{2007ApJ...655..275W}. If monitoring with approximately daily cadence can estimate the magnetic field (from spectral asymmetry) and velocity (from centroid shifts), and a length scale can be found from VLBI observations, then work by \citet{2020A&A...636A..93K} can be used to invert this data to an estimate of the turbulent magnetic diffusivity and, with additional assumptions, the turbulent viscosity. These are important quantities related to the dynamical effects of the magnetic field in accretion disks and outflows.

\vspace{3mm}
At higher cadence, minutes to hours, there is cyclotron maser emission from low-mass main sequeunce stars. Frequencies in the C-band are common, but can range over L- to K-band, and possibly higher frequencies. Bursts can be periodic \citep{2007ApJ...663L..25H}, for example P = 1.96 hr in TVLM 513-46546, an M9 dwarf, and this was attributed to rotation of the host star. An interesting feature of TVLM 513-46546 is the variation of its polarization from a ‘low’ state with circular polarization varying between 13 and 40 per cent periodically, and an active state, where 100 per cent circular polarization is found \citep{2007ApJ...663L..25H}. Moreover, each periodic burst appears as a 100 per cent LHCP pulse, switching to 100 per cent RHCP pulse. The reason for this is unknown. The cyclotron frequency can be used to recover the magnetic field: at least 3\,kG in TVLM 513-46546. A small catalogue of $>$12 objects of this type has been prepared \citep{2017ApJ...845...66R} with an estimate that 13 times as many remain to be discovered. Although mJy sensitivity is needed for detection of these objects, they exhibit broad-band emission compared to molecular masers, improving the chances of detection by the TNRT.

\vspace{3mm}
Very rapid variability of 22.2-GHz H$_2$O masers at timescales tens of minutes or faster has been detected towards a number of sources, including Cep A, W49 W3(OH) and Orion A \citep{2010AIPC.1206..346S}. Towards W3(OH) and Orion A, the variability is detected only through changes in linear polarization. This behaviour remains unexplained.

\vspace{3mm}
In summary, polarimetric monitoring of masers with the L and K band receivers (and others as available) with cadences of 100s of seconds to months, over periods of months to years, will provide significant insight into magnetic fields. In particular, strengths, geometrical components and the dynamical importance of the magnetic fields in star forming regions, and the main parameters of turbulence (including the lifetime of eddies). These studies will need to be complemented with collaborative VLBI observations.

\vspace{3mm}
\subsection{Unbiased Thermal Molecular and Maser Line Surveys}\label{subsec3:survey}

\subsubsection{Maser Line Surveys}\label{subsubsec3:masersurvey}

Among the main problems with observations using the very long baseline interferometry (VLBI) technique are strong competition amongst observing proposals and limitations in the time domain. By contrast, observations with a single-dish radio telescope are far less constrained in terms of observing time. An appropriate research technique that exploits this advantage is the \textit{\textbf{unbiased survey}} of radio sources. However, there are many large survey programmes in the northern hemisphere, such as the HI/OH/Recombination line survey of the Milky Way (\textbf{THOR}), and the Southern Parkes Large-Area Survey in Hydroxyl (\textbf{SPLASH}) in the southern hemisphere. One possible starting point, and perhaps the easiest entry to this form of study, is to note the target regions from previous surveys and concentrate new blind surveys on portions of the sky, or specific areas, which the original large programmes could not access, for geographic reasons. The result of this ‘completion surveying’ is similar to filling in missing pieces of a jigsaw that can help astronomers to obtain a better understanding of the underlying populations and characteristics of astrophysical objects.

\vspace{3mm}
Cosmic masers, or simply \textbf{masers} for short, form an important class of targets. They have strong flux densities (i.e. milli-Jy up to kilo-Jy), various source types, and are easy to detect with a single radio dish or with VLBI telescopes. Moreover, studying the physical properties of masers could lead us to solve the mysteries of stellar evolution, and the association between many types of masers could also improve the theoretical understanding of masers, improving consistency with observational data. There are various types of maser that can be described, with their particular uses, as follows. The first molecule detected as a maser was the hydroxyl radical (\textbf{OH}) at radio frequencies in the interstellar medium (\textbf{ISM}; \cite{Weinreb1963}). The \begin{math}^{2}\Pi_{3/2}, J = 3/2 \end{math} ground-state transitions of OH are at 1612-, 1665-, 1667-, and 1720- MHz and their respective relative intensities under local thermodynamic equilibrium (LTE) are 1:5:9:1 (see \cite{Townes1955}). The mainline frequencies i.e. 1665- and 1667-MHz are found mainly in massive young star-forming regions (\cite{Minier2003}, \cite{xu2008}), and the satellite lines 1612-, 1720-MHz, are found in different objects i.e. evolved stars and supernovae remnants or planetary nebulae, respectively. At the excited-state, J = 5/2 of OH, the masers at the frequency of 6035-MHz are generally stronger than those at 6030-MHz, and almost all of them are found in star-forming regions. All these line have large Zeeman splittings, which can be used to calculate magnetic field strengths directly. Ground-state and excited-state OH maser transitions are often associated with the 6668-MHz methanol (\textbf{CH$_3$OH}) maser (e.g. \cite{Caswell1997}, \cite{Caswell1998}, \cite{Etoka2005}). This transition of (CH$_3$OH) is the second-strongest maser line known: only the extensively studied water maser (\textbf{H$_2$O}) transition at 22235-MHz is brighter. The 6668-MHz masers are found to be associated exclusively with very young massive stars (\cite{Minier2003}, \cite{xu2008}), whilst the 22235-MHz H$_2$O masers are found in two types of objects, i.e. the envelopes of evolved asymptotic giant branch (AGB) stars and also in the discs and outflows associated with young stellar objects (YSOs). However, those associated with the massive YSOs are the more powerful type (for example, \cite{Palagi1993}, \cite{Furuya2003}). Moreover, other usefulness of H$_2$O masers are good kinematic tracers, measuring the parallax distance, and H0 measurements.
Likewise CH$_3$OH masers, they are also used in the parallax measurement and mapping the Galactic inventory of massive star formation.

\vspace{3mm}
There are unbiased surveys that form part of large legacy programmes as follows: firstly, \cite{Beuther2016}, \cite{Beuther2019}) used the Very Large Array (VLA) in the C-array configuration to observe the HI/OH/Recombination lines within the Milky Way (\textbf{THOR}). This survey covers the L-band, spanning the frequencies 1.4 - 2.0 GHz. The sky area covered includes the Galactic longitudes from 14.5 to 67.4 degrees, and Galactic latitudes between $\pm$1.25 degrees. In the Southern hemisphere, \cite{Dawson2014} observed all four ground-state maser transitions of OH using the Parkes telescope, which covered the region of sky between the Galactic longitudes 334 to 344 degrees and Galactic latitudes $\pm$2 degrees. This set of observations belongs to a large programme, named the Southern Parkes Large-Area Survey in Hydroxyl \citep[\textbf{SPLASH}: full data release in][]{2022MNRAS.512.3345D}. \cite{Qiao2018} used the Australia Telescope
Compact Array (ATCA) to confirm the accuracy of positions of OH masers obtained from the results of \cite{Dawson2014}.  These C-band observations were started by \cite{Caswell2010}. They used a seven-beam receiver on the Parkes telescope (called the methanol multibeam receiver: MMB) and observed the CH$_3$OH transition at 6668 MHz along the Galactic plane (Galactic latitude range $\pm$2 deg) and Galactic longitudes from 186 to 360 and from 0 to 60 degrees. They released five unbiased catalogues, including the original set of observations, with a typical rms noise level of 0.070 Jy (see more details in the 2nd until the 5th catalogue as follows: \cite{Green2010}, \cite{Caswell2011}, \cite{Green2012}, and \cite{Breen2015}). From those catalogues, the authors detected a total of 972 methanol maser sites of which about 33\% were new detections. Under the same sky area as the MMB survey, \cite{avison2016} reported the unbiased survey of the Galactic plane for 6035-MHz excited-state OH masers. \cite{2011MNRAS.416.1764W} and \cite{Walsh2014} used the 22-m Mopra telescope to observe in K-band over the range of frequencies 18 - 28 GHz, which included H$_2$O and ammonia (NH$_3$) masers. The sky area covered by the Mopra survey is along the Galactic plane (Galactic latitude range $\pm$0.5 deg) and covers Galactic longitudes 290-360 degrees and 0-30 degrees.

\vspace{5mm}
\noindent{\textbf{What could the 40-m TNRT do?}}

\begin{figure}[h!]
\centering
\includegraphics[scale=0.45]{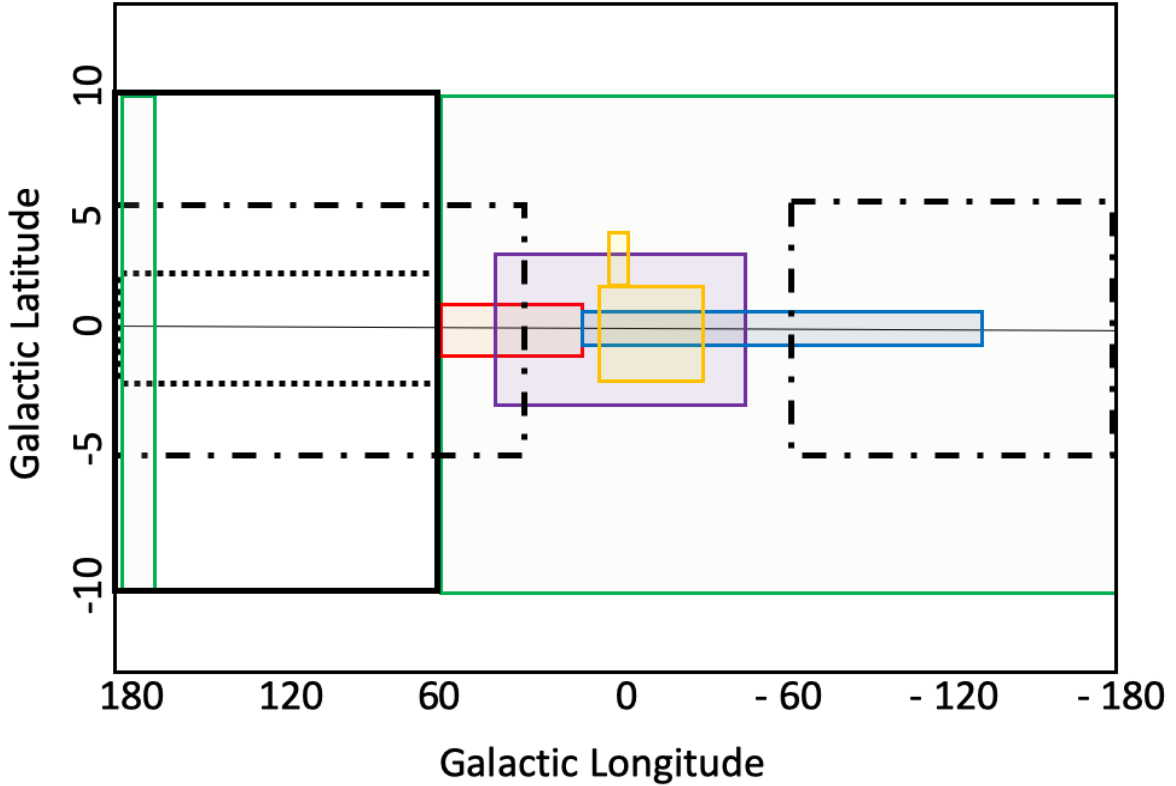}
\caption{This diagram shows the tentative areas for the unbiased survey of maser emission by using the 40-m TNRT in the galactic coordinate system. Solid, dashes and chain boxes in black colour represent the tentative area for observing in L-, C-, and K-band respectively. Other filled colour boxes refer to the large survey programmes: Galactic Australian SKA Pathfinder Survey (GASKAP) in green, THOR in red, SPLASH in yellow, \cite{Caswell1998} in blue and \cite{Sevenster1997a},\cite{Sevenster1997b},\cite{Sevenster2001} in violet.}
\label{fig:area}
\end{figure}

\vspace{3mm}
There are a few options to do the unbiased survey of maser emission as follows: 1) in order to reach a minimum requirement of an rms noise (e.g.$\sim$70 mJy), long integration time is needed ($\sim$20 to 25 mins for C-band). Therefore, it will take more time in each scan observation to cover a particular region of sky than the large programmes. However, although the 40-m TNRT takes longer than a bigger dish, e.g. Parkes, Effelsberg, to do a scan, this is partially compensated for by the larger beam of the TNRT: each scan covers more sky. 2) select some portions of the sky which the large programmes (see the filled colour boxes of Figure~\ref{fig:area}) have not covered in the high latitude, and 3) due to the flux variability of maser emission, try to search in the same areas which have been found or detected the maser signal before and re-measure the flux again otherwise select in the bright targets such as the evolved stars which are bright enough for observing by the 40-m TNRT.  
Here are the tentative sky areas to fill in, missed by those large programmes discussed above, starting from L-band, should be the Galactic longitudes 180 to 60 deg, and the Galactic latitude $\pm$10 deg (the black solid box in Figure~\ref{fig:area}). In the case of C-band, the area not covered by the MMB is the Galactic longitude range 180 to 60 deg and the Galactic latitude $\pm$2 deg (the dashes box). Finally, to complete all sky areas of K-band in the large programme, we need to observe in the ranges 30 to 180 deg (Galactic longitude) and $-$180 to $-$60 deg along the Galactic plane (the chain boxes).

\vspace{-2mm}

\subsubsection{Molecular Lines Surveys}\label{subsubsec3:molecularsurvey}

Stars form in molecular clouds. The main ingredients of these clouds, dust and gas, are found to be distributed along the galactic plane. Some of the key questions in star formation research are how stellar birth and evolution in the Galaxy is connected to the molecular clouds? what are the physical and chemical processes involved in formation in new stars?  In order to understand the detailed picture of star formation there has been, and planned, numerous multiwavelength surveys.  

One of the advantages of radio spectral line observations is to achieve high spectral resolution, providing a velocity information of target sources at an order of 0.1~km~s$^{-1}$. 
Such high velocity resolution observations play essential roles to understand dynamics of the astrophysical processes. 
Furthermore, observed radio spectral lines can be identified to various kind of transitions from different atomic and molecular species in interstellar matter (ISM) and circumstellar envelopes (CSE). 
By combining multiple transitions, one can estimate not only physical properties such as temperature, density, ionization degree, turbulent motions, but also chemical composition (column density or fractional abundance) of the target sources. 
Thus, radio spectral line observations are powerful tools for astrochemistry and future astrobiology.

\vspace{5mm}

\noindent
\textbf{3.2.2.1 Unbiased line survey}

\vspace{3mm}

\vspace{3mm}
At radio wavelengths, atomic lines have been observed from the early phase of radio astronomy to very recently\footnote{The most important atomic transition is a well known hydrogen 21~cm transition. However, we will not discuss this transition to focus on more complex chemical species. 
Hyperfine structure of the carbon atom (CI) has two transitions at submillimeter wavelengths, although these are out of the scope of the TNRT target.}. 
At the observational frequency of TNRT, hydrogen and other atoms at high electron excitation states close to the ionization energy emit so-called radio recombination lines (RRL) from the ionized gas. 

\vspace{3mm}
Radio spectral lines are emitted mainly from molecular species at centimeter and (sub)millimeter wavelengths. 
At a recent count, more than 200 interstellar and circumstellar molecules have been identified \citep[][for comprehensive review]{2018ApJS..239...17M,2022ApJS..259...30M}. 
Molecular emission lines arise mostly from rotational transitions and some are different transitions such as those caused by torsional/vibrational excitation and hyperfine structure. 

\vspace{3mm}
In the case of high-mass star-forming regions, the Orion Kleinmann-Low (KL) nebula and the Galactic center molecular cloud Sagittarius B2 (Sgr B2) have been most extensively studied in unbiased molecular lines surveys mainly in millimeter wavelengths \citep{1989ApJS...70..539T, 1991ApJS...76..617T}. 
They are known to show strong spectra from various molecular and atomic lines including masers. 
A number of new molecular species have been discovered in these sources, and recent searches for prebiotic molecules are also conducted, in particular for Sgr B2(N) \citep[e.g.][]{2012ApJ...755..153N}. 
There have been unbiased line surveys at almost the same frequency range of TNRT toward the well studied source Orion KL \citep{2009ApJ...691.1254G, 2015A&A...581A..48G}. 
At K-band, 164 RRLs from hydrogen, helium, and carbon, and 97 molecular lines from 23 species are detected \citep{2015A&A...581A..48G}. 
Ammonia lines are detected including non-metastable ($J>K$) transitions and they are used to estimate temperature structure (see sections of ammonia survey: ``3.2.2.2 Unbiased Amomonia Survey" and ``3.2.2.3 Targeted Ammonia Surveys"). 
Even though single-dish observations cannot provide images of the compact target sources, the velocity information (peak velocity and linewidth) provides dynamical information of the target sources combined with the follow-up interferometer data. 
One of the advantages of lower-frequency observations compared with most of the above millimeter wave observations (e.g. at K-band) is the lower line density. 
The molecular spectra with the rotational constant $B$ in MHz appear at the constant spacing of $2B$, but the separation in the velocity domain, $c/(J+1)$ in which $c$ is the speed of light and $J$ is the rotational quantum number, is much smaller in higher-$J$ or higher frequency lines. 
This is more critical in line-rich sources with broader line widths such as high-mass hot cores. 

\vspace{3mm}
As for low-mass star-forming regions, a prototypical  dark cloud core, Taurus Molecular Cloud 1 (TMC-1) has been recognized as the good laboratory of interstellar chemistry thanks to its proximity to the Sun (140~pc) and abundant molecular species. 
In particular, TMC-1 is characterized by  extremely rich chemistry of abundant carbon-chain molecules including the longest species of HC$_{9}$N and aromatic molecule Benzonitrile c-C$_{6}$H$_{5}$CN \citep{2018Sci...359..202M}.  
These are important species for understanding of formation of the polycyclic aromatic hydrocarbons (PAHs) and prebiotic molecules in interstellar clouds.  
One of the largest unbiased molecular line survey was conducted with the Nobeyama 45~m radio telescope in NAOJ  \citep{2004PASJ...56...69K} with the wide frequency coverage from 10 to 50~GHz. 
The lowest frequency line survey observations at $<$10~GHz has been carried out toward TMC-1 with the 305~m Arecibo telescope \citep{2004ApJ...610..329K}. 
Recently, unbiased line survey projects with the GBT 100-m ``GOTHAM"\footnote{\href{https://greenbankobservatory.org/science/gbt-surveys/gotham-survey/}{https://greenbankobservatory.org/science/gbt-surveys/gotham-survey/}} and Yebes 40-m antennas ``NANOCOSMOS"\footnote{\href{https://nanocosmos.iff.csic.es}{https://nanocosmos.iff.csic.es}} have discovered various carbon chains and rings toward TMC-1 at almost the same frequency bands with the TNRT 40-m antenna. 
Along with the high spectral resolution, the survey allows us to identify the molecular species and to accurately estimate excitation temperature and molecular abundances. 
TMC-1 has been recognized as the first molecular cloud detected in various carbon-chain molecules such as CCS which has been identified in a laboratory experiment prior to radio astronomical observations \citep{1987ApJ...317L.115S}. 
Carbon-chain molecules now become important tools as they are turn out to be useful tracers of chemical evolution of molecular clouds during star-formation processes. 
Because of the chemical reaction network starting from the ionized carbon (CII), carbon-chain molecules are more abundant at the early phase of star-formation. In contrast, some other molecules such as NH$_{3}$, N$_{2}$H$^{+}$, and deuterated molecules show an opposite trend in which abundances increase at the later phase due to a slow production reaction \citep{1992ApJ...392..551S}. 
Thus, a molecular line census would also be essential for understanding the complete picture of molecular clouds, which restore initial conditions of the potential sites of star and planet formation. 

\vspace{3mm}
At the more evolved phase of low-mass star-formation, newly born Solar-type young stellar objects (low-mass protostars) can affect their environments by internal heating by radiation and outflow shocks. 
In such regions, molecular species which are frozen out onto dust grains during the cold dark cloud phase start to sublimate and/or new molecular species are formed by formation/destruction reactions under high temperature. 
Consequently, these sources show rich chemistry. 
The chemical compositions of early phase of Solar-type stars would provide clues for understanding of the origin of organic molecules in planetary systems. 
In the last decade, chemical diversity in various low-mass protostars have been discovered \citep{2013ChRv..113.8981S}. 
The compact regions around the central protostar, called hot corinos, are rich in saturated organic molecules and another type of region is known for its warm-carbon-chain chemistry (WCCC) where unsaturated carbon-chain molecules are more dominant.  
Unbiased line surveys clearly show such striking differences between a hot corino source IRAS~16293-2422 \citep{2011A&A...532A..23C} and WCCC L1527 \citep{2019PASJ...71S..18Y}. 
Possible origins of this diversity could be related to the timescale of dynamical evolution in the starless phase. 
If the timescale of the starless phase is enough long to lock carbon atoms into CO molecules, saturated organic molecules are more efficiently produced by grain surface reactions from adsorbed CO prior to protostar formation \citep[][for more details]{2013ChRv..113.8981S}. 
Chemical diversity seen in saturated organic molecules and unsaturated carbon-chain molecules is also reported for high-mass star-forming regions \citep{2018ApJ...866..150T}. 
Along with the origin of organic molecules in planetary systems, the chemical diversity in protostars will provide hints of different evolutionary history of dynamical and chemical evolution in low-mass star-formation processes.

\vspace{3mm}
Scientific value: High sensitivity K-band observations will be able to detect carbon-chain molecules and aromatic hydrocarbons. This will be helpful in understanding the formation on PAH and prebiotic molecules in star forming clouds. These observations will answer questions related to the physical and chemical evolution of molecular clouds. 

\vspace{2mm}
Bands: With the first generation K-band receiver unbiased line survey can be done between 18-26.5 GHz. The frequency ranges can be extended between $\sim$8 and $\sim$100 GHz when later generation planned receiver are available at TNRT.

\vspace{5mm}
Similar Galactic plane unbiased surveys with large instantaneous bandwidth have been conducted with the 22~m Mopra telescope at K-band \citep[HOPS][]{2011MNRAS.416.1764W} and Q-band \citep[Malt-45][]{2015MNRAS.448.2344J}. 
For the K-band HOPS survey, they covered a large field of the Galactic plane covering 100~degrees$^{2}$ with the frequency coverage of 19.5-27.5~GHz. 
Although it focuses on H$_{2}$O masers, the data can be used for an unbiased molecular line survey in space and frequency. 
The molecular line unbiased survey with TNRT can be done in the course of other surveys for masers and ammonia lines (see sections \ref{subsubsec3:masersurvey} and 3.2.2.2 \& 3.2.2.3). 
It will be complementary with the other unbiased and targeted survey in terms of wider spatial coverage and frequency in northern hemisphere.

\vspace{5mm}
\noindent
\textbf{3.2.2.2 Unbiased Ammonia Survey}

\vspace{3mm}
Hydrogen is the most abundant element in the universe. In the diffuse interstellar medium where number density is very low, hydrogen is found in atomic form. Hydrogen atoms emit at 1.4 GHz at radio wavelengths, associated with a hyper-fine transition due to spin flip of electron and proton within the atom. However, at higher densities where star formation takes place, Hydrogen is found rather in its molecular form and direct observations of H$_2 $ are extremely hard. 
The next most abundant molecule in the interstellar medium (ISM) is CO. The CO being a hetero-nuclear molecule has a dipole moment and hence is easier to observe in its rotational transitions. But at further higher column densities where star forming cores and clumps are formed, CO is optically thick, an therefore cannot be used as a dense gas tracer. At such densities polyatomic NH$_3$ has been proved to be a more sensitive diagnostic of dense gas \citep{1989ApJS...71...89B} and therefore has been most widely used as a higher-density gas tracer. 

In the past, there have been few dedicated NH$_3$ surveys (K-band, unbiased and targeted) of the Galactic plane. The first systematic unbiased NH$_3$ (and H$_2$O) survey was called \textbf{HOPS} (H$_2$O Galactic Plane Survey) during 2008-2010 using the Mopra 22m radio telescope \citep{2011MNRAS.416.1764W}. 
This survey covers 100 sq. deg. of the galactic plane \textbf{($l$ = 290-30 deg, $|b|$=0.5 deg)} mostly in the southern hemisphere. The aims of this survey were to i) provide an un-targeted survey of water masers toward the inner Galaxy, ii) To map the high density gas component of inner Galaxy using (1,1), (2,2) and (3,3) inversion transitions of ammonia, and iii) to search for emission from unusual emission lines. The second unbiased ammonia (and water) survey, The Radio Ammonia Mid-Plane Survey (\textbf{RAMPS}) was carried out with the 100m GBT \citep{2018ApJS..237...27H}. This survey was complementary to HOPS and covered an area of 24 sq. deg of the Galactic plane \textbf{($l$ = 10-40 deg, $|b|$=0.4 deg)}. The observations were made at NH$_3$(1,1), NH$_3$(2,2), and H$_2$O transitions. The aims of the RAMPS were i) how do high-mass star forming clumps evolve, ii) what is the role of filaments in high-mass star formation, and iii) what is the Galactic distribution of star forming regions and evolved stars. Another targeted ammonia survey, The Green Bank A Ammonia Survey (\textbf{GAS}) is being carried out for all the northern Gould Belt star forming regions with $A_V$ $>$ 7 \citep{2017ApJ...843...63F}. The Gould Belt being the location of nearby star forming regions with varying scale of star formation activity, the GAS aims to understand how star formation proceeds in different environments. 

The composite CO survey of the entire Milky Way by \citet{2001ApJ...547..792D} reveals a plethora of dense molecular emission concentrated along the Galactic plane with spread of few vents of low density gas.  As mentioned before, a large part of the Galactic plane is not covered in earlier  NH$_3$ surveys.  To complement previous Galactic plane NH$_3$ observations, we propose to extend these observations for \ang{30} $\leq$ $l$ $\leq$ \ang{95}; $|b|$ = $\pm$\ang{0.5} and \ang{108} $\leq$ $l$ $\leq$ \ang{114}; $|b|$ = $\pm$\ang{0.5} of the Galactic plane which is achievable from TNRT. 

\vspace{3mm}
Scientific values: This survey will provide a catalog of dense ammonia star forming clumps within an area of $\sim$100 deg$^2$ of the Galactic plane. 

\vspace{2mm}
Bands: There can be two settings for the survey using the K-band receiver. Standard setting is designed to cover 22.2$-$24.2~GHz (bandwidth 2 GHz) covering 4 ammonia inversion transitions NH$_3$(1,1)(2,2)(3,3)(4,4) and H$_2$O(6$_{16}$-5$_{23}$). A complementary setting can be designed to cover 23.6 - 25.6 GHz which includes NH$_3$(1,1)(2,2)(3,3)(4,4)(5,5)(6,6).

\vspace{5mm}
\noindent
\textbf{3.2.2.3 Targeted Ammonia Surveys}\\

The \textit{Herschel} Hi-Gal survey \citep{2010A&A...518L.100M,2016A&A...591A.149M} has identified about 10$^5$ high density cold dust clumps which covered the inner part of the Galactic Plane ($-$\ang{70}~$ \leq  l  \leq $ \ang{68} and $|b|$ $\leq$ \ang{1}) using PACS \citep{2010A&A...518L...2P} 70 and 160 $\mu$m and SPIRE \citep{2010A&A...518L...3G} at 250, 350, and 500 $\mu$m simultaneous imaging in all five bands. These clumps, as they appear, are associated with filaments \citep{2010A&A...518L.102A}. These filamentary structures,  when they intersect one another, can form junctions termed hubs by Meyers (2009). Recently, \citet{2020A&A...642A..87K} presented a new paradigm for formation of low and high mass stars in these hub filament systems (HFSs) wherein low mass stars are associated with filaments and high mass stars connected to junctions of HFSs. Based on their analysis, they identified ~3700 HFSs. These HFSs, as described by \citet{2020A&A...642A..87K}, may evolve from prestellar to evolved star phase with gas density that varies depending on the evolutionary phase. In addition to an unbiased Galactic plane survey we also propose to carry out targeted observations of some of the nearby interesting sources. A part of these HFSs can be covered in our unbiased Ammonia survey, however, given the wider longitudinal coverage of Hi-Gal data targeted observations for some interesting sources can easily be done. 

Given the fact that high mass stars are located relatively at farther distances, it is difficult to resolve cores of high mass star forming regions. For a source 2 kpc away, spatial resolution would be $\sim$33.25 pc which is too high to spatially detect a star forming core as their sizes are of the order of 0.1 pc. However, physical conditions of the surroundings can easily be traced. These sources can later be studied with interferometric observations for better spatial detail. 

\vspace{3mm}
Scientific value: These observation will be helpful in deriving the physical condition in HFSs. In conjunction with multiwavelength observations, a detailed star formation scenario can be developed. 

\vspace{2mm}
Band: K, with a bandwidth of 2 GHz covering NH$_3$(1,1)(2,2) and (3,3) inversion transitions.

\begin{figure}
 \begin{minipage}[b]{0.39\linewidth}
  \centering
  \includegraphics[width=\textwidth]{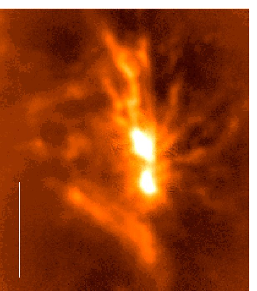}
 \end{minipage}
 \begin{minipage}[b]{0.60\linewidth}
  \centering
  \includegraphics[width=\textwidth]{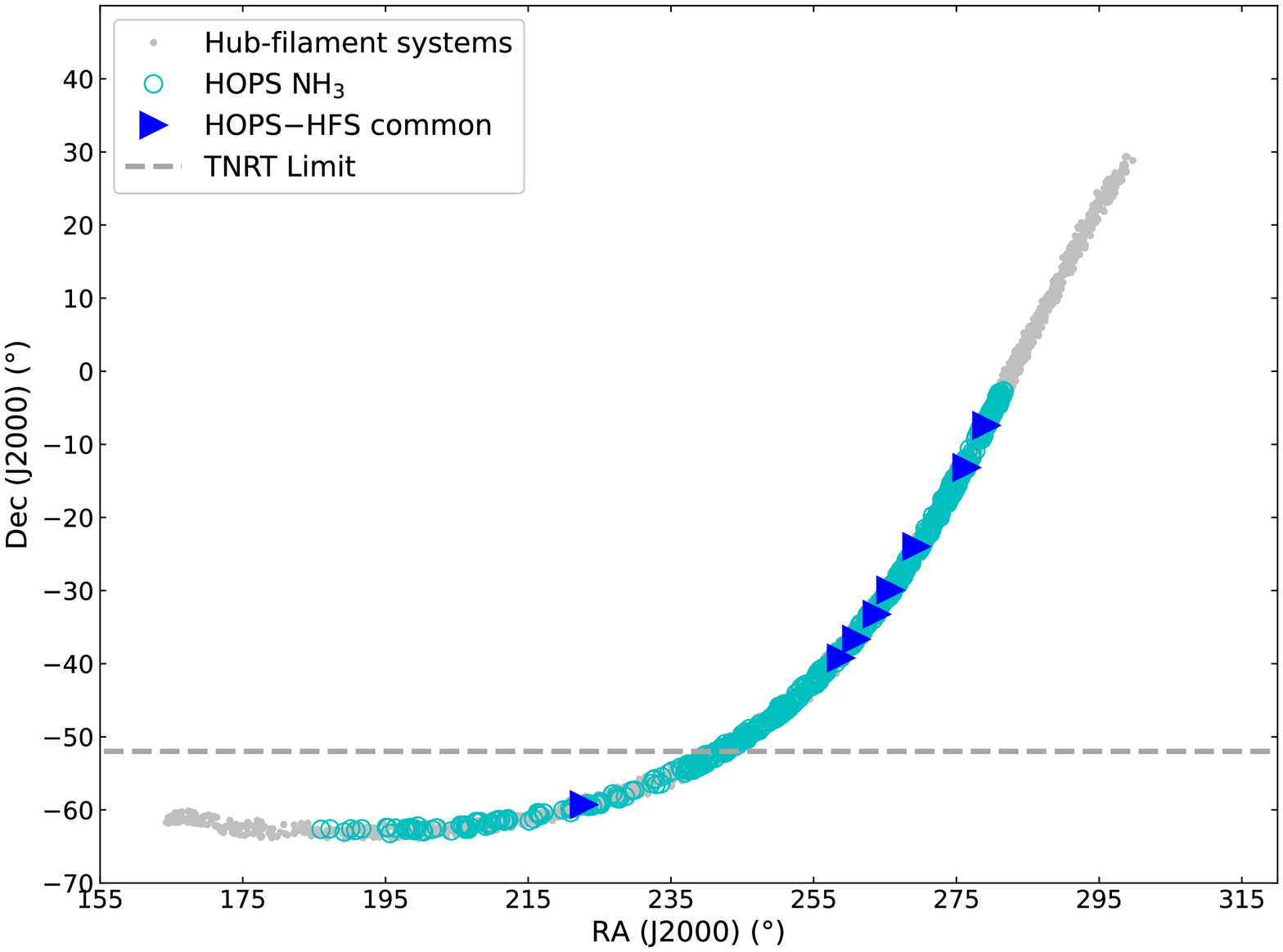}
 \end{minipage}\\
  \begin{minipage}[b]{\linewidth}
  \centering
  \includegraphics[width=\textwidth]{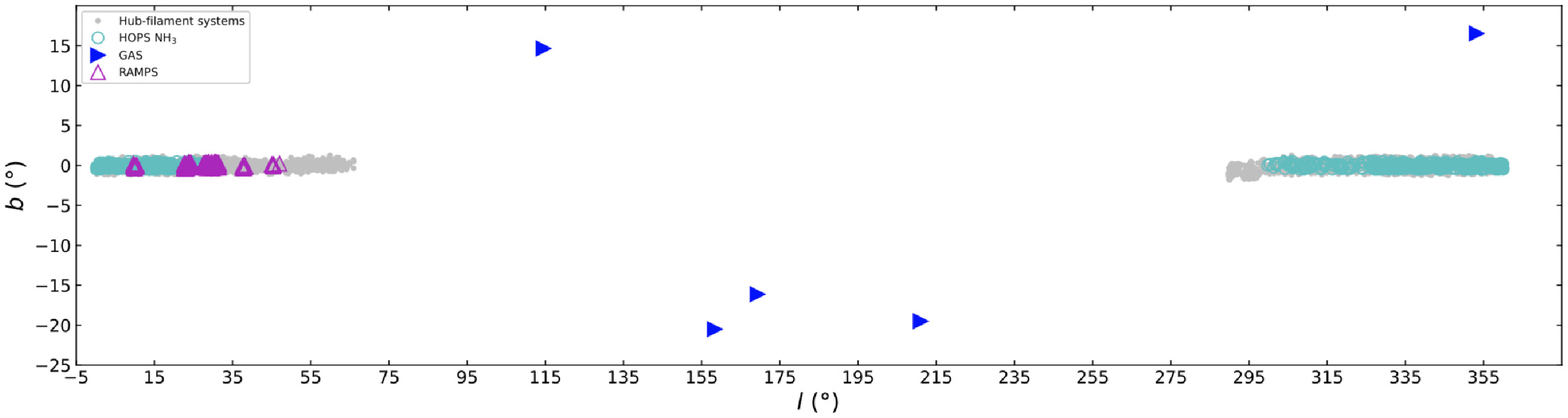}
 \end{minipage}
\caption{Upper-left: Submm continuum image (0.85~$\mu$m) of integral filament associated with the Orion Nebula Cluster \citep{1999ApJ...510L..49J}, taken from \cite{2009ApJ...700.1609M}. The region shows hub-filament feature. The scale bar in this figure indicates 0.5~pc. Upper-right: Distribution of HOPS-Ammonia (cyan open circles) and Hi-GAL HFSs in equatorial coordinates. TNRT declination limit (zenith distance $\sim$ $-$52$^{\circ}$) is shown with gray dash$-$horizontal line. Lower: Distribution of HOPS-Ammonia clumps (cyan open circles),  RAMPS maser sources (magenta open triangles), and GAS-Ammonia sources (filled blue triangles) over-plotted on Hi-Gal HFSs (gray dots), in Galactic coordinates.}
\label{fig:filament}
\end{figure}

\vspace{3mm}
\subsection{Address the Fundamental Maser Physics}\label{subsec3:maser}

It remains a moot point whether astrophysical masers share any of the coherence properties typical of laboratory lasers and masers. \citet{1992ARA&A..30...75E} discusses a couple of related points: In Chapter 3, p 57-58, he argues that astrophysical masers are not coherent across the wavefront because the ratio of the scale size to the wavelebgth, $l / \lambda$ is too large. The phase shift across the wavefront is $\sim \pi \theta^{2} l / \lambda$ for beam angle $\theta$, so a very small beam angle indeed is required to reduce the phase shift to a number $<$1. This argument remains well-founded today \citep{2018MNRAS.477.2628G}, despite the discovery of exceptionally small ($\sim$10$^9$ m) H$_2$O maser cores with space-VLBI \citep{2018ApJ...856...60S}.

\vspace{3mm}
In Chapter 4, p68-71, \citet{1992ARA&A..30...75E} makes a somewhat different argument, based on the bandwidth coherence in a single ray. To obtain coherence, the stimulated emission rate needs to exceed the width of the Doppler envelope (the inhomogeneous line width of the maser transition). A more useful version of this criterion is to say that the power-broadened homogeneous width, which is proportional to the angle-averaged maser intensity, must exceed the Doppler width. \citet{1992ARA&A..30...75E} transforms this requirement to a minimum brightness temperature, $T_b$, of maser emission for a given beam angle. The minimum $T_b$ required is essentially the temperature corresponding to the saturation intensity of the maser, weighted by the ratio of the inhomogeneous to the original un-power-broadened homogeneous line width, and the beaming factor, $4 \pi / \Omega$, where $\Omega$ is the beam solid angle of the maser.

\vspace{3mm}
It is timely to re-investigate coherence in astrophysical masers because of a combination of observational and technical developments. Firstly, \citet{1992ASSL..170.....E} concludes that masers are bandwidth incoherent on the basis of H$_2$O brightness temperatures of order 10$^{14}$ K. However, we now know of examples with significantly higher values of $T_b$. Since 22-GHz H$_2$O masers of 10$^{14}$ K miss the coherence criterion only narrowly, it is very likely that some brighter detected sources satisfy it, for example \citet{1982SvAL....8...86S} measured $8 \times 10^{17}$ K for $T_b$, and a figure of $> 10^{16}$ K \citep{2000aprs.conf..109K} has been recorded during a maser flare in the Orion source. Secondly, after a number of previous attempts to measure coherence from astrophysical masers, with negative results \citep{1972PhRvA...6.1643E,1973IzVUZ..16.1344P,1975SvAL....1...37L,1981BAAS...13..508M} towards OH and H$_2$O sources, a successful coherence measurement has been made towards a small number (W49, W75 and W3(OH)) of bright Galactic 22-GHz H$_2$O maser sources \citep{2016PASJ...68...86T}.

\vspace{3mm}
The method employed by \citet{2016PASJ...68...86T} is not the standard one for measuring radiation statistics (see below), but made use of technical developments in a software autocorrelator to bring about the new technique of ‘cross-correlation spectroscopy’. The new method uses an FX-style autocorrelator to correlate each amplitude-in-frequency not just with its complex conjugate from the same time sample, but also with other shifted samples, where no correlation would normally be expected. In the \citet{2016PASJ...68...86T} experiment samples were expected to be statistically independent for shifts $>$5\,$\mu$s. However, correlation was detected for significantly longer shifts, corresponding to coherence times of 10-30\,$\mu$s, depending on source and spectral feature. Moreover, the decay of the coherence with shift is not gaussian. In W3(OH), the measured coherence time of 18\,$\mu$s converts to a coherent bandwidth $1 / t_{\mathrm{co}}$ of 56\,kHz. This bandwidth corresponds approximately to the power-broadened homogeneous line width, and is a substantial fraction of the Doppler width (96\,kHz).

\vspace{3mm}
The cross-correlation spectroscopy method is directly relevant to the TNRT. It was originally developed as a single-dish technique, and can simultaneously analyse many spectral features. A suggested programme might be to confirm previous observations, survey a variety of bright maser sources to see how widespread the phenomenon is, and to study the development of coherence with maser power, noting that homogeneous profiles for most astrophysical masers without power broadening are typically of order 1\,Hz.

\vspace{3mm}
The more traditional method of detecting maser coherence is to test the statistics of the radiation for departures from gaussianity. Such departures may be quantified through the function $L_n = (\langle I_{n}^{2} \rangle/\langle I_{n}\rangle^{2}) - 1$, where the subscript n refers to a statistically independent Fourier component of width equal to the reciprocal of the sampling time. We note that $L_n$ involves a correlation in intensity, rather than electric field amplitude, and measurement of this function is therefore somewhat analogous to the the optical Hanbury-Brown and Twiss intensity interference experiments. If the statistics of the radiation are gaussian, all higher-order correlations may be expressed in terms of the first two, and we may write $\langle I_{n}^{q}\rangle = q! \; \langle I_{n}\rangle ^{q}$ \citep[][Loudon 2000]{1978PhRvA..17..701M}. Therefore $L_n$ has the limiting value 1 for gaussian radiation; it falls below one, most strongly at line centre, for light with residual coherence from stimulated emission \citep{1978PhRvA..17..701M,2009MNRAS.396.2319D}.

\vspace{3mm}
The underlying physics of non-gaussian radiation statistics is again related to power broadening of the homogeneous line shape function in very intense masers. An unsaturated maser has a homogeneous line shape with a width controlled by the rate of processes that destroy coherence, including collisions, spontaneous emission and the absorption of pumping radiation. This original width is independent of the maser intensity. As the maser saturates, the Stark effect due to the electric field of the maser radiation distorts the molecular response, resulting in a broadening of the width. Once this effect becomes dominant, the homogeneous width becomes proportional to the maser intensity. The molecular response in this regime is very different from that of an isolated molecule: it strongly favours emission in the direction of an existing beam, and the effective Einstein coefficients are multiplied by the photon occupancy number, proportional to the maser intensity. The system favours emission in a series of pulses. Pulsation may be directly detectable on timescales corresponding to the reciprocal of the power-broadened width (microsecond to second timescales, dependent on saturation state, species and transition).

\vspace{3mm}
The radiation statistics method is also directly applicable to the TNRT as a single-dish method, though it can also be usefully applied to interferometry when the TNRT is part of a VLBI network. Polarization measurements are not required.

\vspace{3mm}
Scientific value: Answers question of whether astrophysical masers share any of the coherence properties associated with laboratory lasers. If previous results are confirmed, extend source list to include a range of massive Galactic star-forming regions with 22-GHz water masers.

\vspace{2mm}
Bands: K, for 22-GHz H$_2$O masers, which have the highest brightness temperatures. Timescales: deep one-off observations of each source, narrow channels (1~Hz).

\vspace{3mm}
\subsection{VLBI Vision with TNRT in High-mass SFRs}
\label{subsec3:vlbi}

Interstellar masers reveal three-dimensional (3-D) velocity structures around (proto-)stars and Galaxies on the basis of proper motion measurements on the sky plane. In star-forming regions, for instance, H$_{2}$O masers that are collisionally pumped are well known to trace the 3-D velocity structure of a high-velocity jet or a low-velocity outflow \citep[e.g.,][]{1981ApJ...244..884G,1981ApJ...247.1039G,1992ApJ...393..149G,2000ApJ...534..781L,2002PASJ...54..741I,2003ApJ...598L.115T,2011MNRAS.410..627T,2005PASJ...57..595H,2005A&A...438..889M,2006A&A...447..577G,2007PASJ...59..897H,2008MNRAS.390..523M,2016PASJ...68...69M,2008PASJ...60..183N,2010PASJ...62..287S,2010ApJ...720.1055S,2012ApJ...748..146C,2020PASJ...72...54C,2021ApJ...908..175C,2016MNRAS.460..283B,2017MNRAS.467.2367B} and these were promised through 3-D velocity measurements towards a statistically large sample of 40 high-mass YSOs by The Protostellar Outflows at EarliesT Stages (POETS) survey project, resulting in that H$_{2}$O masers trace the YSO position within a few 100 au and their proper motions align with the YSO's jet direction within 30$^{\circ}$ \citep[e.g.,][]{2018A&A...619A.107S,2019A&A...631A..74M,2020A&A...635A.118M},
while OH masers are expected to be associated with the edge of an expanding HII region \citep[e.g.,][]{1992MNRAS.254..501M}. The CH$_{3}$OH masers classified into the class I that is represented in 36.2, 44.1, and 95.2 GHz are thought to be excited at a shock region formed by a low-velocity outflow and surrounding ambient gases \citep[e.g.,][]{2004ApJS..155..149K,2009ApJ...702.1615C,2010A&A...517A..56F,2014MNRAS.439.2584V}, while CH$_{3}$OH masers classified into the class II that is represented in 6.7 and 12.2 GHz are expected to trace a rotating and/or expanding/infalling disk on the basis of their spatial distribution and 3-D velocity structure \citep[e.g.,][]{1993ApJ...412..222N,1998MNRAS.300.1131P,2000A&A...362.1093M,2009A&A...502..155B,2020A&A...637A..15B,2010A&A...517A..71S,2010A&A...517A..78S,2011A&A...535L...8G,2014PASJ...66...31F,2014A&A...566A.150M,2014A&A...562A..82S,2017ApJ...849...23M} and these were also verified towards multiple target sources presenting accretion bursting activities, such as S255IR-NIRS3 and G358.93-00.03 MM1 \citep[e.g.,][]{2017A&A...600L...8M,2020NatAs...4..506B}. 
The SiO masers that are rarely detectable in high-mass SFRs \citep[e.g.,][]{2009ApJ...691..332Z,2015ApJ...815..106H,2016ApJ...826..157C} might trace a disc wind emanating from the surface of the circumstellar disc with its diameter of $\sim$100~au surrounding Source I in the famous high-mass SFR Orion KL \citep{2008PASJ...60..991K,2009ApJ...698.1165G,2010ApJ...708...80M,2012A&A...548A..69N,2013ApJ...770L..32G,2017A&A...606A.126I,2019ApJ...872...64K}, as in the case with a recent discovery of direct observational proof for a magnetohydrodynamic disc wind traced by an extraordinary H$_2$O maser in IRAS~21078$+$5211 \citep{2022NatAs...6.1068M}.

\vspace{3mm}
On the basis of their tracing a variety of physical phenomena, spatial scales, and difference phases in high-mass star evolution, in which some of them are overlap and the masers could be thus probe for chronology as well \citep[e.g.,][]{2007IAUS..242..213E,2010MNRAS.401.2219B,2012ApJ...760L..20C}, simultaneous VLBI monitoring toward multi-species masers in the same source would enable us to reliably estimate the evolutionary state of a source from its 3-D velocity and spatial structures: How does a protostar grow through accretion?, does infall onto the accretion disc follow free-fall or spiral?, how can we follow the development and ejection of jets/outflows?, what are their sizes and time-scales?, and so on. 
Such investigation has been initiated with the KVN and VERA Array (KaVA) as the Large program\footnote{\href{https://radio.kasi.re.kr/kava/large_programs.php}{https://radio.kasi.re.kr/kava/large{\_}programs.php}} since 2016 toward the 22.2~GHz H$_{2}$O and the 44.1~GHz CH$_{3}$OH masers in 87 HMSRFs targets, of which evolutionary phases are different from each other (Hirota, Kim, et al. and  KaVA Star-Formation Science sub-Working Group). They have published a result of the VLBI map of the 44.1~GHz CH$_{3}$OH maser in IRAS~18151$-$1208 for the first time in the world \citep{2014ApJ...789L...1M}, a result of the VLBI map of the 22.2~GHz H$_{2}$O maser in G25.82$-$0.17 for discussion of its maser associated with large-scale outflows and related to the location of an exciting source via comparing mm continuum and molecular lines with ALMA \citep{2020ApJ...896..127K}, and the 2nd yr program has been already initiated to complete the measurements of 3-D velocity structures. 

\vspace{5mm}
\noindent{\textbf{What could the 40-m TNRT do?}}

\vspace{3mm}
As presented in figure~\ref{fig1.2:uveavn} and \ref{fig1.2:beameavn} of section~\ref{subsec1.2:vlbi}, TNRT will drastically improve via filling gaps of uv-holes in the UV-coverage on the basis of its unique location and reducing side-lobe of the synthesized-beam of the EAVN, in which TNRT forms the longest baseline in the KaVA, providing $\sim$2 times longer than one at current. Besides TNRT is one of the largest radio telescope with its diameter of 40 m in KaVA/EAVN, providing a high-sensitivity together with wide-range of frequencies that are observable for all species of masers. This improvement will thus raise a stage of the KaVA Large program with higher spatial resolution and baseline/imaging sensitivities, achieving more complete images including weaker maser spots and measurements of more tiny proper motions on the sky plane as well. 
Furthermore, C-band receiver will be installed on TNRT within a few years at phase 2 of the development. This upgrade will strongly contribute to a complementary VLBI observations of the 6.7~GHz CH$_{3}$OH masers in the KaVA Large program targets with the EAVN (Sugiyama, Hirota, Kim, et al., on-going). 

\vspace{3mm}
Simultaneously, this improved array together with TNRT will upgrade the world-wide network for immediate following-up of accretion bursting activities in masers, entitled ``Maser Monitoring Organization" (M2O)\footnote{See M2O website at \href{https://www.masermonitoring.com/}{https://www.masermonitoring.com/}} since 2017, via observing multi-species masers in wide-range of their frequencies. These follow-up observations are achievable by immediate trigger by M2O to all the members once after discover and/or detection of flux flaring or bursting increment actions in any masers monitored with their radio telescopes. Such immediate follow-ups with not only VLBI but also Infrared instruments have been achieved at bursting maser source G25.65+1.05, G358.93-00.03 MM1, and S255IR-NIRS3 \citep[e.g.,][]{2020MNRAS.491.4069B,2020NatAs...4..506B,2021A&A...646A.161S,2020PASJ...72....4U,2021A&A...647A..23H}, which were triggered by radio telescopes in M2O \citep[e.g.,][]{2017ATel10728....1V,2017ATel10853....1V,2019ATel12446....1S,2015ATel.8286....1F}.


\clearpage
\section{Galaxy and AGNs}\label{sec4:galaxy}

\textit{Led by Kazuhiro Hada, Malcolm D. Gray, Richard Dodson, Maria Rioja, Koichiro Sugiyama, et al.}

\begin{table}[h]
\centering
 \caption{Summary of basic parameters for each topic in this section.}
  \begin{tabular}{lp{80mm}|lccc} \hline\hline
    Sec. & Topic & \multicolumn{1}{c}{Band} & Single-dish? & VLBI? & Pol.?\\
    \hline
    \ref{subsec4:jetvlbi} & VLBI Monitoring of AGN jets & LCXKuKQW & $\times$ & $\bigcirc$ & $\bigcirc$ \\
    \ref{subsec4.2:megamaser} & Search for Cosmologically Important Megamasers & LC \hspace{5.8mm} K & $\bigcirc$ & $\times$ & $\times$ \\
    \ref{subsec4.3:astrometry} & Development for Galactic Astrometry & LC \hspace{5.8mm} KQW & $\times$ & $\bigcirc$ & $\times$ \\
    \hline
  \multicolumn{6}{p{175mm}}{\small Note.-- Columns~1, 2. number and name of sub-sections; Column~3. frequency bands to be used for each topic; Columns~4--6. necessity of ways to observe as single-dish, VLBI, and/or polarization, respectively.}
  \end{tabular}
\end{table}

\subsection{VLBI Monitoring of AGN jets}\label{subsec4:jetvlbi}

\noindent
Active galactic nuclei (AGN) are powered by the accretion of material onto their central supermassive black holes (SMBH), and approximately 10\% of AGN exhibit powerful relativistic jets. AGN jets are one of the most spectacular phenomena in the Universe and can span observable scales from subpc to Mpc scales. Understanding the formation processes of jets from accreting SMBH is one of the primary goals in high-energy astrophysics. AGN jets transport a significant fraction of BH accretion energy back into their host galaxies and intergalactic space, so studies of AGN jets are also important for understanding the formation and cosmological evolution of galaxies. According to the recent general relativistic magnetohydrodynamic simulations, powerful relativistic jets can be ultimately launched from a spinning black hole or an inner part of magnetized accretion flows~\citep[e.g.,][]{2022NatAs...6..103C}. To observationally access such tiny scales, high-resolution VLBI that achieves angular resolution of mas to $\mu$as scales are necessary. The best example is probably the recent Event Horizon Telescope (EHT - a global VLBI network operated at 230\,GHz) observations of the nearby radio galaxy M87, where the shadow of SMBH at the jet base was successfully imaged~\citep{ehtc2019}. Nevertheless, VLBI at millimeter wavelengths is generally less sensitive to the extended emission surrounding the black hole, preventing us from studying the detailed connection between the central SMBH and the large-scale jet or surrounding accretion flows.  

\vspace{3mm}

In this context, VLBI observations at longer wavelengths (1--43/86\,GHz) are optimal to study the morphology, dynamics and polarization of AGN jets, complementing mm-VLBI. In fact, various important jet physics and questions such as the collimation and acceleration are associated with a wide range of distances above subpc/pc scales (roughly corresponding to $\sim$100--10$^5$ gravitational radii from BH), which are challenging to probe at mm wavelengths. Recent extensive cm-VLBI observations of nearby AGN jets reveal that an increasing number of sources show morphological transitions in their jet shape from parabolic to conical around the Bondi radius that are roughly 10$^{4-6}$ gravitational radii from SMBH~\citep[e.g.,][]{asada2012, park2021, boccardi2021}. On one side, multi-epoch VLBI monitoring of AGN jets indicate that the observed jet speed gradually accelerates in the region where the jet shape is parabolic, suggesting that the collimation and acceleration are cospatial, which is in agreement with the expectation from magnetically-driven jet models~\citep[e.g.,][]{park2019}. Moreover, some AGN jets show a bright stationary, highly polarized shocked feature at the location of the jet transition, which could be a potential site of high-energy emission up to TeV $\gamma$-rays~\citep[e.g.,][]{asada2012, hada2018}. However, the number of samples in which the jet shape and velocity fields in ACZ were measured are still limited due mainly to the limited angular resolution and sensitivity of existing VLBI array.  

\begin{figure}[t]
  \begin{minipage}{0.49\linewidth}
    \centering
    \includegraphics[clip,width=\textwidth]{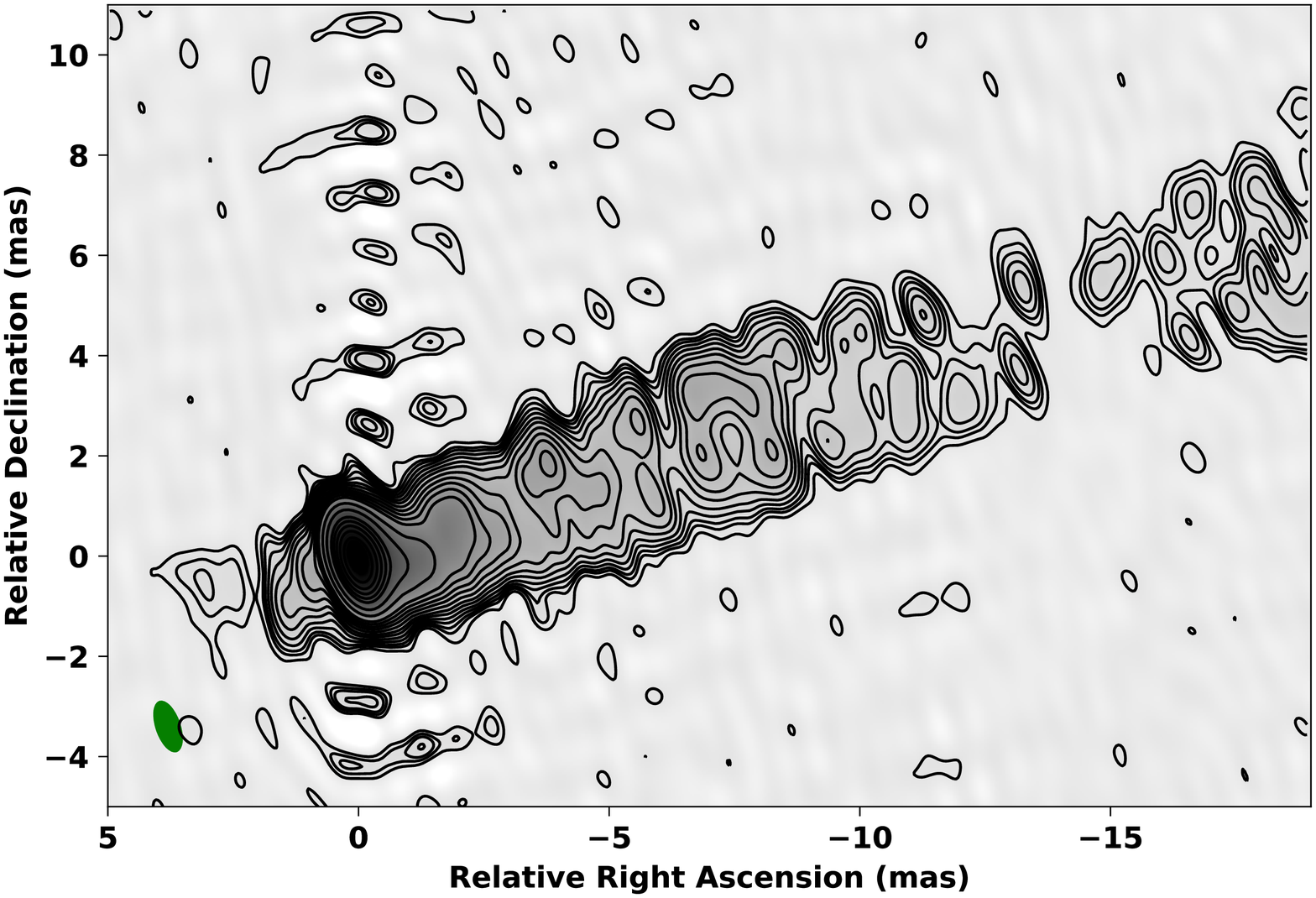}
  \end{minipage}
  \begin{minipage}{0.49\linewidth}
    \centering
    \includegraphics[clip,width=\textwidth]{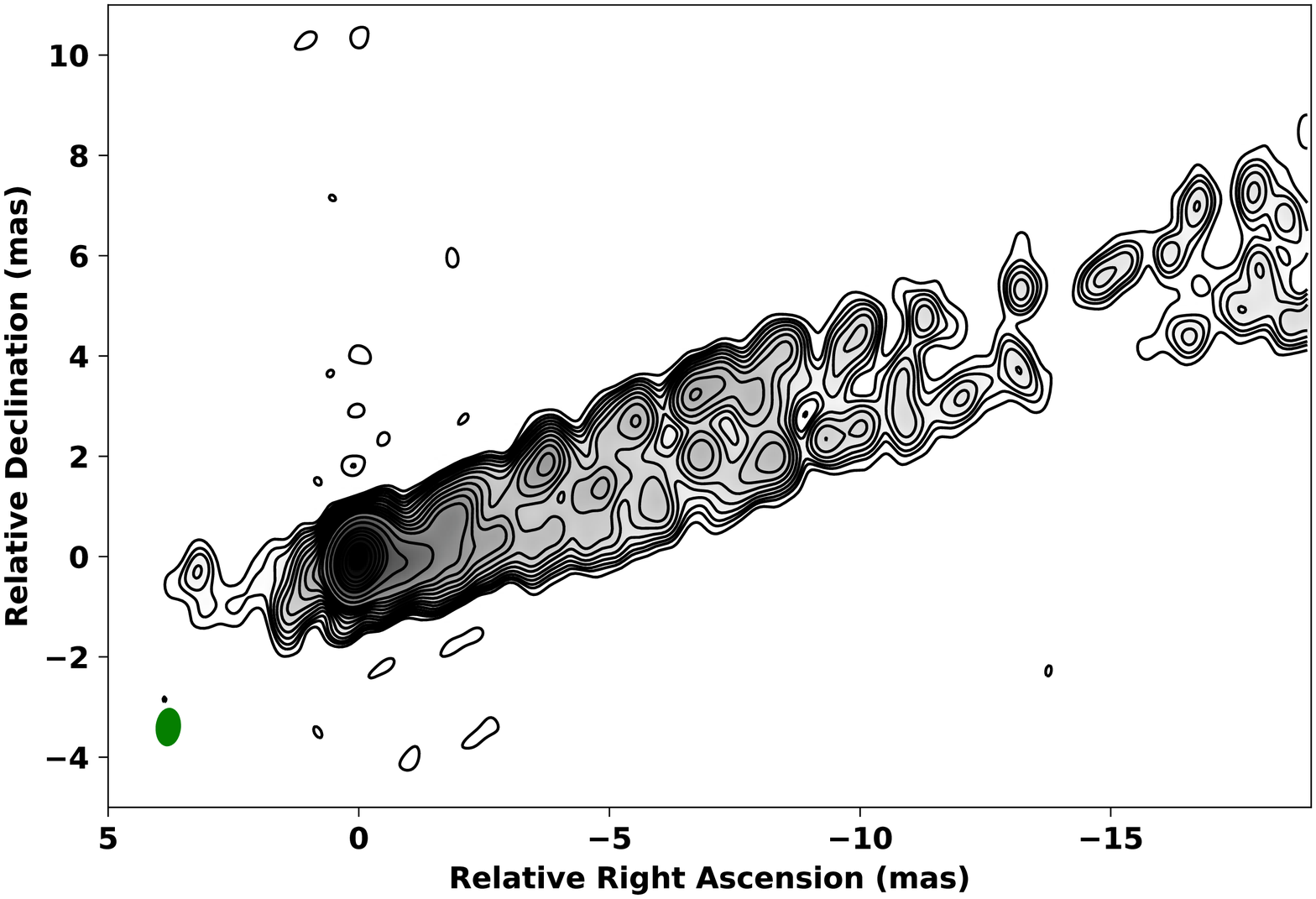}
  \end{minipage}
  \caption{Simulated VLBI CLEAN (uniformly weighted) images of M87 at 22\,GHz. (Left) image reconstructed with EAVN (KaVA+Tianma+Nanshan+Takahagi). (Right) image reconstructed with EAVN+TNRT. Beam size is shown as a green ellipse in the bottom left corner of each panel. For both images, the contours start from $-$1, 1, 2... times 240~$\mu$Jy/beam and increase by factors of $\sqrt{2}$.}
  \label{fig:m87vlbi}
\end{figure}

\vspace{3mm}
Thanks to its unique geographical location, the addition of TNRT to existing cm-VLBI networks (e.g., EAVN at K band and EVN at L band for a commissioning phase) offers an excellent opportunity to resolve and monitor a larger number of AGN jets. As a representative example, in figure~\ref{fig:m87vlbi} we show simulated EAVN 22\,GHz images of the M87 jet, one without TNRT and the other with TNRT. Thanks to a large improvement of (u,v) coverage along the north-south direction, one can see $\sim$30\% improvement of the north-south angular resolution. This enables us to better resolve the core at the jet base and the transverse structure of the extended jet. Besides, the addition of TNRT greatly improves the image dynamic range of EAVN thanks to the large collecting area of TNRT as well as the improved (u,v) coverage. Similarly at L band, we can also expect a significant improvement of both angular resolution and sensitivity when TNRT is connected to EVN. TNRT also plays a unique role in bridging VLBI stations between the northern and southern hemispheres. TNRT nicely fills the (u,v) gap between East Asia and Australia, which will greatly reduce the sidelobes in the image domain for VLBI with $\sim$8000\,km north-south baselines. Such a VLBI array is especially powerful for observations of AGN in the southern hemisphere such as Centaurus A, Sombrero galaxy and also the Galactic Center SgrA$^*$.

\vspace{3mm}

Beyond L and K bands with which TNRT will start its commissioning, further addition of QW and CXKu receivers at TNRT will further enhance our science capability on AGN jets. This allows us to obtain spectral information over a wide range of frequency coverage, which is important to accurately constrain the opacity and magnetic field strength of the innermost jet regions. Also, dual-polarization capability of TNRT with these multi-frequency receivers additionally enable us to study the Faraday rotation measure of the jet in great detail, which is the key to reveal the topology of magnetic fields as well as to constrain the origin of the Faraday screen. We also stress the importance of single dish dense monitoring of AGN targets, which is complementary to high-resolution VLBI. This is especially relevant to active $\gamma$-ray blazars, which are highly variable also in radio bands due to the strong Doppler beaming effect with small jet viewing angles. Blazars are also suggested as a potential candidate of high-energy neutrino emission. Cross-correlation analysis of densely sampled light curves between TNRT, Cherenkov Telescope Array (CTA) and IceCube will help us to localize the site of active particle acceleration.

\subsection{Search for Cosmologically Important Megamasers}\label{subsec4.2:megamaser}

\noindent
The Hubble constant, $H_0$, is perhaps the best-known cosmological parameter, and an accurate value is vital to constrain the range of possible cosmological models. At present, the {\it formally} most accurate value is 67.4$\pm$0.5~km~s$^{-1}$~Mpc$^{-1}$, provided by results from Planck observations \citep{2018A&A...617A..48P}. However, the Planck result is dependent upon a cosmological model, specifically the $\Lambda$CDM paradigm, so it is important to continue investigating the Hubble constant via methods that are independent of such assumptions and, if possible, based solely on geometry.

\vspace{3mm}
One of the best independent methods is based on observations of H$_2$O megamasers, with NGC4258 being the original object of this type. This method is purely geometric, and based on local Universe measurements, fulfilling the requirements of independence from cosmological models. The best estimate of $H_0$ using this method is from the Megamaser Cosmology Project \citep[MCP:][and works cited within]{2020ApJ...891L...1P}, with a value of 73.9$\pm$3~km~s$^{-1}$~Mpc$^{-1}$. It is interesting that this figure is in statistical tension with the Planck result, with only a 2 per cent probability that the values of $H_0$ are consistent \citep{{2020ApJ...891L...1P}}.

\vspace{3mm}
The results in \citet{{2020ApJ...891L...1P}} are based on the original galaxy, NGC4258, and five more distant objects. NGC4258 is too close to us for a typical peculiar velocity, of order 250~km~s$^{-1}$, to be negligible compared to the recessional velocity from the Hubble flow. It would therefore obviously be beneficial to discover more objects in this class with recessional velocities in the range 2500-12000~km~s$^{-1}$ in order to improve the statistics of the megamaser method. It is desirable to extend the data set to more distant objects, but this is less important than finding more in the Hubble flow range presented in \citet{{2020ApJ...891L...1P}}.

\vspace{3mm}
The TNRT would be able to help with providing a larger sample of magamaser galaxies for the MCP, by detecting megamasers with the right characteristics (cosmological red-shift, broad three-component spectrum) that indicate a distant, nearly edge-on, maser disc, of  approximately parsec scale, orbiting a super-massive black hole. We note that surveys for galaxies in this class have been carried out, mainly with the 100-m Greenbank telescope (GBT), for example \citet{2010ApJ...718..657B}, and that similar observations are ongoing \citep[reference to a forthcoming paper in ][]{2017ApJ...834...52G}. We would expect to use similar methodology with the TNRT. That is making a targeted survey, based on optical identification of candidate galaxies as type-2 Seyferts, or based on IR colours of galaxies. Obviously the TNRT is not as sensitive as the GBT: if the sensitivity is goverened solely by dish area, the TNRT would need to observe for 6.25 times longer to achieve the same limiting flux density. However, the TNRT has the advantage that it has a considerably better view of the Southern sky, and is likely to be able to detect some sources not visible to the GBT. It should be emphasized that the principal aim is to secure more galaxies in the current range of recessional velocities, rather than to extend the sample to greater red-shift.

\vspace{3mm}
Scientific value: One problem with NGC4258 itself is that it is too close to have a velocity dominated by the Hubble flow. It would be highly beneficial to increase the number of galaxies where a measured distance and red-shift can be converted directly to an estimate of H$_0$. 

Bands: K. Timescales: very deep one-off observations of candidate sources based on IR colours or identification as Seyferts.

\subsection{Development for Galactic Astrometry}\label{subsec4.3:astrometry}

\noindent
In addition to high quality direct imaging of interesting compact objects the TNRT will contribute greatly to accurate astrometric surveys of these objects. 
The conventional imaging approach of self-calibration has the consequence of losing information on the source position on the sky, as this is absorbed into the calibration solutions. Phase referenced VLBI retains this astrometric information, relative to the target source.
Conventional phase referencing requires interleaved measurements of the target source in comparison to a known (i.e. reference) source, and this provides a host of astrometric information that allows us to tackle multiple types of astronomical questions.
Astrometry is routinely a hundred-fold more precise than the resolution, and this allows for accurate distance determination for more or less any detectable source in our Galaxy.

\vspace{3mm}
The instrumental requirements are a sensitive telescope, with rapid slewing. The sensitivity is required to improve the signal to noise on the measurements, to provide the required dynamic range. The rapid slewing is to allow for the tracking of the rapid phase variations introduced by the atmosphere. 
The TNRT design is optimal, as it is large and therefore sensitive yet not so large that rapid slewing is not possible.

\subsubsection{Commissioning Phase}

\paragraph{L-band}
The L-band feed at the prime focus covers the OH-maser lines, and these are suitable targets for an astrometric program.
Conventional phase referencing at L-band is limited by the systematics from the Ionosphere, in most cases.
For example, with a calibrator at 1$^o$ the astrometric errors would about 1\,mas. Even with an in-beam calibrator at, say, 6$^\prime$ the astrometric errors would be 100$\mu$as.
This would limit the measurement of parallaxes (at the nominal 10\% accuracy) to sources within 1\,kpc.
However the new astrometric method of MultiView \citep{2017AJ....153..105R} would allow astrometric errors potentially as low as 10$\mu$as \citep{2020A&ARv..28....6R} .
A MultiView campaign of observations of OH-masers, in conjunction with the Australia Long Baseline Array (LBA) would be able to populated the Period-Luminosity (P-L) plane for sources across the Galaxy, and particularly in the direction of the Galactic centre.

\paragraph{K-band}
The BeSSeL survey \citep[e.g.][]{2014ApJ...783..130R,2019ApJ...885..131R} with the VLBA of the distance to 22\,GHz water masers in HMSR around the Galaxy has allowed us to accurately measure the rotation curve of our Galaxy, and deducing a significantly closer distance for the Sun to the Galactic centre and thus a increase in the mass of our Galaxy.
These results are derived from measurement of the maser parallaxes combined with the radial velocity of the maser emission.
The VERA network of radio telescopes has been performing high quality PR observations of water masers, as a compliment to the BeSSeL survey. Adding TNRT to the this and the whole East Asia VLBI network would double the baseline length, increasing the resolution.

\subsubsection{Phase Two}
\paragraph{C-band}
\typeout{Then multi view phase referencing}

TNRT could, alternatively, join the EAVN for observations of the 6.7GHz methanol maser, using MultiView PR. MultiView will allow the systematic errors in the astrometry to be dramatically reduced from about 50$\mu$as to less than the thermal limits ($\sim$10$\mu$as), for an observation with a Dynamic Range of 100.
Whilst it would be hard to better the VLBA observations at 22GHz, high precision 6.7GHz parallaxes should be possible.
There is an opportunity to apply the recently developed astrometric method of MultiView \citep{2017AJ....153..105R}, that should deliver an order of magnitude improvement in astrometric accuracy at 6.7GHz. 

Note that TNRT, or the Geodetic antennas, would be able to make a significant contribution to the AuScope BeSSeL-South project at these frequencies.

\typeout{Then frequency phase referencing}
\paragraph{Lambda ($\lambda$) Astrometry at K,Q and W bands}

The proposed simultaneous tri-band receiver packages on the TNRT will allow for some extremely innovative calibration methods. 
When one is able to observe multiple frequencies simultaneously the calibrations from one frequency can be scaled and applied to another \citep{2011AJ....141..114R}.
This is particularly useful when the goal is not to reference the astronomical images to an external frame, but to each other, as is often the case. Common examples of this are for multi-frequency polarisation and rotation measure studies. 
In particular the relative brightness of the H$_2$O and SiO masers around AGB stars is a sensitive probe of the physical and astro-chemical conditions. 
However the observations of these different transitions must be accurately registered allow for the correct interpretation. 
This is a major program of the Korean VLBI Network, but this is limited by their $\sim$500km baselines. The addition of a $\sim$3,000km baseline to TNRT would increase the resolution by nearly an order of magnitude.


\clearpage
\section{Evolved Stars}\label{sec5:evolved}

\textit{Led by Sandra Etoka, Anita M.S. Richards, Hiroshi Imai, Bannawit Pimpanuwat, and Malcolm D. Gray}

\vspace{-6mm}
\begin{table}[h]
\centering
 \caption{Summary of basic parameters for each topic in this section.}
  \begin{tabular}{lp{80mm}|lccc} \hline\hline
    Sec. & Topic & \multicolumn{1}{c}{Band} & Single-dish? & VLBI? & Pol.?\\
    \hline
    \ref{subsec5:photometry} & Spectro-photometry of Masers around Evolved Stars & LC \hspace{0.3mm} KuKQW & $\bigcirc$ & $\triangle$ & $\bigcirc$ \\
    \ref{subsec5:CSE-maser-vlbi} & Circumstellar Dynamics and Galactic Kinematics Revealed in VLBI & \hspace{12.9mm} KQW & $\bigcirc$ & $\bigcirc$ & $\bigcirc$ \\
    \ref{subsec5:SiOandContinuum} & Thermal SiO Line and Stellar Continuum Emissions & \hspace{12.9mm} KQW & $\bigcirc$ & $\times$ & $\triangle$ \\
    \hline
  \multicolumn{6}{p{175mm}}{\small Note.-- Columns~1, 2. number and name of sub-sections; Column~3. frequency bands to be used for each topic; Columns~4--6. necessity of ways to observe as single-dish, VLBI, and/or polarization, respectively. circle: necessity, triangle: useful, cross: not required}
  \end{tabular}
\end{table}

\vspace{2mm}
\noindent
Studies of evolved stars such as asymptotic giant branch (AGB) and post-AGB stars and red-supergiants will investigate copious stellar mass loss in the final stages of stellar evolution, interaction of binary systems and their evolution into close binary or symbiotic star systems, and stellar explosions such as novae, supernovae, and their remnants. In those objects, thermal and non-thermal emission sources are expected, with rapid variation or evolution within an observatory's lifetime (typically $\sim$50~yr). 

\vspace{3mm}
Typical visible target spectral lines are listed for evolved stars as follows.

\begin{figure}
    \centering
    \includegraphics[width=0.9\textwidth]{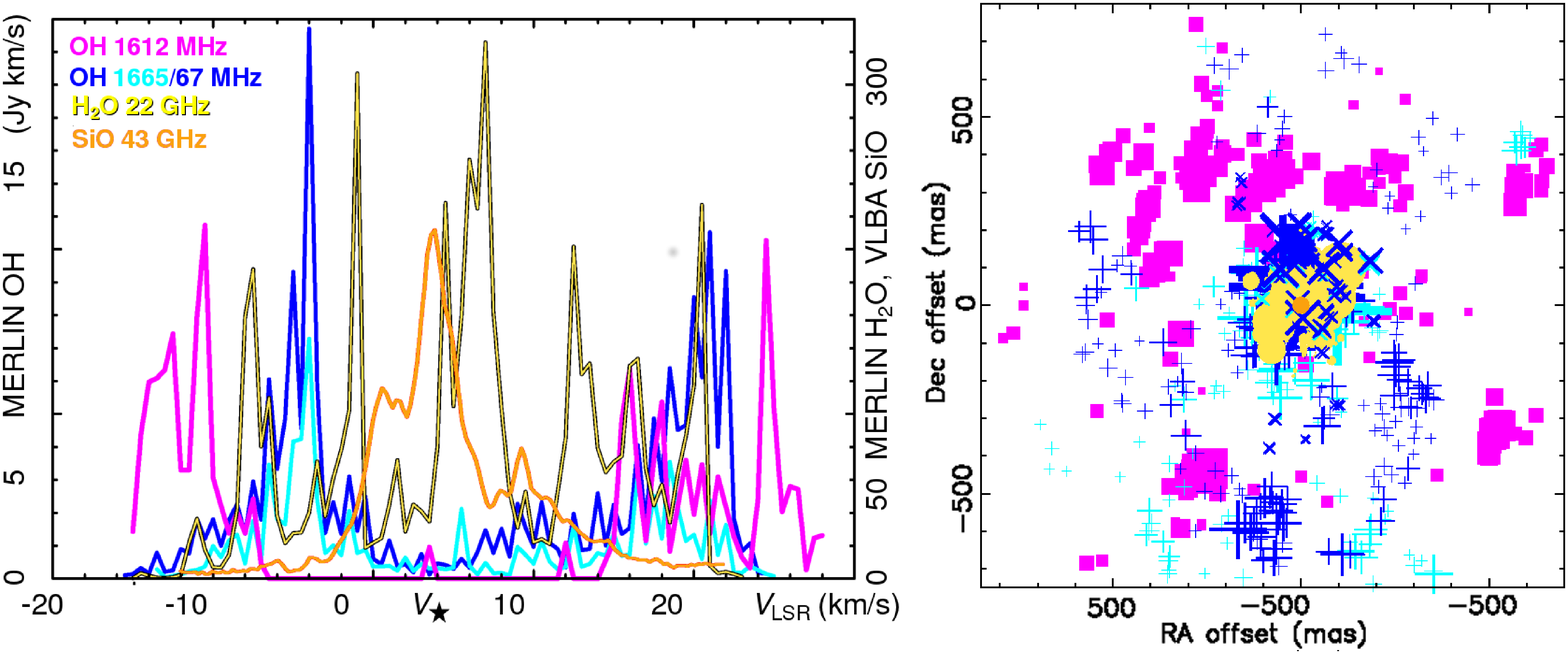}
    \caption{Spectra and positions of maser components around VX Sgr. Data from \citet{1997MNRAS.288..945S} (OH 1612 MHz), Bartkiewicz priv. comm. (OH mainlines), \citet{2003MNRAS.344....1M} (H$_2$O), \citet{2007IAUS..242..312C} (SiO)}
    \label{fig:VXall.png}
\end{figure}

\begin{enumerate}
    \item A large number of circumstellar maser lines are observable such as: OH (at 1612, 1665, 1667, 1720, and 6035 MHz), H$_2$O (22235 MHz), SiO (42519, 42821, 43122, 43423, 86243 MHz), isotopologues, such as $^{29}$SiO (42880 MHz) HCN (20181, 88631 MHz). They are located in circumstellar envelopes (CSEs) but at different radial distances from the central star, from 2 to many hundreds of stellar radii.
    \item Thermal lines of SiO $v=0$ $J=1\rightarrow 0$ (43423 MHz), which are weak but detectable (and eventually $J=2\rightarrow 1$ at 86846 MHz). Thermal SiO, being radiatively excited, highlights interaction regions if the stellar wind is asymmetric or irregular.
    Comparison of the line profile of SiO with that of CO taken with other large telescopes such as Nobeyama 45m and Mopra 20m telescopes enables us to extract the velocity field in the CSE and measure its radial acceleration. The CO $J=1\rightarrow0$ line is also a target of the TNRT, dependent on the TNRT performance at this frequency band and the current progress of CO mapping for nearby evolved stars with the JCMT and Nobeyama 45m telescope.
    \item Detection of circumstellar HI emission at 1420~MHz is considered but challenging.
    \item Radio recombination lines from planetary nebulae is also considered but challenging.
    \item Non-thermal emission such as that in Algol but at later evolutionary stage provides diagnostics of magnetic activity of evolved stars in interacting binary systems.   Non-thermal emission from interacting binaries such as Wolf-Rayet stars is also variable but for reasons of space is not covered here. Most stars are only weak radio continuum emitters, but monitoring thermal emission from nearby supergiants could be possible.  See Sec. 6 for the TNRT's potential for other radio star studies. 
\end{enumerate}
{Fig.~\ref{fig:VXall.png} shows the relative location of the main O-rich masers around RSG VX Sgr} and the relationship between expansion velocity and greater angular separation from the star - with some exceptions.

\subsection{Spectro-photometry of Masers around Evolved Stars}
\label{subsec5:photometry}

Masers originating from several molecular species are commonly found in the CSEs of AGB stars and red supergiants. In O-rich AGB stars, there is evidence, summarised for example in \citet{2012msa..book.....G}, Ch.~6.2, that circumstellar masers form in concentric radial zones, where each of these typically corresponds to a particular maser 
transition (for example of SiO, H$_2$O, OH) and therefore emits under distinct physical conditions. 
The CSEs of O-rich red supergiants superficially resemble such AGB winds, scaled up roughly in proportion to the order of magnitude larger spatial scale of the star itself.
In C-rich stars, the masers are instead emitted from HCN, CS, SiS and CO molecules, including  two HCN maser transitions at 88.6 and 89.1\,GHz, observable with TNRT when W-band receivers are installed (see above). From now on, however, we will mainly discuss the masers in the CSE of O-rich stars.

SiO masers are abundant in the region closest to the stellar surface where the density, kinetic temperature and radiation field favour their excitation, out to 4-5 stellar radii \citep{2009MNRAS.394...51G}. See Sec.~\ref{subsec5:SiOandContinuum} for thermal SiO.
The strongest SiO maser transitions are those between rotational levels within the vibrational levels $v=1$ and $v=2$, but have been observed from vibrational states up to $v=5$. 
On average, stellar cycle-wise, masers with low rotational levels are correlated to the optical cycle, albeit with a phase lag of approximately 0.1 periods. In terms of the mechanisms responsible for generating SiO masers, both radiative and collisional pumping schemes have been shown to play a significant role, where the former is evidently favoured in the $v=0$ lines while the latter is more dominant in $v>0$ masers implying predominantly tangential amplification. Linear polarization, in the range of 10-30 per cent, with the upper limit as high as 40 percent, appears to be the norm for SiO masers, although circular polarization has also been observed at a typical level of a few per cent, but up to $\sim$ 20 per cent, for example \citet{2019ApJ...871..189T}. 

The next concentric radial zone in the CSE is the site of 22 GHz water masers, forming at 
10-100 stellar radii; the velocity profile at these radial distances shows more consistent, accelerating expansion outside the attenuation of pulsation shocks.
The best-studied line
which will be one of the main targets of TNRT, is at 22.235 GHz, but other higher-frequency H$_2$O maser transitions may be included in future upgrades to the telescope, for instance, 96 and possibly 68 GHz. 
H$_2$O maser features appear spectrally narrow, having typical widths of $\leq$1 km/s. VLBI shows spots with apparent, beamed sizes of a few mas or less forming features extending $\sim$0.1-0.2 stellar radii. 
Water masers are collisionally pumped and often associated with shocks.
Monitoring (e.g. \citealt{2018IAUS..336..393B} and references therein) has shown some 
periodicity (maser delay from optical of order 0.2 periods) and erratic behaviour of the emission, including sudden, dramatic flares. Polarization in H$_2$O masers is generally $\leq$ 2 per cent (linear) and up to 20 per cent (circular).

Ground-state OH masers need cooler, less dense conditions as found further out in the CSE, although mainline OH masers (1665 and 1667 MHz) sometimes appear to overlap the outer 22-GHz maser zone. The mainline masers can be very variable but in general appear as double peaks in thicker-shelled stars.
The 1612 MHz OH maser transition is typically observed at very large (hundreds to thousands) of stellar radii  where H$_2$O is photodissociated by external radiation. Pumping is mostly radiative and spectral peaks are often seen at close to the terminal velocity of shell expansion. Distance measurements can be performed using the phase lag (Section~\ref{subsec5:OH-phase-lag}). OH masers often show strong Zeeman splitting with some linear polarization.

\subsubsection{Phase-lag measurements and distance determination towards OH/IR stars}
\label{subsec5:OH-phase-lag}

Stars of intermediate mass (i.e., M $\le$ 8~M$_{\odot}$), when reaching the AGB, see their mass-loss rate increase dramatically, to values as high as $\dot{M}=10^{-4}$~M$_{\odot}$~yr$^{-1}$. This produces a (typically) spherical CSE, which eventually can obscure the central star. For such embedded stars, referred to as OH/IR objects, (reliable) optical (e.g. in particular Gaia) distance determination is not possible. For the shorter period Miras, the so called Period-Luminosity (PL) relation can be used to infer their distances, but it has been shown that this relation does not hold for variable stars having period greater than $\sim$450~days \citet{1991MNRAS.248..276W}
and hence cannot be used for OH/IR stars which periods are typically ranging between 1 to 5-6 years. Other distance determination such as kinematic distances and maser parallaxes are in theory possible but, the former can be substantially imprecise due to peculiar motions \citep{2009ApJ...700..137R},
while maser parallaxes requires the persistence of (compact) maser spots over the duration of a period which can be quite difficult to achieve especially for the objects exhibiting (very) long periods. 

OH/IR stars have the property of emitting a very strong (typically tens of Jy) 1612-MHz maser emission. The layer where this emission comes from in the radially expanding CSE is such that acceleration is not longer expected and the maser layer itself is of very small extent compared to the diameter of the shell itself. These basic properties of the CSE in terms of velocity field and structure and the need of velocity coherence for amplification of the maser emission to be detected in a given line of sight, produce the characteristic double-peak spectrum observed towards these objects with the ``blue(-shifted)'' peak emanating from the front cap of the CSE while the ``red(-shifted)'' peak from the back cap.

Phase-lag distance determinations rely on 2 independent measurements of the diameter of the OH-maser shell and are possible because of the very nature of the CSE itself and the fact that the OH maser emission, being radiatively pumped, follow the (optical) stellar pulsation albeit with a small delay (typically of 15-20\%).  

\begin{figure}[h]
    \centering
    \includegraphics[width=0.95\textwidth]{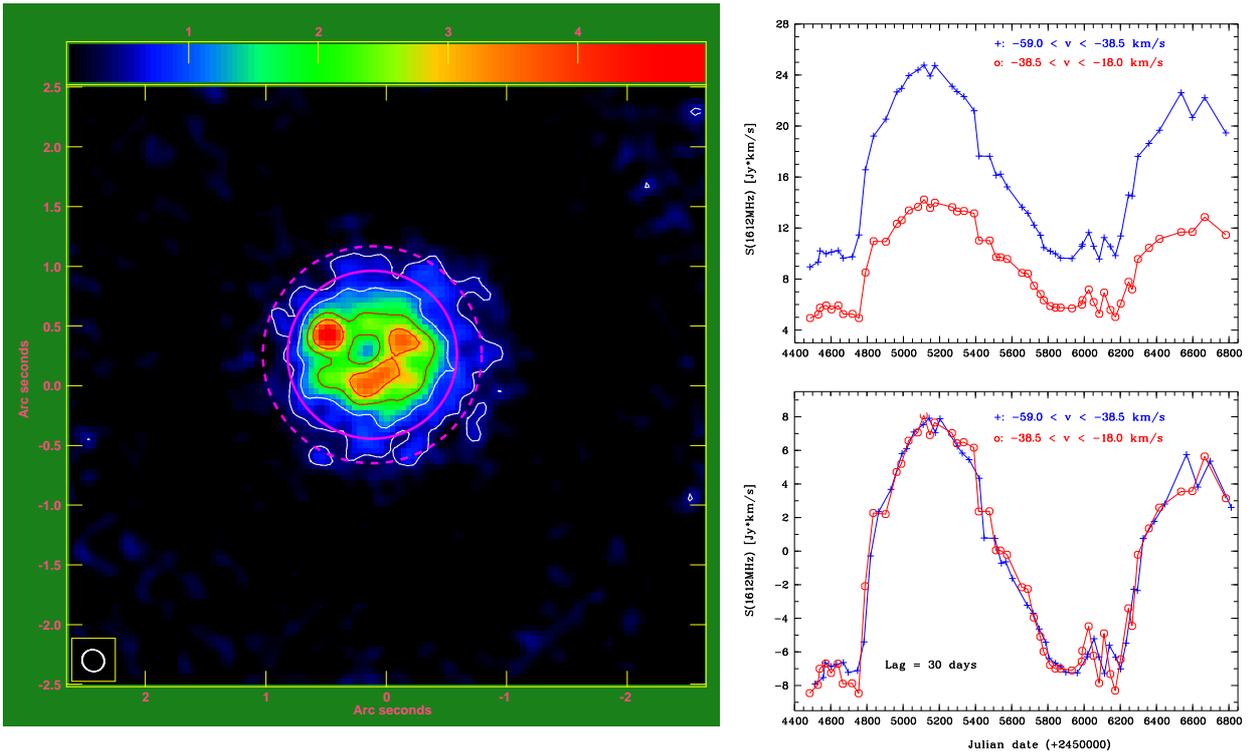}
    \caption{{\emph top:} eMERLIN map of the integrated emission over the inner part of the spectrum of OH~83.4-0.9 covering the velocity range $\sim$[$-55$;$-23$]~km~s$^{-1}$. The magenta full and dotted circles delineate the outer part of the shell. {\em middle:} ``raw'' blue-peak and red-peak light curves of OH 83.4-0.9 obtained from the NRT monitoring over a period of 7 years. {\em bottom:} scaled and shifted light curves for phase-lag determination (\citealt{2018IAUS..336..381E}).}
    \label{fig:PhaseLag}
\end{figure}

The first ingredient: the phase-lag is the measurement of the delay in the light curves of the ``red'' peak with respect to the ``blue'' peak simply due to the light travel time across the CSE, which ranges typically between several days to months for the most sizable OH/IR stars.
The second ingredient: the angular diameter is obtained by (a single epoch) interferometric mapping, providing the angular diameter of the shell (cf. Fig.~\ref{fig:PhaseLag}).

The first phase-lag measurements were performed by \citet{1978A&A....63L...5S}, while the first attempt at actual distance determinations were made in the 1980's by \citet{1985A&AS...59..523H} and \citet{1990A&A...239..193V}. 

The method has been recently revisited with the phase-lag measurements of a sample of 20 such objects performed with the {Nan\c cay} Radio Telescope (NRT) (\citealt{2018IAUS..336..381E}; \citealt{2015ASPC..497..473E}).
A detailed analysis of the method, and the identified caveats of its use, show that an accuracy of $<10$\% in the distance determination can be achieved (\citealt{2021evn..confE..12E}).
In order to achieve this precision, a well-sampled, i.e., regularly spaced with an adequate number of measurement points to determine the shape of the light curves with enough precision to infer the delay is needed along with a high-sensitivity interferometric map.

Single-dish monitoring of these objects alone not only provides the determination of their phase-lag, and hence the linear diameter of their CSEs but also an accurate measurement of their periods for free.  
Radio data are crucial to determine the stellar pulsation periods of OH/IR stars which are heavily obscured by their CSEs and invisible in the optical and even in near infrared bands.  To measure the period, it is essential that the monitoring program shall cover at least a few pulsation periods (typically 5 years or more). 
Taking plotting on period-luminosity diagrams of long period variables as a target for the period determination accuracy to be achieved, 
7 days is enough for a star with a pulsation period greater than 500 days. 

Beyond this, the monitoring cadence must be such that the light curves are well sampled so that their shape is well defined as it is paramount for the lag-determination. It should consequently be a fraction (i.e., ideally less than 50\%) of the lag to be measured. 

Due to its location, the TNRT monitoring in single-dish mode will allow such measurements for  OH/IR stars at lower declinations than achievable with in particular the NRT and hence in a complementarity fashion especially with respect to the Galactic center.

To determine the distance of these objects punctual interferometric observations are needed.

\subsubsection{Maser Monitoring to Investigate the Shaping and Driving of Evolved Star Winds}
\label{subsec5:periodic}

It is not well-understood how mass is lost from the atmospheres of cool, evolved stars, although once dust has formed its role in driving the wind is well-modelled \citep{2015A&A...575A.105B}.  Mass loss may be affected by exceptionally deep pulsations, thermal pulses (\citealt{2012A&A...546A..16R}; \citealt{2021A&A...651A..82H}) and  episodic ejecta \citep{2021AJ....161...98H}.  Current mass loss models differ by orders of magnitude (e.g. \citealt{2019A&A...625A..81S} ; \citealt{2021ARA&A..59..337D}).
Subtle asymmetries developing during the AGB lead to the shaping of post-AGB water fountains and PNe, see section~\ref{subsubsec5:CSE-maser-vlbi-WFs}. Evolutionary changes may be seen even in a few years e.g.  \citep{2014A&A...569A..92V}.

The kinematics inside $\sim$5 $R{\star}$, traced by SiO masers, are pulsation dominated and complex \citep{2019A&A...626A.100B}. 
Shocks further out in the 22-GHz maser shell ($\sim$5$R{\star}$ to tens $R{\star}$) might be caused by rapid expansion of denser  regions leading to a pressure difference.  22 GHz masers are concentrated in clumps where the dust-gas collision rate and heating are more effective. 
OH mainline masers appear less strongly accelerated than  22-GHz maser clumps at apparently similar radii, consistent with tracing cooler, less dense, interleaving gas (e.g. 
\citealt{1999MNRAS.306..954R}). Very high-velocity OH 1667 MHz and/or H$_2$O 22-GHz  emission is sometimes seen (exceeding the 1612 GHz expansion speed), possibly related to exceptional, directed mass loss.  
Some sources show an equatorial density enhancement and biconical outflow  as modelled by \citet{2001MNRAS.322..280Z}; potential origins include magnetic fields and interactions with companions (planetary mass upwards) e.g. \citet{2020Sci...369.1497D}, \citet{2020A&A...644A..61H}

Maser variability arising from different causes will propagate away from the star at different speeds depending on the cause: direct, radiative effects;
the timescale of heating and cooling (e.g. ionisation/recombination), shock speed, the wind velocity and localised inhomogeneities.   Out to 5 $R{\star}$ the light-travel time is $\sim$1 hr for an AGB star of radius 1 au but will be slowed by factors of a few in IR-optically thick clumps. The timescale for shock propagation at $\le$10 km s$^{-1}$ is $\le$2.5 yr and the net wind outflow timescale could be as long as 50 yr \citep{2018ApJ...869...80A} due to non-linear trajectories  (see e.g. \citet{2001ASPC..223.1603H} for timescales around RSG).   Further out, most O-rich winds follows radial streamlines with slow acceleration.  22-GHz maser fluctuations may follow direct heating by the IR period (lagging the optical) or via locally-induced shocks propagating through clumps and density inhomogeneities at sound speed (up to years), or as long-term variations in the mass loss rate affect the wind density (e.g. 'superperiods', \citealt{2001A&A...376..928L}). Disruption of masing by planetary companions around a solar-mass star would occur every (few) year(s) at a few $R{\star}$. 
Monitoring OH and H$_2$O close in time is needed to reveal features with coordinated variability (ruling out projection effects)  and investigate the small-scale structure of the wind in the region where it attains escape velocity (e.g. \citealt{2014AstL...40..212C}).  Monitoring will catch brief high-velocity outflows and establish whether H$_2$O activity is connected to extreme velocity OH.

Monitoring tracks velocity drifts  due to complex morphology and dynamics.  
A red-blue velocity offset at one epoch may appear in an opposite sense at another epoch hinting at asymmetric mass loss, or show velocity drifts indicating a preferred axis of mass loss, distinguishing between an equatorial density enhancement versus a rotating disc. For example, asymmetries were deduced from single-dish monitoring for VX Sgr, \citet{1999ARep...43..311P}, refined by interferometry \citep{2003MNRAS.344....1M}.

Polarisation monitoring of SiO \citep{2006A&A...450..667H} and OH (e.g. \citealt{2013MNRAS.430.2499W}) will investigate the magnetic field dependence on radius (hence distinguishing between dipole, solar and toroidal configurations) and its variability and role in shaping the wind or confining clumps. There are also polarisation phenomena at the clump scale, probed by VLBI, where rotations of the electric vector position angle through 90\,degrees, apparently within a single clump, have been observed in TX~Cam \citep{2019ApJ...871..189T} and R~Cas \citep{2013MNRAS.431.1077A}.
In-depth studies of specific objects e.g. examining the effects of low-mass companions or the aftermath of a thermal pulse, can be complemented by interferometry (sec~\ref{subsec5:CSE-maser-vlbi}).  
H$_2$O also survives around objects which have only recently become C-rich e.g. \citet{2020ApJ...890...34Z} and references therein and SiO masers also occur around S-type (C/O $\sim$ 1) stars.  The 89-GHz HCN maser is found close to C-rich stars (\citet{2014MNRAS.440..172S}  and references therein) and, like SiO, shows rapid variability. Monitoring and comparisons with O-rich stars are of great interest in examining how the chemical composition of the star and nascent dust may fundamentally determine mass loss dynamics and wind acceleration (e.g. \citealt{2022A&A...660A..94G}).

{\bf{Single-dish monitoring observations of circumstellar masers are one of the key topics where the TNRT will have high impact.}}
A database of  OH, H$_2$O and SiO masers in O-rich CSEs has recently been compiled \citep{2019RAA....19...34S}.
H$_2$O-SiO studies have been made using Yonsei and KVN antennas (e.g. \citealt{2016JKAS...49..261K}). These cover several hundred lower-mass stars, mostly single-epoch. A large-scale SiO maser survey such as the Bulge Asymmetries and Dynamical Evolution (BAaDE) project (see e.g. \citealt{2018ApJ...862..153S}; \citealt{2020ApJ...892...52L}) has observed over 28,000 colour-selected red giants in the Galactic plane to explore the dynamical structures of the inner Galaxy, which provides a wealth of information on the statistics of SiO maser transitions and their sources. Selected targets can thus be investigated further with single-dish observations. Long-term monitoring e.g. using Medicina and Pushchino at 22 GHz or {Nan\c cay} at 1.6 GHz is only coordinated for a few objects. Interferometric observations are not practical for frequent monitoring of large samples, rapid ($<$few hr) variability or rapid responses  and much flux is often resolved out on 100s-km baselines. 

{\bf{Dedicated TNRT monitoring programs will provide consistent, multi-frequency coverage of a large sample of evolved star masers, at ~1 Jy sensitivity, allowing the statistical analysis crucial for testing the hypothesis mentioned above.}} This kind of intensive monitoring at three frequencies is  globally unique apart from a connected project using the  Yebes 40 m telescope, where $\sim$800 hours will be spent per year.   The TNRT can observe stars down to Dec. --55$^{\circ}$ at 15$^{\circ}$ elevation, for example covering almost all the ATOMIUM \citep{2020Sci...369.1497D} and NESS \citep{2022MNRAS.512.1091S} samples, and giving easy access to high mass stars concentrated towards the Galactic plane.  

Sampling intervals are determined by sensitivity,  rates of maser variability and survival of distinct features, depending on location with respect to the star, typically from a few weeks in the SiO maser zone to months at greater distances, but the TNRT could also investigate poorly-understood hints at much shorter maser flickering. 1-2 Jy sensitivity, sufficient to monitor bright SiO and H$_2$O sources, can be achieved in minutes. Detection experiments, and polarimetric observations (also requiring high spectral resolution) may need a few hr for individual targets. Significant monitoring results will emerge in $<$ a year for short-period stars; several years or more will build up a clearer picture for longer-period objects and reveal changes in objects undergoing rapid evolution. Occasional interferometric imaging (e.g. per stellar period) will locate features with respect to the star and inform the interpretation of more frequent spectral monitoring (e.g. \citealt{2012A&A...546A..16R}) and can be coordinated with the TNRT in VLBI (Sec.~\ref{subsec5:CSE-maser-vlbi}).

\subsubsection{Signposts/Signatures of transitional stages on the AGB}
\label{subsec5:transitional-phases}

In optically-thin-CSE evolved stars, dramatic flaring events can occur in OH. The first report of an OH maser flaring event was reported by \citet{1981ApJ...249..118J}. A census and statistical analysis of such events was presented by \citet{1997Ap&SS.251..193E} solely on the basis of long-term monitoring which constrained several aspects of these flaring events. These objects are all Miras, with CSE properties putting them in the ``(OH) maser'' and ``(OH) non-maser'' O-rich AGB transitional phase, hence heralding changes in the CSE. The most important/striking characteristics are that this type of emission is substantially polarised and that their velocities point to restricted locations close to the star. The latter points were indeed confirmed recently by VLBI mapping; additionally, confrontation between contemporaneous OH and water maser monitoring indicates that OH flaring seems to be associated with fainter water maser emissivity\citep{2017MNRAS.468.1703E}.

While OH/IR stars show a very low percentage of polarisation (typically less than 10\%), highly polarised emission (up to 100\%) is sometimes observed towards objects with an intermediary-thick CSE, reminiscent of what is observed toward the optically-thin-CSE flaring Miras.  The strong polarisation observed towards these objects higher up on the AGB, is suspected to be linked with them being in the transitional phase Mira $\rightarrow$ OH/IR star \citep{2004A&A...420..217E}.
How this highly polarised emission (i.e., the influence of magnetic field) is linked to the evolutionary changes undergone by the CSE when an object evolved towards the [P]PN stage is not yet understood.
  
The TNRT capability of monitoring in both species transitions ``quasi simultaneously'' and retrieve at the same time the polarimetric information, will allow to investigate the possible relation between the OH and water emissivity of transitional objects (i.e., both in the ``early'' Mira stage and in the Mira $\rightarrow$ OH/IR star stage) so as to identify the CSE changes that the OH maser flaring/strong polarisation ``flashes'' are the sign post of. 
  
This requires a dual monitoring approach: a "searching" low-cadence monitoring (i.e., typically monthly or even every other month) of the selected candidates (at a sensitivity of a few 100 mJy) and an "event-recording" high-cadence monitoring (i.e., as fast as a few days -- weekly) towards those objects showing the aforementioned transitional signatures.   

Punctual inteferometric imaging is needed in order to localised the impacted regions of the CSE while the monitoring itself provide the crucial dymanical evolution of these events.  

\subsubsection{Episodic Events of Circumstellar Masers}
\label{subsubsec5:CSE-masers-episodics}

Here they are defined as events of the CSE masers that cannot be explained by repeated or stable patterns in spectra and interferometry maps, different from those that are likely attributed to regular stellar pulsations and stable radial accelerations in the CSEs. They may include episodic appearances and flares of the masers on time scales shorter than a typical pulsation period of long-period variable stars (100--2000~d). Some of them might be associated with  non-recurrent events such as a nova (e.g. V407 Cyg, \citealt{2011PASJ...63..309D}). Including such an extreme case and recurrent events over a stellar pulsation period, thye may be associated with binarity of an evolved stellar system forming a ``common envelope" such as a symbiotic star. \cite{2022NatAs...6..275K} suggest that a group of the stellar systems hosting high velocity ($\geq$100~km~s$^{-1}$) H$_2$O maser sources, so-called ``water fountain" sources (WFs, e.g., \citealt{2017MNRAS.468.2081G}) should be experiencing such a common envelope evolution during a very short period ($\leq$100~yr). Therefore, it is also expected to see a secular evolution of SiO/H$_2$O/OH maser sources in the same stellar group over 
a human lifetime,
namely a discovery of a new maser source and a permanent death of an existing maser source though the evolution of a fast bipolar molecular jet (e.g. \citealt{2020ApJ...890L..14T}).

One of the science goals in this topic is to reliably determine the timescales of key events described above, namely the period of the recurrent events and total duration of the WF phase. Catching new highest velocity components of H$_2$O masers associated with such events may need a high cadence (an epoch spacing of 1--2 weeks) monitoring program such as FLASHING (Finest Legacy Acquisitions of SiO-/H$_2$O-maser Ignitions by Nobeyama Generation, \citealt{2020PASJ...72...58I}). Tracing decadal evolutions of the individual WFs, even a small number ($\sim$15) of such sources, is a tough work for an observatory because of the weakness (as weak as 0.1~Jy) of the H$_2$O masers, especially the highest velocity components. Detections of SiO ($v=1$ $J=1\rightarrow 0$) masers from the WFs \citep{2005ApJ...622L.125I,2022AJ....163...85A} are challenging because of their rareness and faintness (as weak as 0.2~Jy) but scientifically brilliant because they should locate the central stellar system and trace the spatio-kinematics of the root of the WF jet. Eventually, such single-dish monitoring programs need collaborations of large telescopes including TNRT, similarly to M$_2$O (Maser Monitoring Organization). 

It has been expected that the episodic event of H$_2$O masers should be associated with a new launch of a fast jet \citep{2020PASJ...72...58I} and/or an intercation between the jet decelerated and the ambient CSE entrained by the jet \citep{2020ApJ...890L..14T}. This may be able to be traced by finding some systematic trends of line-of-sight velocity drifts in the H$_2$O maser spectral peaks. Breaking a record of the jet' top speed will improve our understanding about how deep gravitational potential can launch such a fast jet.  

Definitely, collimated jets found in WFs should be driven by a magneto-hydrodynamical force \citep{2006Natur.440...58V}. Measurement of linear polarization and the Zeeman effect of H$_2$O masers should provide key information on the configuration and the strength of the magnetic field in the jet (e.g. \citealt{2019IAUS..343...19V}). Because the degree of circular polarization is quite low ($<<$1\%), a spectral stacking using the monitoring data may be useful for the polarization detection. 

\subsection{Circumstellar Dynamics and Galactic Kinematics Revealed in VLBI}
\label{subsec5:CSE-maser-vlbi}

Thanks to compactness of the individual features of CSE masers and their associations with unique objects in astrophysics, they are interesting targets for VLBI animation synthesis and high-accuracy astrometry. Note that, because of dominant flux contributions from more extended structures of the CSE masers, their mapping also needs relatively short baselines (100--1000~km).   

\subsubsection{Propagation of Shock and Heat Waves in CSEs}
\label{subsubsec5:CSE-maser-vlbi-monitoring}

Animation synthesis of CSE maser sources is a much more straightfoward way to reveal the dynamics of the CSE than the spectroscopic monitoring mentioned above (e.g. \citealt{2013MNRAS.433.3133G}). One of the key  issues about the CSEs is the mechanism of a stellar wind acceleration that may be enhanced by propagation of pulsation-driven shock waves. Another is the origin of asymmetric mass loss. The maser animations definitely visualize the possible accelerations/decelerations of maser gas clumps (e.g. \citealt{2003ApJ...590..460I}) and will shed light on these issues. 

However, the change in the physical conditions of maser excitation should be taken into account when intepreting the observed spatio-kinematics. This is tracable in only simultaneous monitoring of multiple maser lines, especially SiO masers ($J=1\rightarrow 0$, $2\rightarrow 1$, and $3\rightarrow 2$) located closely to each other so as to shape a common maser ring (e.g. \citealt{2018NatCo...9.2534Y}). Based on these motivations, large projects for intensively monitoring CSE masers have been conducted. As their first phase, the KaVA ESTEMA (Expanded Study on Stellar Masers)\footnote{See KaVA Large Programs website at \href{https://radio.kasi.re.kr/kava/large_programs.php}{https://radio.kasi.re.kr/kava/large{\_}programs.php}.} was conducted during 2015--2018, aiming to statistically explore CSE masers towards $\sim$80 stars and find the targets suitable for the intensive VLBI monitoring \citep{2018IAUS..336..341I}. The second phase of ESTEMA (EAVN Synthesis of Stellar Maser Animations) has realized the required intensive (a time spacing shorter than 1/20 of the stellar pulsation period) VLBI monitoring observations of H$_2$O and SiO masers in four frequency bands (22/43/86/129~GHz) for two stars with different periods (BX~Cam and NML~Cyg) since 2018. 

The number of telescopes participating in the monitoring is always a key factor for its success. Taking over six hours per mapping session every two weeks is challenging in terms of time allocation and scheduling.  Imaging simulations (Imai et al.\ in private communication) suggest that a snapshot image needs at least a few consecutive hours with seven telescopes of KaVA. Unless the EAVN telescopes, except KVN, can simultaneously observe H$_2$O and SiO masers, the requested hours should be doubled as seen in ESTEMA for the first two stars. Even KaVA can fully provide seven telescopes in a limited fraction of the whole year. Therefore, other EAVN telescopes including TNRT are always desired to join such monitoring observations, especially for the shorter-period Mira variables (including Mira a.k.a. $o$ Ceti) with pulsation periods $\leq$350~d. 

\subsubsection{Ignitions and Secular Evolutions of Water Fountain jets/flows}
\label{subsubsec5:CSE-maser-vlbi-WFs}

Mapping H$_2$O (especially very high velocity components) and SiO masers associated with WFs is challenging without large telescopes such as TNRT. Internal proper motions of the maser features in WFs will determine the speed and the kinetic timescale of the jet. These parameters are a key to elucidate the mechanism of the jet launch and predict the final CSE evolution leading to the variety of shapes observed in planetary nebulae. The spatio-kinematics of the H$_2$O masers have been revealed for most of the WFs with interferometers. However, it is crucial to revisit those sources. In subsequent interferometric observations, we may able to locate new maser components that are possibly associated with the episodic events mentioned above. We also may be able to trace the secular evolution of the WFs over decades (e.g.  \citealt{2020PASJ...72...58I}).

\subsubsection{The Astrometry for CSE Masers around the Galactic Center}
\label{subsubsec5:CSE-maser-astrometry}

Trigonometric observations of CSE masers provide a strong support for the determination of physical parameters, such as the stellar luminosity and the mass-loss rate, of the accompanying stars. In addition, the period--luminosity relation of long-period variable stars in the Milky Way Galaxy will be more accurately determined with dozens of the trigonometric and infrared photometric samples. All of OH/H$_2$O/SiO masers can be used for the trigonometry. 
OH masers, in particular, will become the main targets in the SKA era for high accuracy astrometry \citep{2020A&ARv..28....6R}.  \citet{2017AJ....153..119O} demonstrated such astrometry for CSE OH masers. TNRT may build a VLBI intermediate length ($\sim$1100~km) baseline with FAST in L(1.6~GHz)-band. With other EAVN telescopes such as TMRT, XAO, Usuda, one can promote the OH maser trigonometry.   

The Galactic Center is a unique region including the Galactic Bulge and the Nuclear Stellar Disk and Bulge, whose mass distribution and formation histories will be elucidated through a large sample of the three-dimensional velocities of associating stars. CSE maser sources associated with these stars should be interesting astrometric targets because they are obscured by clouds in the Galactic Plane and some of them are invisible in optical and near-infrared. TNRT has a good access to the Galactic Center.

\subsection{Thermal SiO Line and Stellar Continuum Emissions}
\label{subsec5:SiOandContinuum}

The depletion of SiO into dust formation or later accretion has been variously estimated at 'almost all' to $\sim$50\% (e.g.  \citealt{1993AJ....105..595S} and references therein), as the dust chemistry is rather uncertain (e.g. \citealt{2017AJ....153..176L}). Accurate estimates of the mass of gas-phase SiO provide vital constraints but the higher-J transitions, whilst good tracers of the local radiation field and shocks, are thus less reliable for density measurements of gas extending further from the star. The low surface brightness makes thermal $J=2\rightarrow 1$ $v=0$ inaccessible to current interferometers and $J=1\rightarrow 0$ is barely studied even by single dish. 

In some objects, $J=5\rightarrow 4$ and $J=6\rightarrow 5$ thermal SiO transitions reach greater apparent outflow velocities than CO, although they have a smaller angular size, \citep{2022A&A...660A..94G}, attributed to the wind interaction with a companion.  Measuring the $J=2\rightarrow 1$ and $J=1\rightarrow 0$ velocities will investigate whether this is influenced by excitation effects.

{\bf{The TNRT is unique in having the sensitivity to search for the two lowest thermal SiO transitions simultaneously}}, measuring the non-dust fraction and revealing any velocity anomalies. In order to detect the line wings,
a few hr per star is needed for a small sample, providing immediate science results by comparison with known dust properties.  One repetition at a different  stellar phase would identify any maser contamination (weak but sometimes present in $v=0$).  These observations will provide an abundance profile for objects with independent higher-J data
and the kinematics can be  compared with CO properties (e.g. from the NESS and Atomium surveys).

Betelgeuse is 10--14 mJy at K-band and ~28 mJy at Q-band, radio spectral index $\sim$1.4.  It is continuously monitored optically but only occasionally at other wavelengths (e.g. \citealt{2015A&A...580A.101O}).  Radio emission represents the flux density at the optically thick surface at that radius, increasing with decreasing frequency (e.g. the C-band radius is 4-6 times the optical stellar radius).  The K-band emission is thought to emanate from just outside the region where pulsation shocks are damped whilst higher frequencies come from nearer the optical photosphere. In 2019 optical dimming appeared to arise from an exceptional mass ejection, such that  Betelgeuse was (partly) obscured by a dust formation event and/or cooled due to internal processes (e.g. \citealt{2021Natur.594..365M}; \citealt{2020ApJ...897L...9D} \citealt{2022ApJ...936...18D}). K-band emission would be unaffected by dust formation further from the star but unusual variability (monitored every few months) would indicate internal changes, potentially related to exceptional convective events.  (e.g. \citealt{2006ApJ...646.1179H})  Coordinated K-Q-W band monitoring will be invaluable in measuring the passage of disturbances out from the photosphere, changes in temperature and propagation mechanisms (e.g. \citealt{2001ASPC..223.1603H}).

{\bf{The TNRT could be unique in providing radio continuum monitoring of the closest supernova precursor.}}
Typical radio fluctuations, whether due to shocks or convection or other causes, are $\sim$10\%, requiring $\sim$1 hr per visit, including accurate flux scale calibration.  Weekly or fortnightly monitoring during the favourable months would sample $\sim$half Betelguese's 400-day period (predicted to re-appear following the Great Dimming). Moreover, piggyback observations at high spectral resolution will reveal if the mass loss event initiates H$_2$O or SiO masers (seen weakly at shorter wavelengths).


\clearpage

\section{Radio Survey and Monitoring of Emission in CP Stars}\label{sec5:CPstars}

\textit{Led by Eugene Semenko \& David Mkrtichian}

\vspace{4mm}
\noindent
In the upper Main Sequence, about 10-15\% of A and B stars showing the abnormally strengthened or weakened lines of some indicative elements in their spectra are recognized as chemically peculiar Ap/Bp or CP stars. This class of hot stars was divided into four subgroups according to the most characteristic chemical elements~\citep{1974ARA&A..12..257P}. Chemical anomalies in such stars are the result of atomic diffusion carrying out in stabilized atmospheres. It is believed that the slow axial rotation, tidal effects, and magnetic field  are the main factors of stabilization. The magnetic field plays a special role. 

\vspace{3mm}
In spectropolarimetric observations, about 10-15\% of all CP stars show the presence of a strong magnetic field which often has a simple dipolar or multipolar configuration and covers the whole surface of the star~\citep{2009ARA&A..47..333D}. The strength of the magnetic field of CP stars ranges from few hundreds to above thirty thousand gauss. In contrast to the Sun and other cool stars, the magnetic field of CP stars remains stable on a timescale of decades, albeit the observed characteristics of the field vary with rotation. 

\begin{figure}[htbp]
    \centering
    \includegraphics[clip,width=9cm]{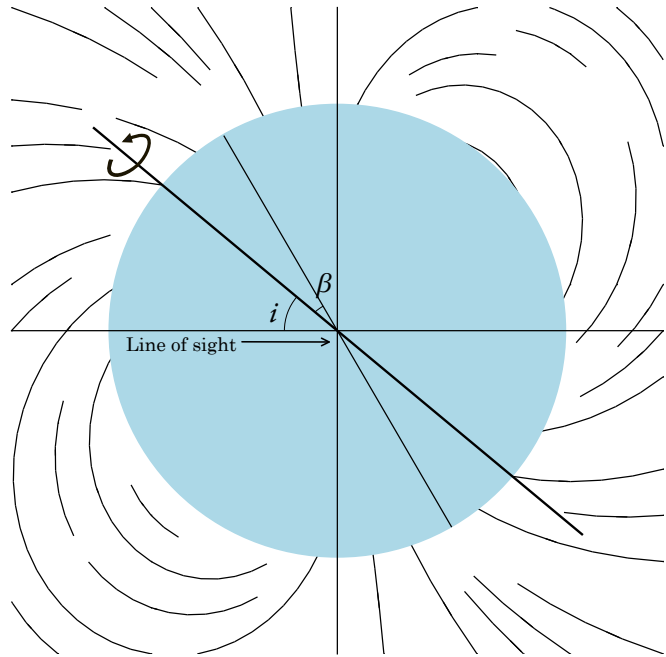}
    \caption{Schematic view of a magnetic oblique rotator.}
    \label{ORM}
\end{figure}

\vspace{3mm}
A synchronous photometric, spectral and magnetic variability is a major feature of peculiar A and B stars. This phenomenon was explained by \citet{1950MNRAS.110..395S} within the framework of a model of oblique rotator. Within this model, the magnetic dipolar or multipolar field is frozen in the stellar plasma and rotates with star. The magnetic axis is inclined to the rotational axis of the star at the angle $\beta$, while the rotational axis is inclined to the line of sight at the angle $i$~(\ref{ORM}). According to the study by \cite{2000A&A...359..213L}, the angle $\beta$ in the slowly rotating magnetic stars with period $P_\mathrm{rot} > 25^\mathrm{d}$ tends to be close to zero, i.e the magnetic and rotational axes are almost co-aligned. The explicit configuration determines the character and amplitude of observed indicators.

\vspace{3mm}
Apart from a regular variability, chemically peculiar stars are mostly quiet in all spectral domains. An exception is a fraction of the hottest CP stars. In their case, a strong magnetic field in combination with relatively fast rotation and powerful stellar winds forms the environment of extended magnetospheres with magnetically confined winds shocks. This phenomenon leads to the variable emission in UV, visible, and IR spectra. In some extreme cases the enhanced signal is observable in X-rays~\citep{2014ApJS..215...10N}. Strongly magnetized atmospheres of Bp stars are also a suitable environment for certain types of radiation in radio waves.

\vspace{3mm}
Radio emission of early type stars was discovered in 1980s, and in the subsequent years the number of identified sources grew very fast. In a paper by \cite{1994A&A...283..908L}, the authors summarized the occurrence of radio sources among different types of CP stars, specifically, 31\% for helium rich stars, 26\% for helium weak stars, 23\% for silicon stars, and zero~--- for cooler Ap stars. The radio emission in these cases usually varies with a rotational period of star, and its amplitude and intensity depend on the wavelength. According to the 3D model proposed by \cite{2004A&A...418..593T}, gyro-synchrotron radio continuum emission arises from the interaction of stellar winds with a magnetic field. Measured radio flux density is determined by the geometry of a magnetosphere and its physical properties. Radio emission at higher frequencies comes from the layers located closer to the stellar surface. Thus, the authors emphasise the importance of multifrequency observations. Such example of a work is a study of hot B star HR\,7355 by \cite{2017MNRAS.467.2820L}. Observing this star with VLA at frequencies 6, 10, 15, 22, 33, and 44 GHz, the authors found variable non-polarized radio flux order of 10--25\,mJy with an amplitude decreasing at higher frequencies.

\vspace{3mm}
In addition to the incoherent radio emission described above, some magnetic CP stars exhibit the radio signal of different nature. This type of emission is characterised by the sharp circularly polarized signals appearing in a narrow range of rotational phases, with fluxes which are almost one order higher than in the case of gyro-synchrotron mechanism. The mechanism of excitation here is similar to Auroral Radio Emission~(ARE) observed on some magnetized planets. A broadband manifestation of ARE is observable in all domains, including X-rays. In radio, ARE appears through the mechanisms of Electron Cyclotron Maser Emission (ECME). A model of this process in hot stars was given by \cite{1982ApJ...259..844M} and \cite{2016MNRAS.459.1159L}. Unlike the incoherent gyro-synchrotron emission which intensity resembles the curve of a magnetic field, sharp impulses of ECME appear only when the star is observed from its magnetic equator.

\begin{figure}[htbp]
    \centering
    \includegraphics[clip,width=14cm]{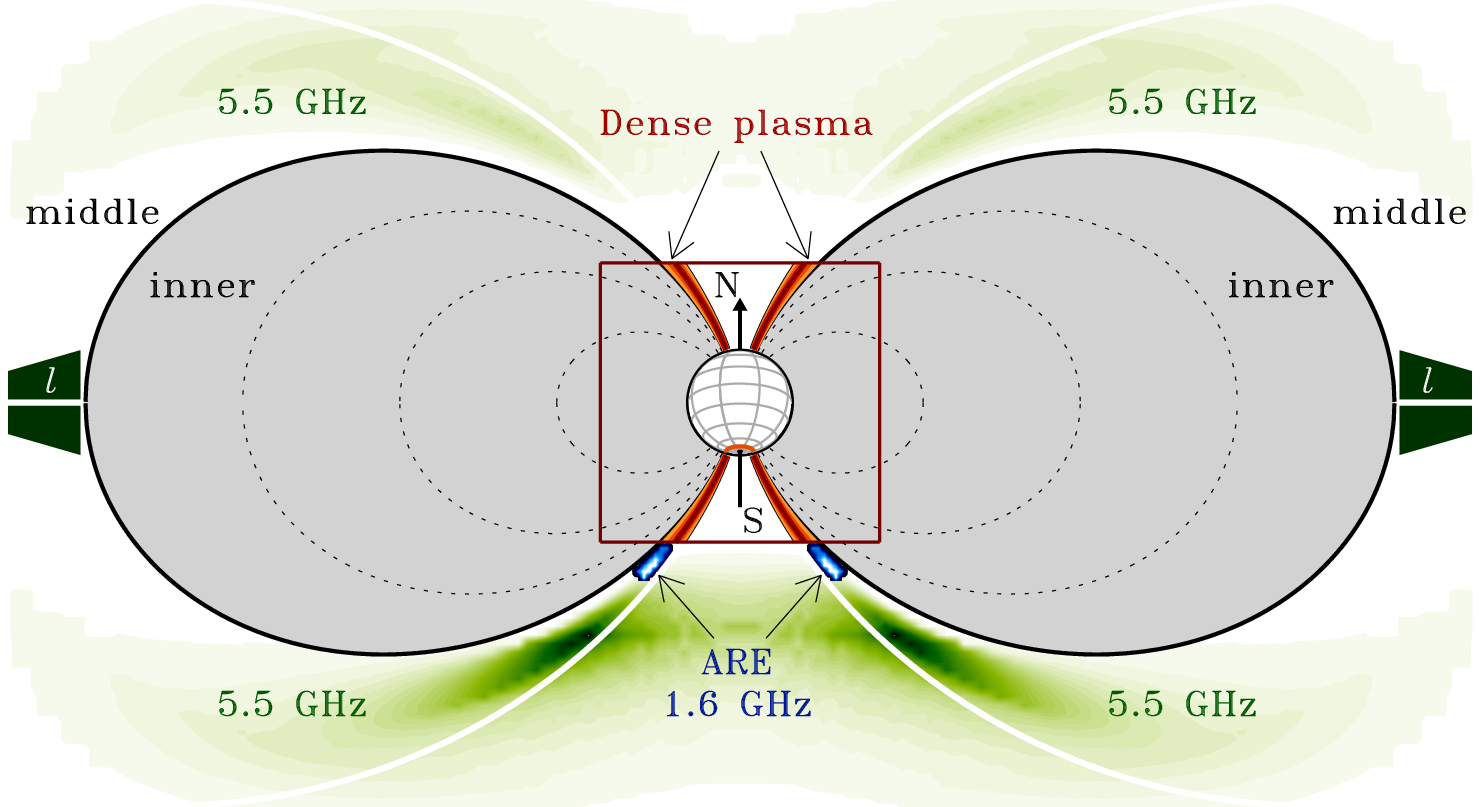}
    \caption{Radio map of a star $\rho$ Oph C with indicated sources of radio emission and corresponding frequencies. Estimated size of the middle-magnetosphere is 15\,$R_{*}$. Originally published in \citet{2020arXiv200902363L}, reproduced with permission of the authors.}
    \label{fig4_by_leto}
\end{figure}

\vspace{3mm}
For the first time, ECME was detected in a famous CP 'pulsar' star CU\,Vir by \citet{2000A&A...362..281T}. Even the number of known sources of ECME among CP stars had been growing in the consequent years, by the end of September 2020, only seven stars showed variable coherent radio emission in observations. These stars are: CU Vir~\citep{2000A&A...362..281T}, HD\,133880~\citep{2015MNRAS.452.1245C, 2018MNRAS.474L..61D}, HD\,142990~\citep{2018MNRAS.478.2835L, 2019ApJ...877..123D}, HD\,142301~\citep{2019MNRAS.482L...4L}, HD\,35298~\citep{2019MNRAS.489L.102D}, $\rho$ Oph A and C~\citep{2020arXiv200902363L, 2020MNRAS.493.4657L}. All items in this list are hot magnetic stars with the effective temperature 12--18 kK and rotational periods shorter than 2 days. For those rare stars where both types of radio emission are observed simultaneously, one can possible to probe the structure of circumstellar envelopes. In a paper by~\citep{2020arXiv200902363L} the authors presented the model of radio emission of the star $\rho$ Oph C. To demonstrate it, we reproduce here Fig. 4 of the original paper~(Fig.~\ref{fig4_by_leto}). 

\vspace{3mm}
In the Main Sequence, magnetic CP stars probably are the only a group of objects where multiwavelength observations open a way to the exploration of entire structure of the star, from the core to the nearest surroundings. With the introduction of a new radio telescope, NARIT can extend its study of chemically peculiar stars in a new dimension. In addition to the photometric and spectroscopic observations, one become possible to monitor the physical conditions in magnetic CP stars in radiowaves.

\vspace{3mm}
We propose the program for the Thai National Radio Telescope aimed at the searches and monitoring of radio emission of selected CP stars in L- and K-bands. Through this research, we plan to constrain the configuration of the magnetic field derived from photometric and spectropolarimetric observations, to probe the presence of magnetospheres which are not recognizable in optical spectroscopy, and to explore their properties in the regions where the radio emission originates.

\begin{figure}[htbp]
    \centering
    \includegraphics[clip,width=14cm]{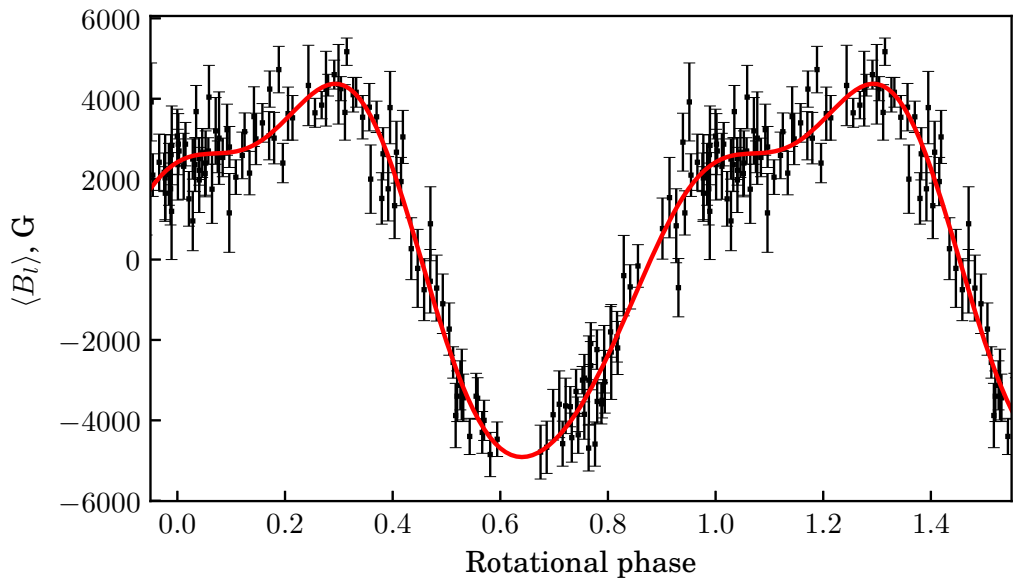}
    \caption{A curve of the longitudinal magnetic field of HD\,34736 phased with a rotational period $P_\mathrm{rot}=1.\!\!^\mathrm{d}2799$.}
    \label{hd34736_bz}
\end{figure}

\vspace{3mm}
In Table~\ref{mcps} we selected mostly bright (and, thus, close) late B and early A candidate stars with notable peculiarities of Si and He-wk type~\citep{2009A&A...498..961R} with available curves of a magnetic field and TESS photometry. The photometric light curves give the accurate values of rotational period (column 8 of the table) and their shape indicates the level of homogeneity of the surface. Rotational periods all of these stars are below or around two days, excepting HD\,40312 where $P_\mathrm{rot}$ is 3.618 days. The reason why this star was included is the Lorenz force, detected for this star by \cite{2007A&A...464.1089S}. The longitudinal magnetic field of all listed stars changes its polarity and has a well determined curve, sometimes, like in the case of HD\,34736, of a non-sinusoidal shape (Fig.~\ref{hd34736_bz}). Extrema of the longitudinal field are presented in column 7 of the Table.

\begin{table}[bthp]
    \caption{List of chemically peculiar stars, possible sources of radio emission.}
    \label{mcps}
    \centering
    \begin{tabular}{l c c l c c c c}
    \hline\hline
    Star ID       & $\alpha_\mathrm{J2000}$ & $\delta_\mathrm{J2000}$ & Sp    & $m_\mathrm{V}$  &  $\pi$, mas  &  $\langle B_{l} \rangle$, kG  &  $P_\mathrm{rot}$, d  \\
    \hline
     HD\,11503   & $01^\mathrm{h} 53^\mathrm{m} 31.\!\!^\mathrm{s}8$ & $+19^{\circ} 17'\,37.\!''9$ & A1\,SiCrSr  & 4.5  & 19.98  & $-1/+0.5$  & 1.6083   \\
     HD\,12447  & $02\phantom{^\mathrm{h}}02\phantom{^\mathrm{m}}02.8$ & $-02\phantom{^{\circ}} 45\phantom{'\,} 49.6$ & A2\,SiSrCr  & 4.1  & 19.79   & $-0.5/+0.4$ & 1.4326  \\
     HD\,12767  & $02\phantom{^\mathrm{h}}04\phantom{^\mathrm{m}}29.4$ & $-29\phantom{^{\circ}} 17\phantom{'\,} 48.5$ & A0\,Si  & 4.7  & 8.79   & $-0.25/+0.25$ & 1.8897  \\
     HD\,19832  & $03\phantom{^\mathrm{h}}12\phantom{^\mathrm{m}}14.2$ & $+27\phantom{^{\circ}} 15\phantom{'\,} 25.1$ & B8\,Si  & 5.8  & 8.04   & $-0.3/+0.4$ & 0.728  \\
     HD\,22470  & $03\phantom{^\mathrm{h}}36\phantom{^\mathrm{m}}17.4$ & $-17\phantom{^{\circ}} 28\phantom{'\,} 01.5$ & B9\,Si  & 5.2  & 7.65   & $-1.6/+1.9$ & 1.9298  \\
     HD\,30466  & $04\phantom{^\mathrm{h}}49\phantom{^\mathrm{m}}16.0$ & $+29\phantom{^{\circ}} 34\phantom{'\,} 16.8$ & A0\,Si  & 7.3  & 6.51   & $-0.25/+2$ & 1.4062  \\
     HD\,34736  & $05\phantom{^\mathrm{h}}19\phantom{^\mathrm{m}}22.2$ & $-07\phantom{^{\circ}} 20\phantom{'\,} 50.0$ & B9\,Si  & 7.8  & 2.75   & $-5/+5$ & 1.2799  \\
     HD\,36526  & $05\phantom{^\mathrm{h}}32\phantom{^\mathrm{m}}13.1$ & $-01\phantom{^{\circ}} 36\phantom{'\,} 01.7$ & B8\,Si/He-wk  & 8.3  & 2.44   & $-1/+4$ & 1.5417  \\
     HD\,37017  & $05\phantom{^\mathrm{h}}35\phantom{^\mathrm{m}}21.9$ & $-04\phantom{^{\circ}} 29\phantom{'\,} 39.0$ & B2\,He  & 6.6  & 0.45   & $-2/0$ & 0.9012  \\
     HD\,40312  & $05\phantom{^\mathrm{h}}59\phantom{^\mathrm{m}}43.3$ & $+37\phantom{^{\circ}} 12\phantom{'\,} 45.3$ & A0\,Si  & 2.6  & 19.7   & $-0.3/0.3$ & 3.6181  \\
     HD\,45583  & $06\phantom{^\mathrm{h}}28\phantom{^\mathrm{m}} 10.7$ & $-04\phantom{^{\circ}} 53\phantom{'\,}56.5$ & B9\,Si & 7.9  & 3.06  & $-3/+4$  & 1.1770 \\
     HD\,63347  & $07\phantom{^\mathrm{h}}54\phantom{^\mathrm{m}}20.4$ & $+71\phantom{^{\circ}}04\phantom{'\,}45.2$ & B8\,SrCrEu  &  7.3   &  5.66  &  $-0.8/+1$  & 1.7495 \\
     HD\,108945  & $12\phantom{^\mathrm{h}}31\phantom{^\mathrm{m}}00.6$ & $+24\phantom{^{\circ}}34\phantom{'\,}00.2$ & A3\,SrCr  &  5.4   &  11.99  &  $-0.3/+0.25$  & 2.0513 \\
     HD\,170000  & $18\phantom{^\mathrm{h}}20\phantom{^\mathrm{m}}45.4$ & $+71\phantom{^{\circ}}20\phantom{'\,}16.1$ & A0\,Si  &  4.2   &  10.77  &  $-0.3/+0.5$  & 1.7165 \\
    \hline
    \end{tabular}
\end{table}

\vspace{3mm}
Given the information about the magnetic field and stellar rotation from spectroscopy and spectropolarimetry, a provisional observational strategy might look as follows. To detect the coherent radio emission from candidate stars, the observations must cover the time around the moments when the longitudinal field changes the sign and carry out in L-band. 
Initially, observations with TNRT will be made to focus on monitoring of flux density without polarization studies. These monitoring should be continuously in whole a day covering the periods of each target source. Sources in which the radio emissions are detected with TNRT will be followed-up by another receiver at phase 1, in K-band, and a forthcoming polarization mode on TNRT to study their magnetic field activities. 
Unlike ECME, incoherent emission reach its maximum when the star exposes its magnetic poles. In terms of the model of oblique rotator, this corresponds to the maxima of the curve of the longitudinal field $\langle B_\mathrm{z} \rangle$. The signal of incoherent gyro-synchrotron emission is expected in both, L and K, radio bands. Because of the corresponding moments of time can be predicted from known ephemerides, the telescope time can be used in more efficient way.


\clearpage
\section{Geodesy: Geodetic VLBI Observation in K-band }\label{sec7:geodesy} 

\textit{Led by Nattaporn Thoonsaengngam and Koichiro Sugiyama}
\vspace{3mm}

\noindent
Geodesy is the Earth science that has objectives of accurate measuring to determine Earth's geometric shape and its change with times. That includes the orientation of the Earth in the spaces and the gravitational field. To obtain expected results nowadays, sources external to the Earth become references of the measurements, artificial satellites, retroreflectors on the Moons, and also quasars, these techniques are so-called the space geodesy.

\vspace{3mm}
In space geodesy, the extragalactic or quasars are measured with the VLBI technique to obtain baseline length between two (or more) radio telescopes that operate the measurement with millimetre-accuracy. The results can be used to determine the station coordinates that are contributed to the international terrestrial reference frame (ITRF). VLBI is the only one of the space geodetic techniques referenced to the kinematically non-rotating frame and the only one that provides all Earth Orientation Parameters (EOPs), i.e., polar motion, universal time, and celestial pole offsets \citep{2008evn..confE..51H}. This lead to global objectives; obtaining precise positioning, monitoring plate tectonics, etc. In addition, VLBI technique achieves residual delays of a few mm or better, which are essential orders of accuracy to measure parallaxes for astronomical sources at beyond 10~kpc distances. This kind of geodetic VLBI observations has been organized as ``The International VLBI Service for Geodesy and Astrometry" (IVS)\footnote{See IVS website at \url{https://ivscc.gsfc.nasa.gov/}}.

\vspace{3mm}
Since the 1970s, radio telescopes with the legacy frequencies, S- and X-band receivers have been using to perform geodetic VLBI observations. However, many reference sources at the S and X frequencies natively have extended structures which can lead to significant errors in VLBI observations. So that, geodetic VLBI at higher frequencies are considered for reduction of sources core-shift. This will allow construction of the more accurate celestial reference frame (CRF), which will be a benefit to the higher accuracy of geodetic VLBI observables. The CRF at higher frequencies is also advantageous for tying the radio reference frames to optical reference frames such as GAIA \citep{2016A&A...595A...5M}. In 2006, four stations in Australia performed the 1.5-hour long VLBI fringe test experiment at 22 GHz in K-band \citep{2009PASA...26...75P}. Results from the test encouraged the scientists to run a full-scale 22-GHz for geodetic VLBI. K-band experiments, afterwards also showed the good results since the ionospheric contribution at the K-band is one order of magnitude smaller than at the X-band. The ionospheric contribution can be negligible during the period of low solar activity. These characteristics make results from K-band comparable or even better than S- and X-band observations.

\vspace{3mm}
The TNRT will contain the K-band receiver thus, the radio telescope could participate in geodetic VLBI observations at this band. Initial station coordinates for the VLBI observing sessions are determined from the relation of positions between the telescope reference point and a co-location GNSS base station on site. The GNSS station can provide the coordinates with an accuracy up to centimetre level. By performing the geodetic VLBI, higher accuracy up to millimetre level of station coordinates can be obtained. Results of the geodetic VLBI will be contributed to the International Terrestrial Reference Frames (ITRF) to produce the EOPs and other geodetic components, such as velocities of station coordinates which are important for studies of tectonic motions. The precise station coordinates, also benefit usages of the TNRT not only for astrometry or geodesy but astronomy, especially observations at higher frequencies. Nonetheless, TNRT will benefit astrometry by enhancing the K-band CRF. 
In particular, its location strongly contributes to create a nice composition of baselines with Hobart 26-m (Australia) and HartRAO 26-m (South-Africa) for achieving more effective geodetic VLBI observations at K-band in the southern hemisphere following goals of the next generation beyond ICRF-3 \citep{2014ivs..confE...1J}, 
where the number of sources is still a factor 2 less especially in he far-south (declination $\leq$~$-30^{\circ}$) and while the number of sessions per source are roughly the same in the north and the south, the average number of observations per source is also factor of 2 less in the far-south, 
although the number of stations and their specifications have been improved and upgraded mainly at S/X-bands in Southern hemisphere \citep{2020A&A...644A.159C,2021evga.conf...85D}.

\clearpage
\section*{Acknowledgements}

This work is based on the national flagship project entitled Radio Astronomy Network and Geodesy for Development (RANGD) by NARIT. We wish to thank all of the members of the International Technical Advisory Committee (ITAC) and the International Scientific Advisory Committee (ISAC) for their supportive and fruitful suggestions and advise, in which ITAC consists of Hideyuki Kobayashi (Chair, NAOJ), Busaba H. Kramer (Secretariat, MPIfR/NARIT), Do-Young Byun (KASI), Francisco P. Colomer (JIVE), Michael Garrett (JBCA), Yashwant Gupta (NCRA), Mareki Honma (NAOJ), Jinling Li (SHAO), Young Chol Minh (KASI), Zhiqiang Shen (SHAO), Tasso Tzioumis (CASS), Pablo de Vicente (IGN), and Gundolf Wieching (MPIfR), while ISAC consists of Michael Bode (Chair, BIUST), Busaba H. Kramer (Secretariat, MPIfR/NARIT), Hideyuki Kobayashi (NAOJ), and Michael Kramer (MPIfR), respectively. 

\vspace{3mm}
We would like to thank all of engineers and technicians in Radio Astronomy Operation Center (RAOC) for their sincere contributions to construct and develop 40-m TNRT: Apichat Leckngam, Kamorn Bandudej, Nonwarit Borvornsareepirom, Attapon Bunwong, Teep Chairin, Nattawit Chanwedchasart, Pathit Chatuphot, Awut Duangkhiaw, Nattapong Duangrit, Nutdanai Hantankul, Prachayapan Jiraya, Kitipoom Kanjana, Pitak Kempet, Natthaphong Kruekoch, Peerapol Meekhun, Nakornping Namkham, Thanadon Paksin, Attasit Phakam, Natee Pongteerarat, Anya Poonnawatt, Nikom Prasert, Sothaya Prathumsub, Songklod Punyawarin, Haseng Sani, Abraham Sanenga, Spiro Sarris, Nonnadda Silamai, Dan Singwong, and Lalida Tantiparimongkol. 

\vspace{3mm}
All the editors and authors should like to our thanks to the the East Asian VLBI Network (EAVN) for their providing an observation data towards M87 at K-band in 2017 as a model for the simulation shown in figure~\ref{fig:m87vlbi} of section~\ref{subsec4:jetvlbi}. The EAVN is operated under cooperative agreement by National Astronomical Observatory of Japan (NAOJ), Korea Astronomy and Space Science Institute (KASI), Shanghai Astronomical Observatory (SHAO), Xinjiang Astronomical Observatory (XAO), Yunnan Astronomical Observatory (YNAO), and National Geographic Information Institute (NGII), with the operational support by Ibaraki University (for the operation of Hitachi 32-m and Takahagi 32-m radio telescopes), Yamaguchi University (for the operation of Yamaguchi 32-m radio telescope), and Kagoshima University (for the operation of VERA Iriki antenna). We also thank Toshio Terasawa (The University of Tokyo) for his fruitful advice and discussion in section~\ref{sec2:pulsar}. The simulations in figure~~\ref{fig:m87vlbi} were generated by using the Astronomical Image Processing System \citep[AIPS:][]{1996ASPC..101...37V,2003ASSL..285..109G} and the Difmap software \citep{1997ASPC..125...77S}. 

\vspace{3mm}
Simultaneously, we would like to thank for a forthcoming collaboration to the Australia Long Baseline Array (LBA), which is part of the Australia Telescope National Facility, which is funded by the Australian Government for operation as a National Facility managed by the Commonwealth Scientific and Industrial Research Organisation (CSIRO), and to the European VLBI Network as well, which is a joint facility of independent European, African, Asian, and North American radio astronomy institutes.

\vspace{3mm}
K.S. is supported by the grant of PIIF Heiwa Nakajima Foundation in 2019 and MEXT/JSPS KAKENHI (or Grant-in-Aid for Scientific Research) Grant Numbers JP19K03921. K.S., P.J., and K.A. are supported by the research grants of NARIT (Public Organization) under MHESI (Ministry of Higher Education, Science, Research and Innovation) Grant Numbers 75249, 75253, 76820, 30137, 30141, 30146, 30152, 30156, 88313, 88314, 88319, 88320, 88321, 88322, 88324, 168577, 168593, and 168595.

\pagebreak
\appendix

\section{UV-coverages and Specifications Expected in World-wide VLBI Collaborations}\label{specvlbi}

In this appendix, we summarize results of simulations for uv-coverages in the cases that TNRT will collaborate with the LBA and the EVN in L-band and K-band towards astrnomical sources in the Northern and the Southern hemispheres, as shown in figures~\ref{figapp:uvlba} and \ref{figapp:uvevnwith}, respectively, while the simulation results for the collaboration with the EAVN are shown in figure~\ref{fig1.2:uveavn} of section~\ref{subsec1.2:vlbi}. 
Besides, specifications achieved in these VLBI collaborations: baseline lengths and baseline sensitivities as 5$\sigma$, are calculated for the cases of EAVN$+$TNRT (table~11 in K-band), LBA$+$TNRT (tables~12, 13 in L-/K-bands), and EVN$+$TNRT (tables~14, 15 in L-/K-bands), respectively. 
These simulationas and calculations were achieved on the basis of station coordinates, array configurations, and specifications for each telescope in each frequency band released via web-pages of each array and their status reports
\footnote{EAVN Status Report, see tables 1 and 4 there:  \href{https://radio.kasi.re.kr/status_report/status_report.php?site=eavn}{https://radio.kasi.re.kr/status{\_}report/status{\_}report.php{$?$}site=eavn}}$^{,}$
\footnote{LBA, see ``Available Telescopes" and ``Telescope Parameters":  \href{https://www.atnf.csiro.au/vlbi/documentation/}{https://www.atnf.csiro.au/vlbi/documentation/}}$^{,}$
\footnote{EVN Capabilities, see ``EVN Status Table" as well: \href{https://www.evlbi.org/capabilities}{https://www.evlbi.org/capabilities}} 
and/or the station coordinates registered in the SCHED program\footnote{SCHED: \href{http://www.aoc.nrao.edu/software/sched/}{http://www.aoc.nrao.edu/software/sched/}} organized for VLBI scheduling generally led by Craig Walker at National Radio Astronomy Observatory. These uv simulations here were generated by using the SCHED program as well. 
For the Warkworth 30-m radio telescope in New Zealand, its station coordinates were referred from \cite{2015PASA...32...17W}, and we assumed that a system noise temperature and an aperture efficiency of this telescope in K-band was 200~K and 30{\%}, respectively, on the basis of information its receiver was in room temperature.
The specifications of TNRT in L-/K-bands are summarized in table~\ref{tablereceiver} of section~\ref{subsec1:specifications}.


\vspace{2mm}
\begin{figure*}[h]
    \centering
    \includegraphics[clip,width=18cm]{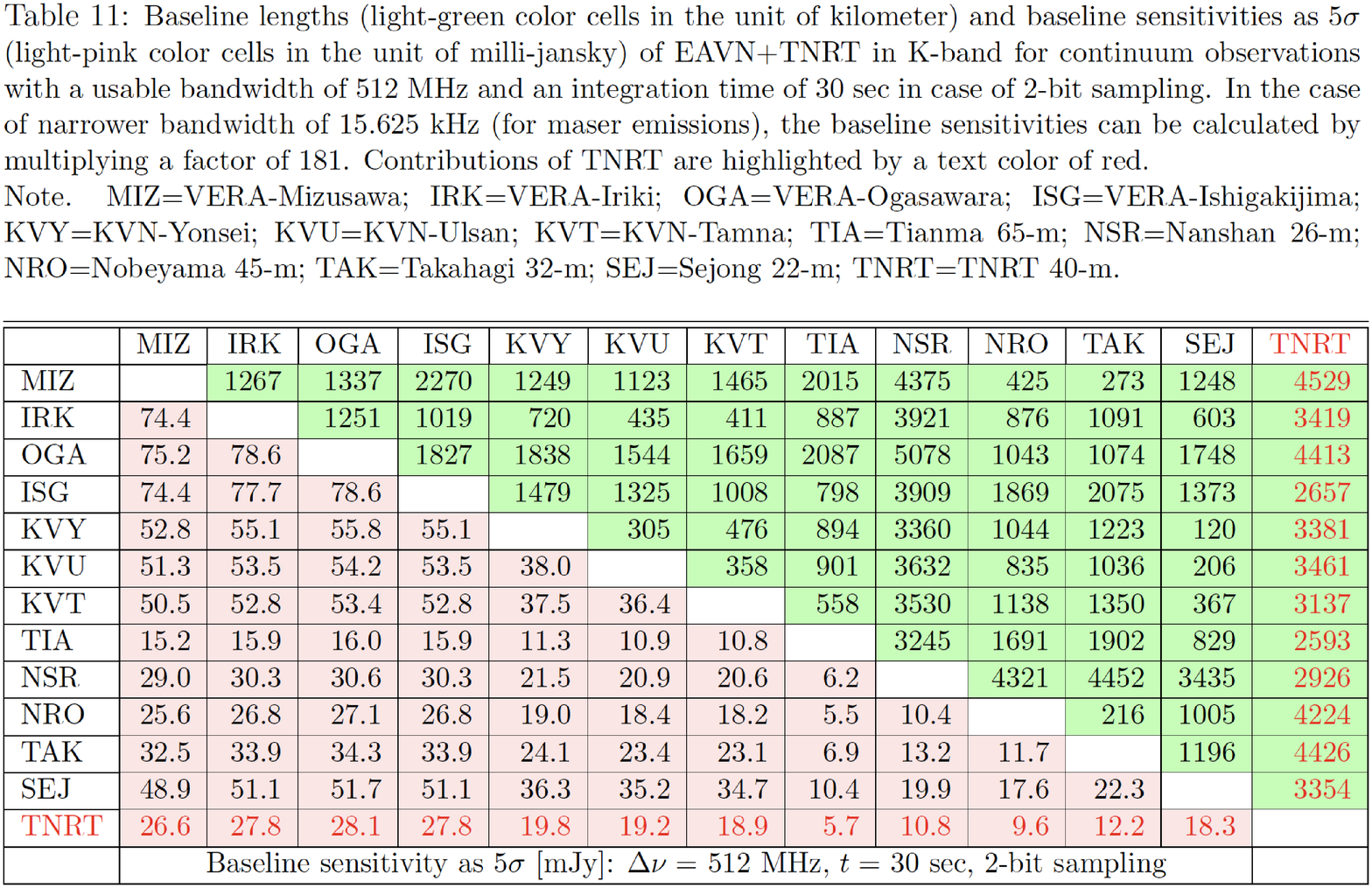}
\end{figure*}


\begin{figure}[htbp]
 \begin{minipage}[b]{0.50\linewidth}
  \centering
  \includegraphics[clip,width=\textwidth]{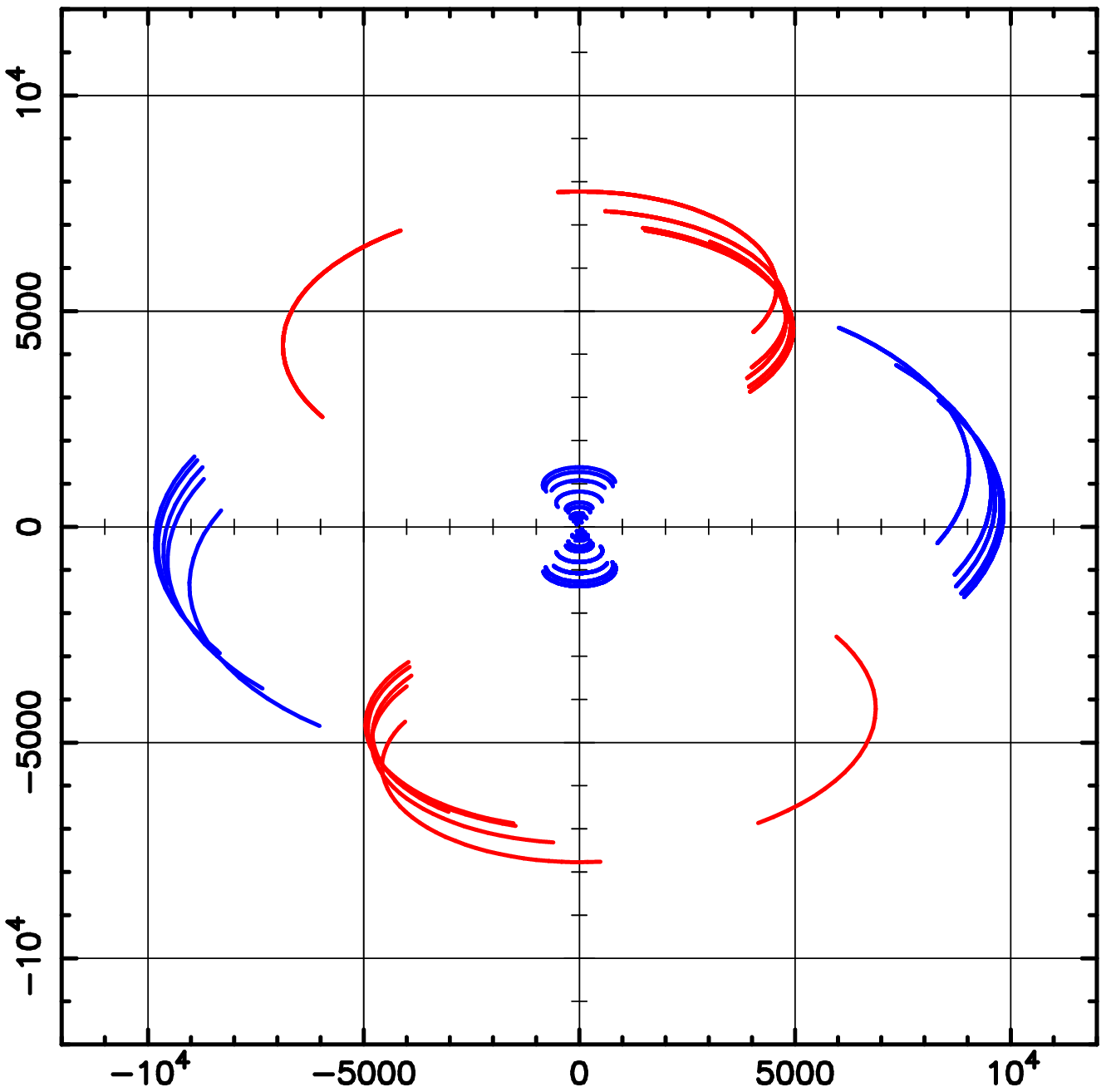}
 \end{minipage}
 \begin{minipage}[b]{0.50\linewidth}
  \centering
  \includegraphics[clip,width=\textwidth]{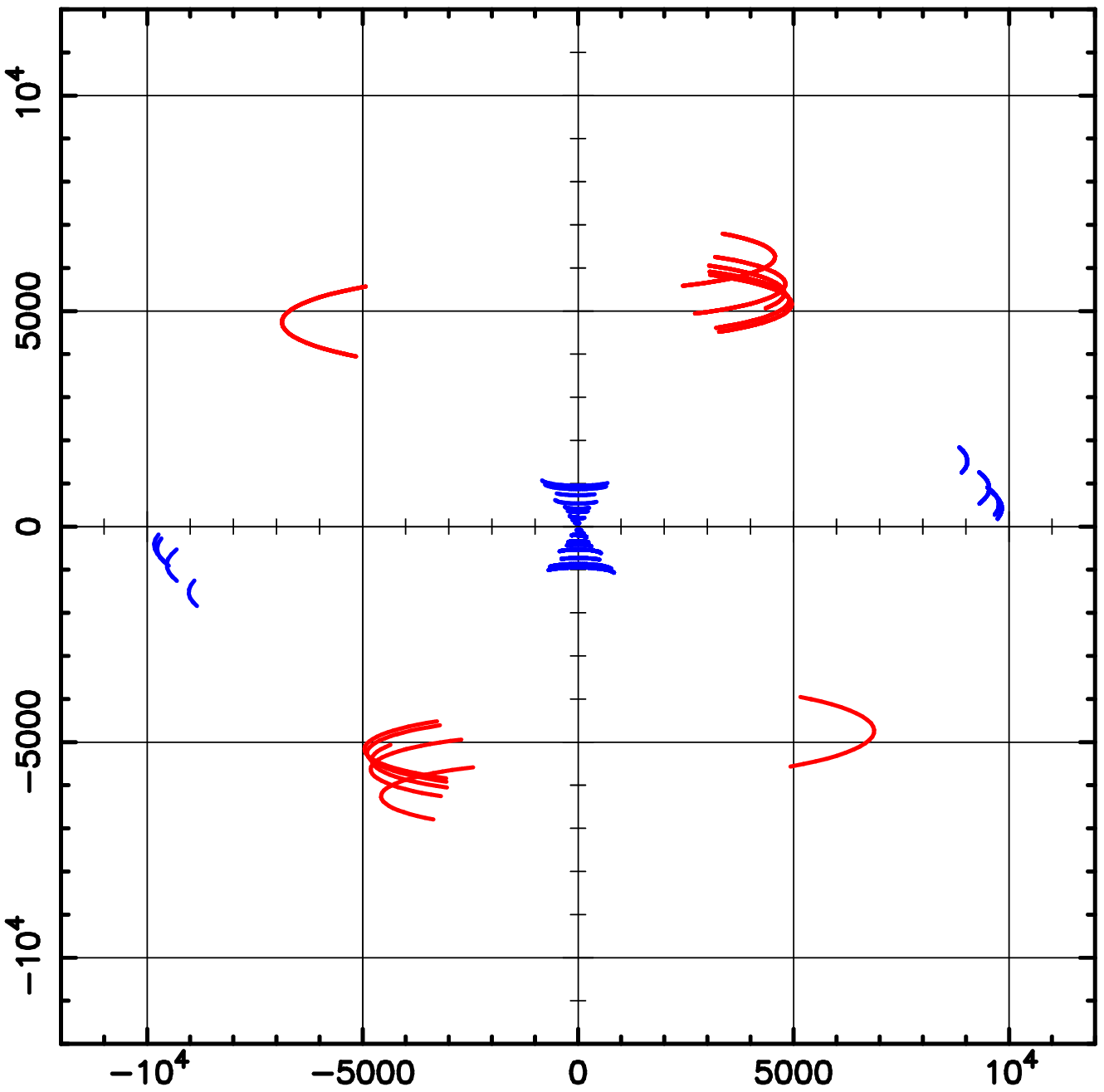}
 \end{minipage} \\
 \begin{minipage}[b]{0.50\linewidth}
  \centering
  \includegraphics[clip,width=\textwidth]{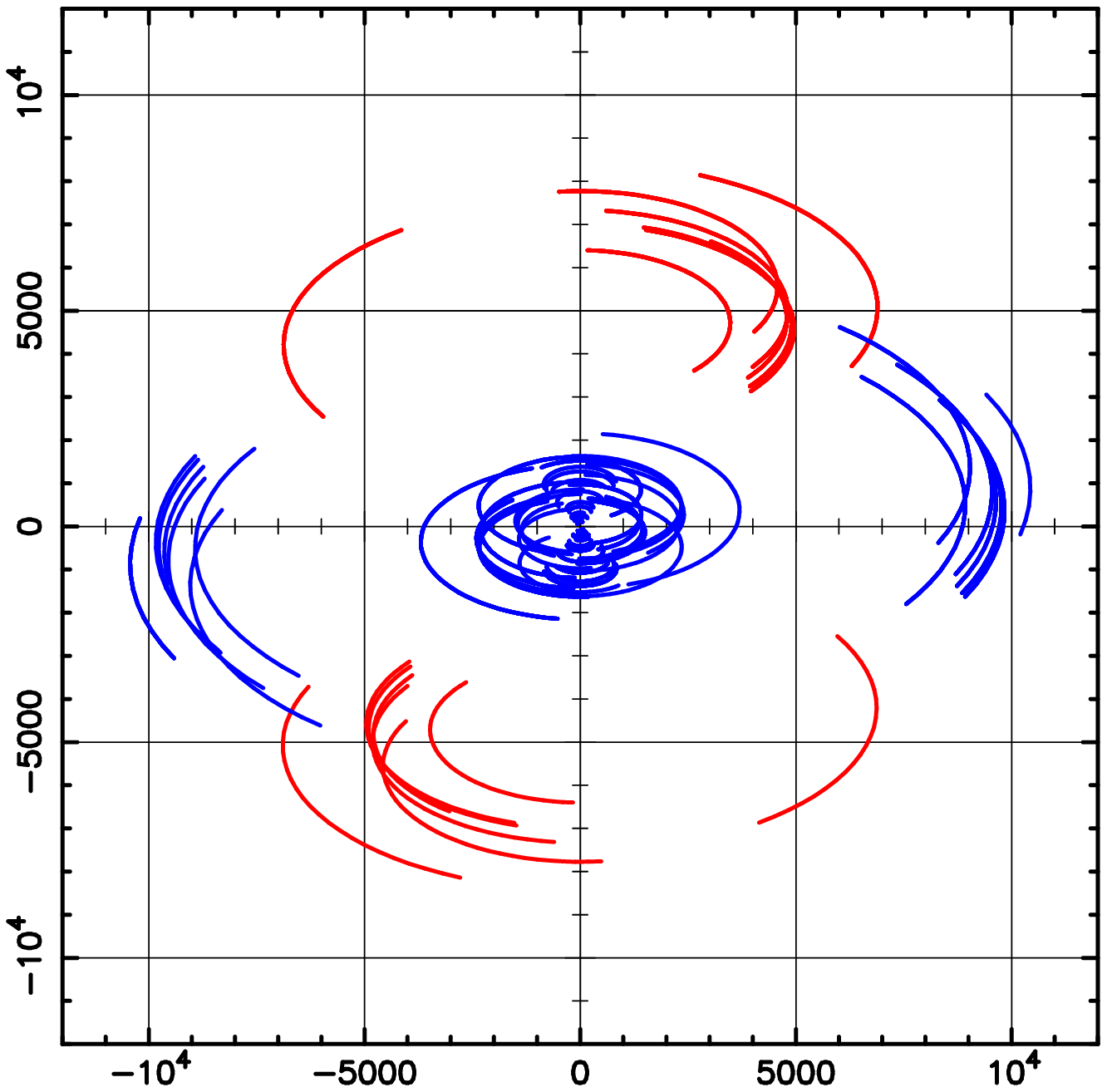}
 \end{minipage}
 \begin{minipage}[b]{0.50\linewidth}
  \centering
  \includegraphics[clip,width=\textwidth]{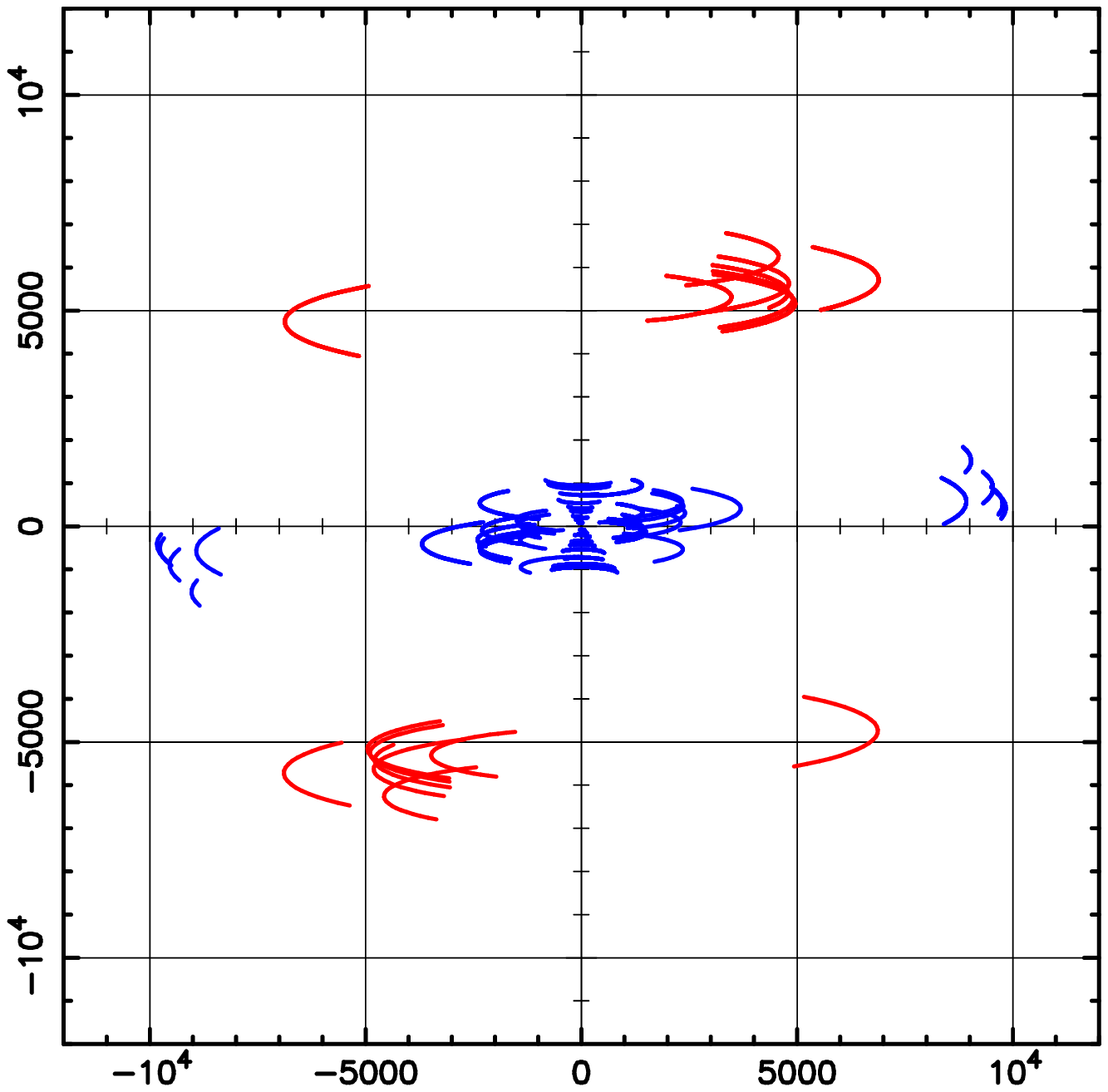}
 \end{minipage}
\caption{Simulated results of uv-coverages for LBA$+$TNRT in the case of observing sources with declination of $-$29 (toward the Galactic Center) and $+$10 degree in L-band (top) and K-band (bottom), respectively, with an elevation limit of 10 degree. The unit of horizontal and vertical axes is kilometer. Contributions of TNRT are highlighted by a color of red.}
\label{figapp:uvlba}
\end{figure}


\begin{figure}[htbp]
 \begin{minipage}[b]{0.50\linewidth}
  \centering
  \includegraphics[clip,width=\textwidth]{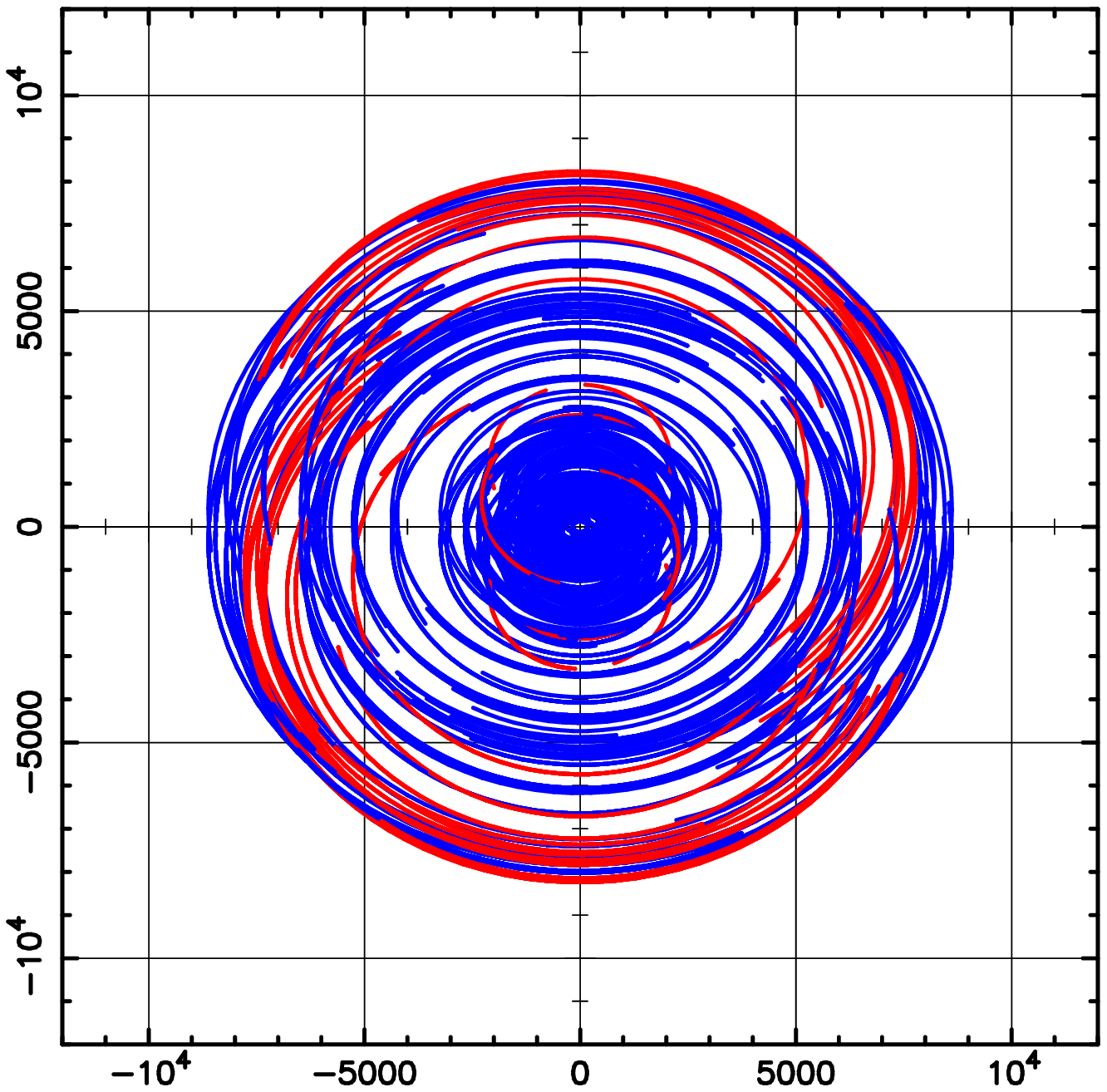}
 \end{minipage}
 \begin{minipage}[b]{0.50\linewidth}
  \centering
  \includegraphics[clip,width=\textwidth]{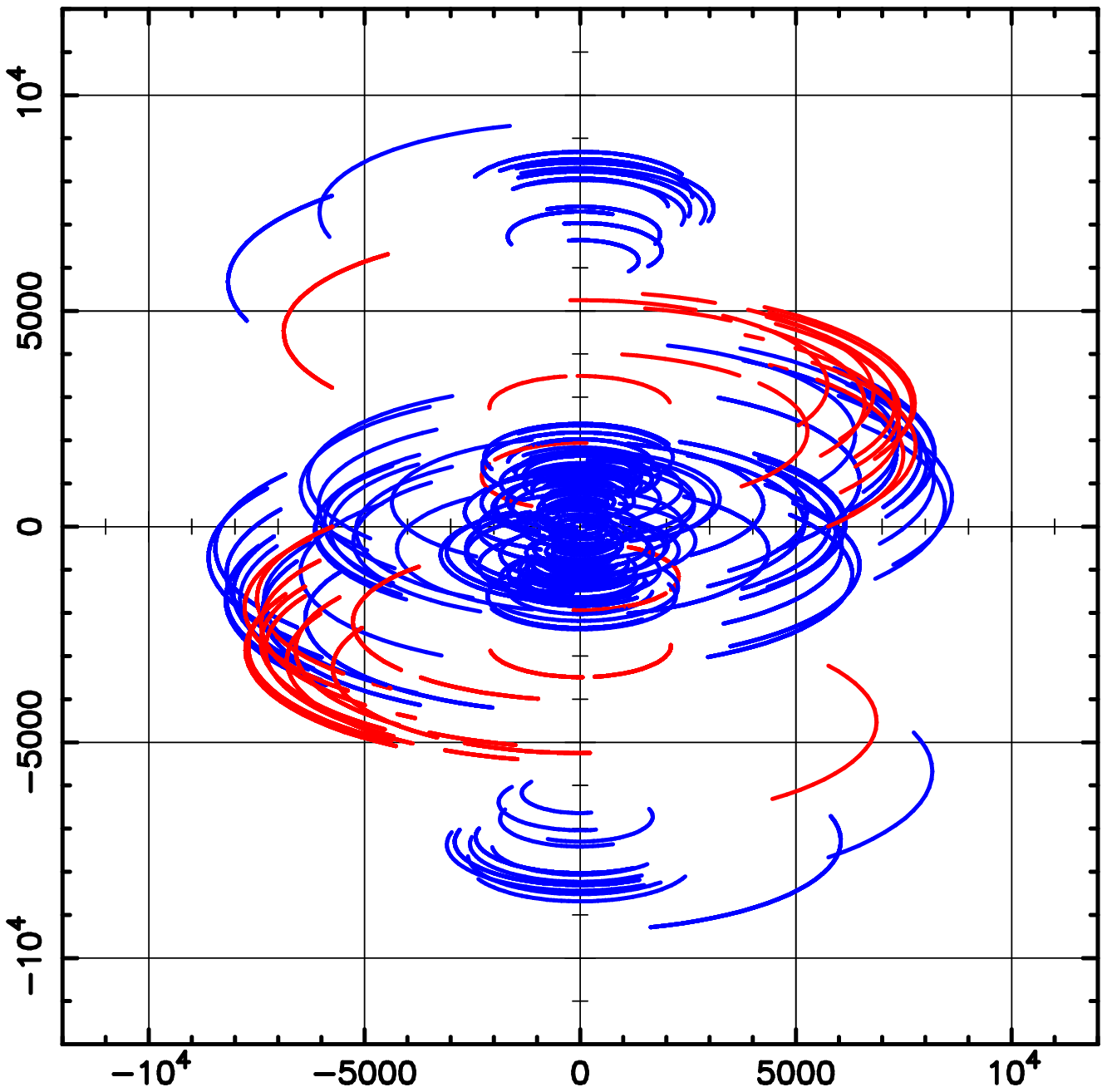}
 \end{minipage} \\
 \begin{minipage}[b]{0.50\linewidth}
  \centering
  \includegraphics[clip,width=\textwidth]{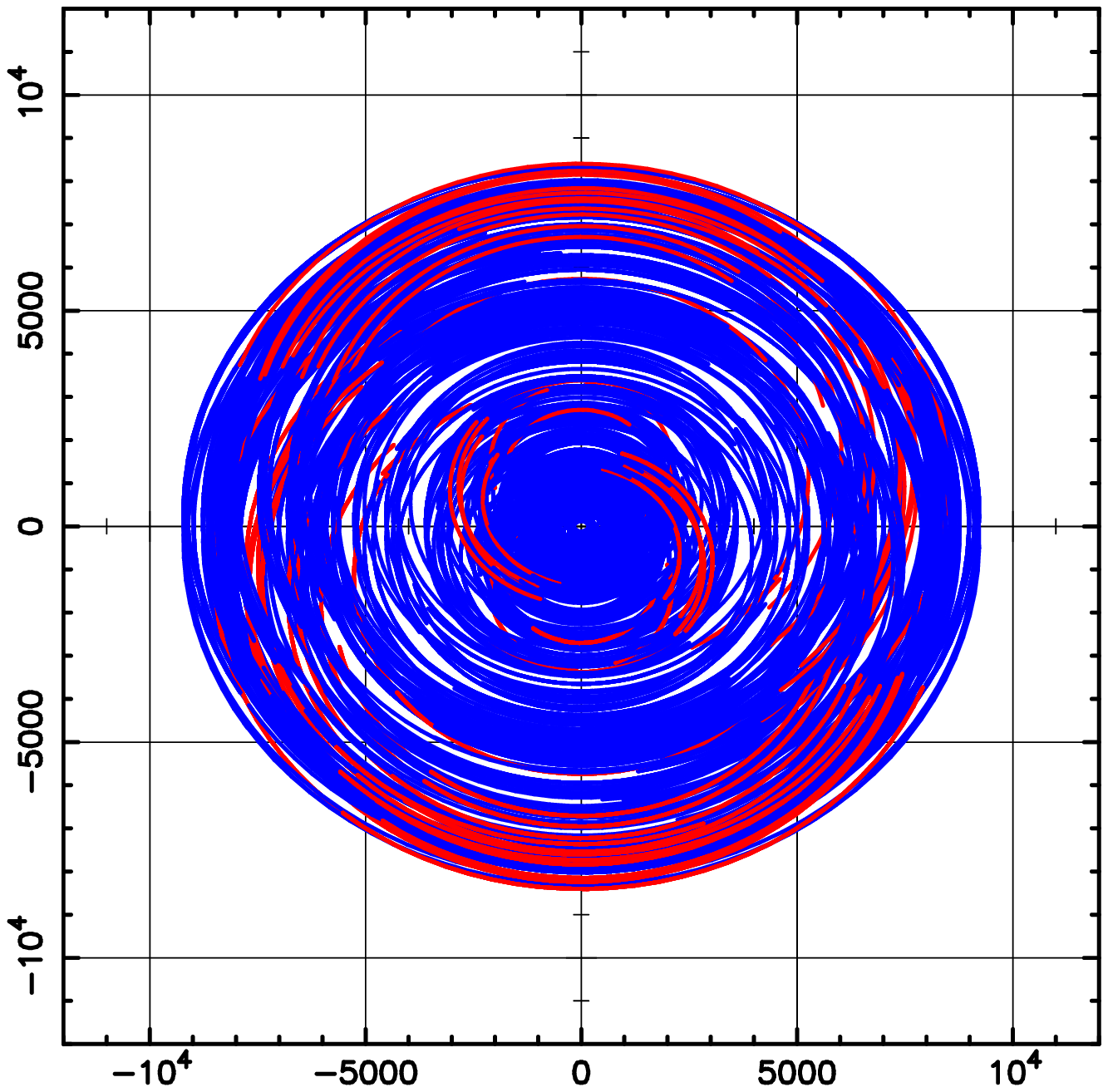}
 \end{minipage}
 \begin{minipage}[b]{0.50\linewidth}
  \centering
  \includegraphics[clip,width=\textwidth]{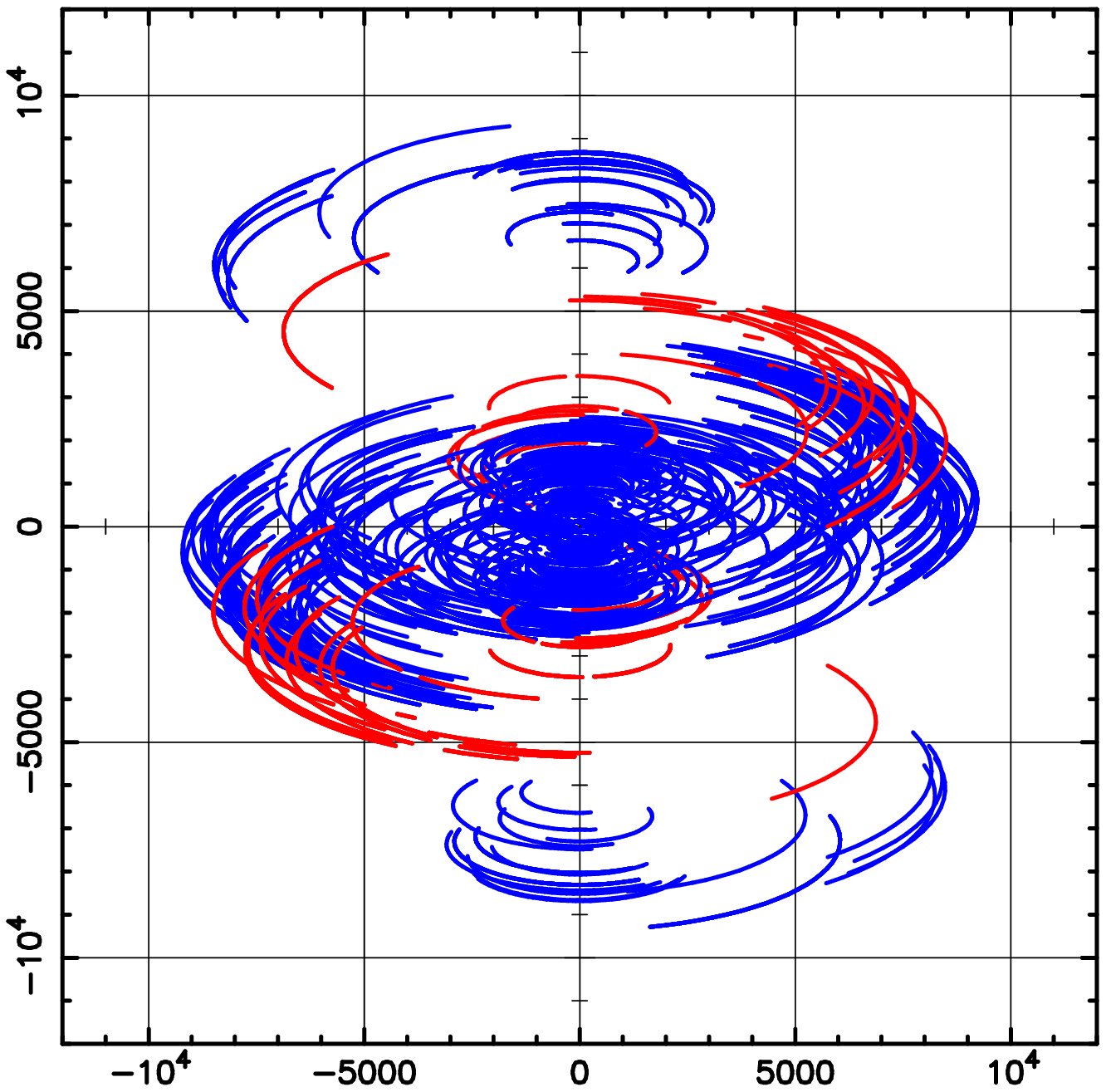}
 \end{minipage}
\caption{Simulated results uv-coverages for EVN$+$TNRT in the case of observing sources with declination of $+$60 and $+$20 degree in L-band (top) and K-band (bottom), respectively, with an elevation limit of 10 degree. The unit of horizontal and vertical axes is kilometer. Contributions of TNRT are highlighted by a color of red.}
\label{figapp:uvevnwith}
\end{figure}


\begin{figure*}[h]
    \centering
    \includegraphics[clip,width=18cm]{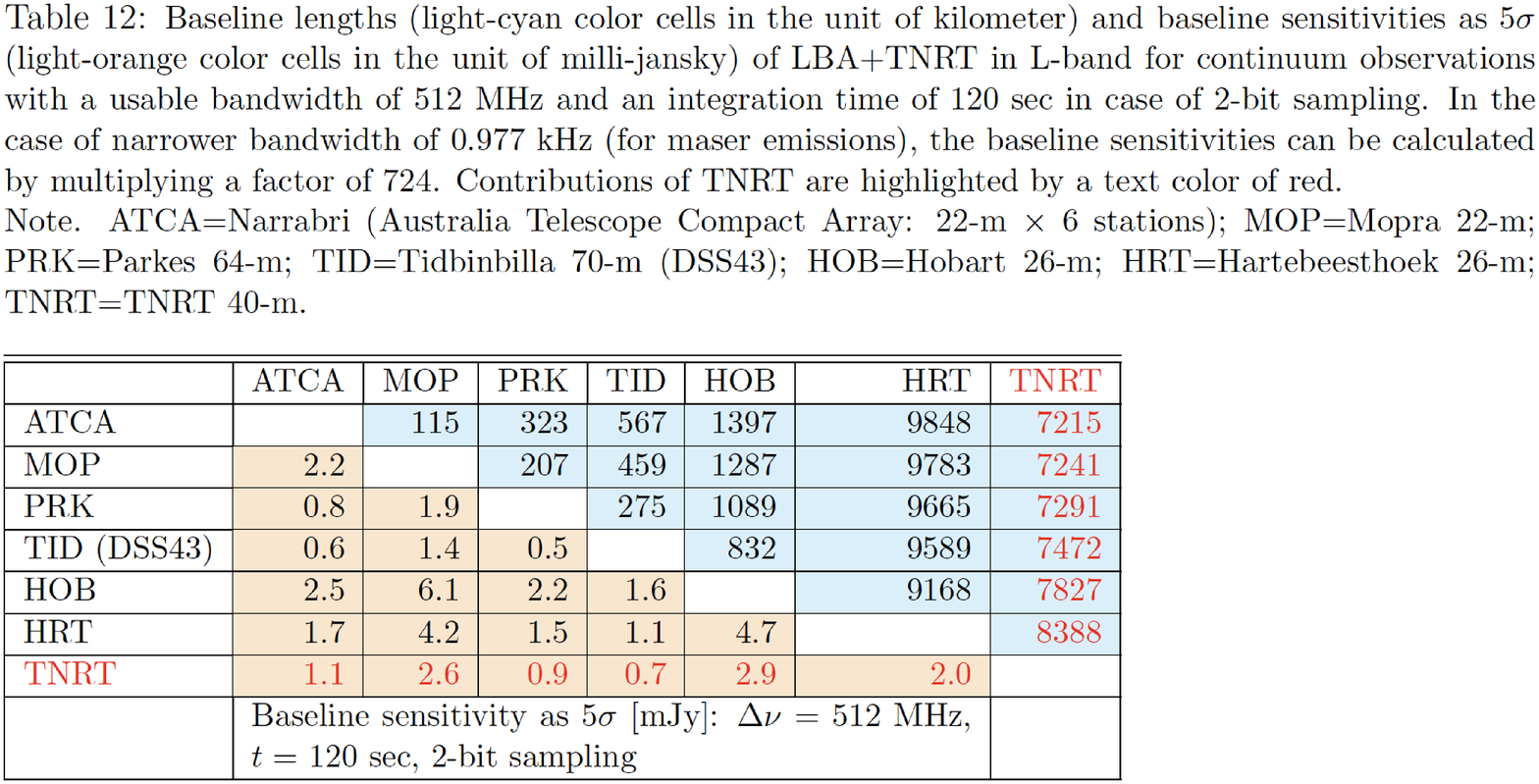}
\end{figure*}


\begin{figure*}[h]
    \centering
    \includegraphics[clip,width=18cm]{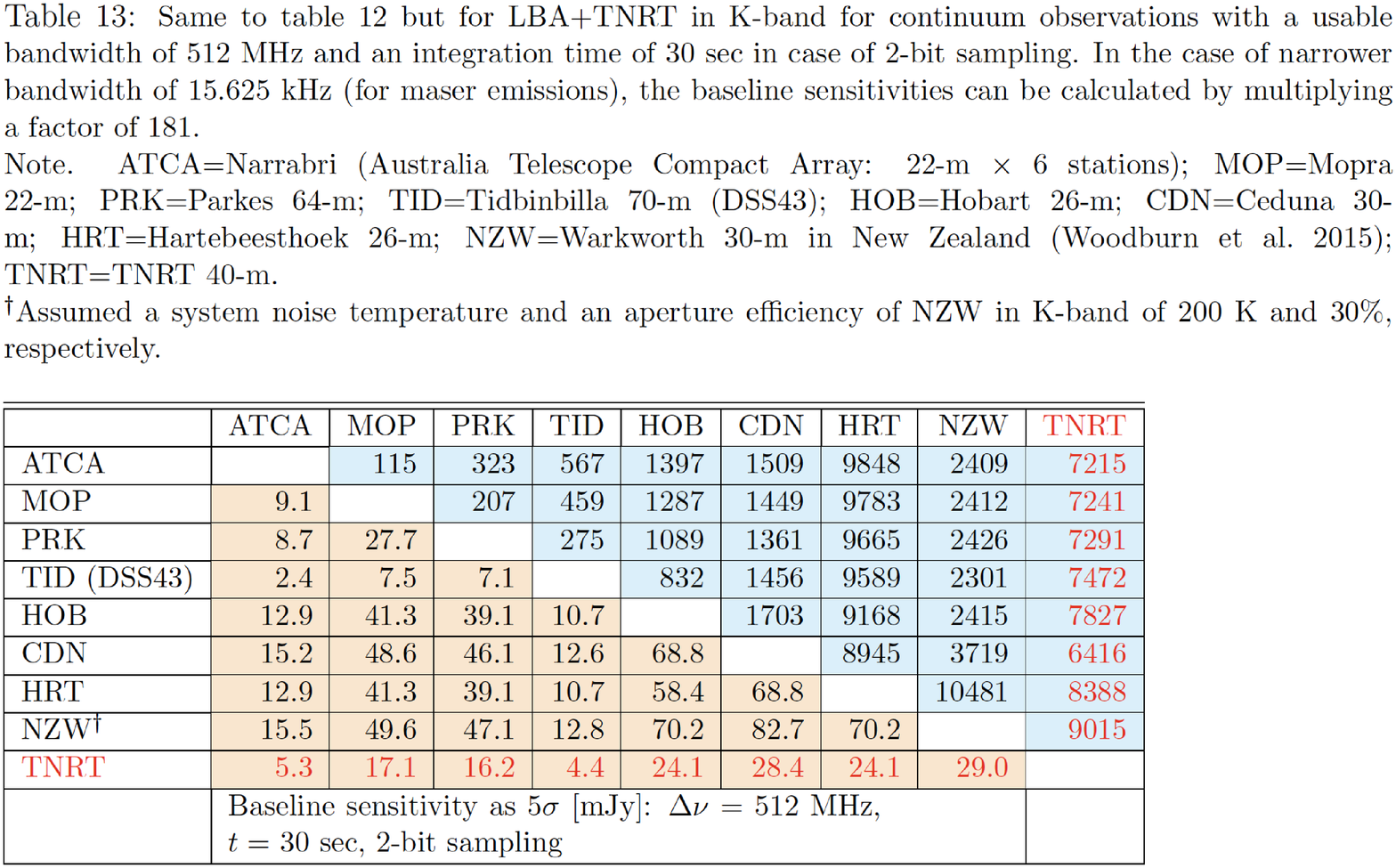}
\end{figure*}


\begin{figure*}[h]
    \flushleft
    \includegraphics[clip,width=24cm,angle=90]{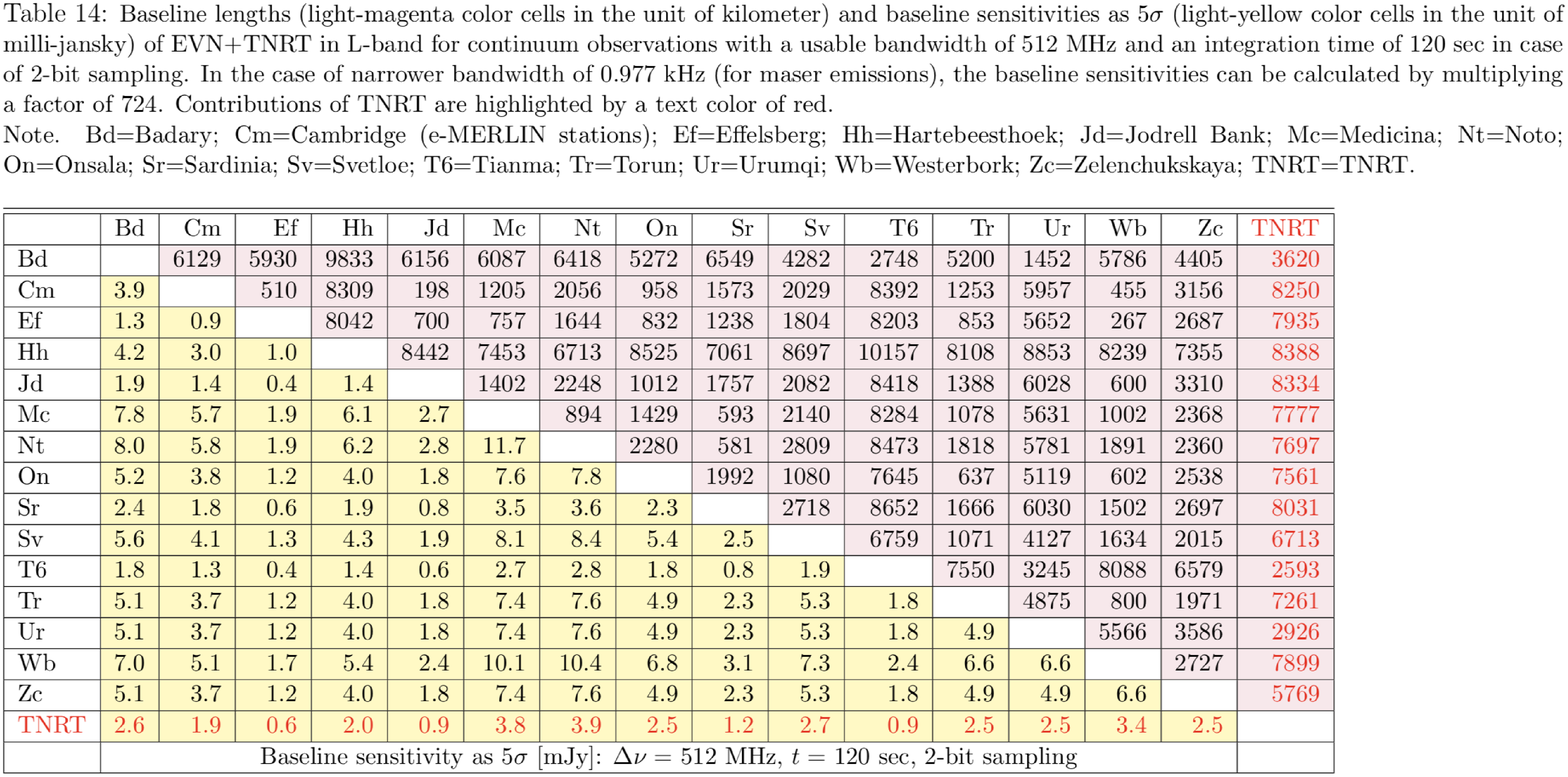}
\end{figure*}


\begin{figure*}[h]
    \flushleft
    \includegraphics[clip,width=24cm,angle=90]{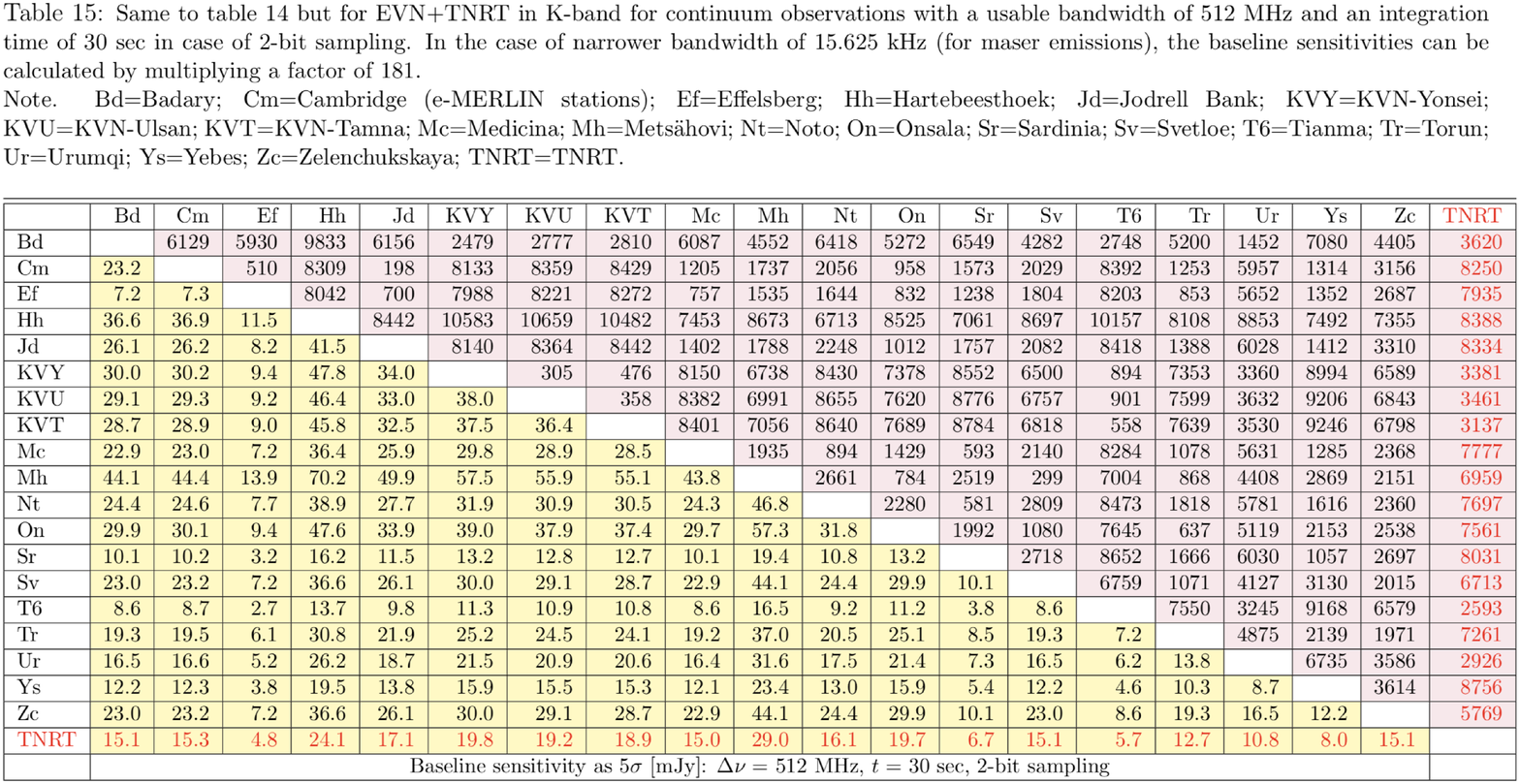}
\end{figure*}

\clearpage
 \bibliography{main}

\end{document}